\documentclass{99-Styles/MICE}
\usepackage{textcomp}
\usepackage{lineno}
\usepackage{bibentry}
\nobibliography*
\usepackage{times}
\usepackage{bm}         
\usepackage{amsmath,amssymb,bm,slashed} 
\usepackage{amssymb}    
\usepackage{graphicx}   
\usepackage{verbatim}   
\usepackage{color}      
\definecolor{RedViolet}{cmyk}{0.60, 0.99, 0.99, 0.0}
\definecolor{BlueViolet}{cmyk}{0.99, 0.99, 0.0, 0.1}
\definecolor{DarkBlue}{cmyk}{0.90, 0.65, 0.0, 0.0}
\usepackage{subfigure}  
\usepackage{hyperref}   
\usepackage{enumerate}

\usepackage{fmtcount}   

\usepackage[square,comma,numbers,sort&compress]{natbib}

\usepackage{color}
\usepackage{ulem}
\definecolor{BlueViolet}{cmyk}{0.90, 0.90, 0.0, 0.0}
\usepackage[defaultlines=2,all]{nowidow}
\usepackage[T1]{fontenc}

\begin{document}

\thispagestyle{empty}

\begin{tabular}{p{0.175\textwidth} p{0.5\textwidth} p{0.225\textwidth}}
  \hspace{-0.8cm}\leftline{\today}                                 &
  \centering{ The ICFA Neutrino Panel}                             &
  \rightline{Final (revision 1)} 
\end{tabular}
\vspace{-1.0cm}\\
\rule{\textwidth}{0.43pt}

\begin{center}
  {\bf
    {\LARGE Roadmap for the international, accelerator-based 
    neutrino programme} \\
  }
  \vspace{0.4cm}
  The ICFA Neutrino Panel \\
  \vspace{-0.2cm}
\end{center}

\makeatletter

\newcommand{\bra}[1]{\ensuremath{\langle #1 |}}   
\newcommand{\ket}[1]{\ensuremath{| #1 \rangle}}   
\newcommand{\bigbra}[1]{\ensuremath{\big\langle #1 \big|}}   
\newcommand{\bigket}[1]{\ensuremath{\big| #1 \big\rangle}}   
\newcommand{\amp}[3]{\ensuremath{\left\langle #1 \,\left|\, #2%
                     \,\right|\, #3 \right\rangle}}  
\newcommand{\sprod}[2]{\ensuremath{\left\langle #1 |%
                     #2 \right\rangle}}  
\newcommand{\ev}[1]{\ensuremath{\left\langle #1 %
                     \right\rangle}} 
\newcommand{\ds}[1]{\ensuremath{\! \frac{d^3#1}{(2\pi)^3 %
                     \sqrt{2 E_\vec{#1}}} \,}} 
\newcommand{\dst}[1]{\ensuremath{\! %
                     \frac{d^4#1}{(2\pi)^4} \,}} 
\newcommand{\tr}{\text{tr}}
\newcommand{\sgn}{\text{sgn}}
\newcommand{\diag}{\text{diag}}
\newcommand{\BR}{\text{BR}}

\renewcommand{\vec}[1]{{\mathbf{#1}}}
\renewcommand{\Re}{{\text{Re}}}
\renewcommand{\Im}{{\text{Im}}}
\newcommand{\iso}[2]{{\ensuremath{{}^{#2}}\ensuremath{\rm #1}}}
\newcommand{\eps}{{\ensuremath{\epsilon}}}
\newcommand{\draftnote}[1]{{\bf\color{red} \MakeUppercase{#1}}}
\newcommand{\panm}[1]{{\color{blue} #1}}
\providecommand{\abs}[1]{\lvert#1\rvert}
\providecommand{\norm}[1]{\lVert#1\rVert}
\newcommand{\gsim}      {\mbox{\raisebox{-0.4ex}{$\;\stackrel{>}{\scriptstyle \sim}\;$}}}
\newcommand{\lsim}      {\mbox{\raisebox{-0.4ex}{$\;\stackrel{<}{\scriptstyle \sim}\;$}}}
\def\simge{\mathrel{%
   \rlap{\raise 0.511ex \hbox{$>$}}{\lower 0.511ex \hbox{$\sim$}}}}
\def\simle{\mathrel{
   \rlap{\raise 0.511ex \hbox{$<$}}{\lower 0.511ex \hbox{$\sim$}}}}

\def\parenbar{\mathpalette\p@renb@r}
\def\p@renb@r#1#2{\vbox{%
  \ifx#1\scriptscriptstyle \dimen@.7em\dimen@ii.2em\else
  \ifx#1\scriptstyle \dimen@.8em\dimen@ii.25em\else
  \dimen@1em\dimen@ii.4em\fi\fi \offinterlineskip
  \ialign{\hfill##\hfill\cr
    \vbox{\hrule width\dimen@ii}\cr
    \noalign{\vskip-.3ex}%
    \hbox to\dimen@{$\mathchar300\hfil\mathchar301$}\cr
    \noalign{\vskip-.3ex}%
    $#1#2$\cr}}}

%
\providecommand{\anmne}{\mbox{$\bar\nu_{\mu} \rightarrow \bar\nu_e$}} 
\providecommand{\nmne}{\mbox{$\nu_{\mu}\rightarrow\nu_e$}} 
\providecommand{\anm}{\mbox{$\bar\nu_\mu$}} 
\providecommand{\nm}{\mbox{$\nu_\mu$}}
\providecommand{\nue}{\mbox{$\nu_e$}} 
\providecommand{\ane}{\mbox{$\bar\nu_e$}} 
\providecommand{\enu}{\mbox{$E_\nu$}}
\providecommand{\piz}{\mbox{$\pi^0 $}}
\providecommand{\pip}{\mbox{$\pi^+$}} 
\providecommand{\pim}{\mbox{$\pi^-$}}

\parindent 10pt
\pagestyle{plain}
\pagenumbering{arabic}                   
\setcounter{page}{1}

\newcounter{nuPanel-RM-Conc-Sect}
\newcounter{nuPanel-RM-Conc-Sect-Conc}[nuPanel-RM-Conc-Sect]
\newcounter{nuPanel-RM-Conc-Sect-Rec}[nuPanel-RM-Conc-Sect]
\newcounter{nuPanel-RM-Conc-Sect-Dec}[nuPanel-RM-Conc-Sect]

\begin{quotation}

\section*{Overview}

The neutrino, with its tiny mass and large mixings, offers a window on
physics beyond the Standard Model.
Precise measurements made using terrestrial and astrophysical sources
are required to understand the nature of the neutrino, to elucidate
the phenomena that give rise to its unique properties and to determine
its impact on the evolution of the Universe.  
Accelerator-driven sources of neutrinos will play a critical role in
determining its unique properties since such sources provide the only
means by which neutrino and anti-neutrino transitions between all
three neutrino flavours can be studied precisely.

In line with its terms of reference \cite{ICFAnuPanel:ToR:2013} the
ICFA Neutrino Panel \cite{ICFA:nuPanelWWWSite} has developed a roadmap
for the international, accelerator-based neutrino programme.
A ``roadmap discussion document'' \cite{ICFA:nuPanel:2016:01} was
presented in May 2016 taking into account the peer-group-consultation
described in the Panel's initial report \cite{Cao:2014zra}.
The ``roadmap discussion document'' was used to solicit feedback from
the neutrino community---and more broadly, the particle- and
astroparticle-physics communities---and the various stakeholders in
the programme.
The roadmap, the conclusions and recommendations presented in this
document are consistent with the conclusions drawn in
\cite{Cao:2014zra} and take into account the comments received
following the publication of the roadmap discussion document.

With its roadmap the Panel documents the approved objectives and
milestones of the experiments that are presently in operation or under
construction.  
Approval, construction and exploitation milestones are presented
for experiments that are being considered for approval.
The timetable proposed by the proponents is presented for experiments
that are not yet being considered formally for approval.
Based on this information, the evolution of the precision with which
the critical parameters governing the neutrino are known has been
evaluated.
Branch or decision points have been identified based on the
anticipated evolution in precision.
The branch or decision points have in turn been used to identify
desirable timelines for the neutrino-nucleus cross section and
hadro-production measurements that are required to maximise the
integrated scientific output of the programme.
The branch points have also been used to identify the timeline for the
R\&D required to take the programme beyond the horizon of the next
generation of experiments.
The theory and phenomenology programme, including nuclear theory,
required to ensure that maximum benefit is derived from the
experimental programme is also discussed.

\end{quotation}

\cleardoublepage
\tableofcontents
\cleardoublepage
\graphicspath{{01-Introduction/Figures/}}

\section{Introduction}
\label{Sect:Intro}

The Standard Model (SM) gives a precise, quantitative description of
the fundamental constituents of matter and the forces through which
they interact.
The study of the properties and interactions of the neutrino has been
seminal in the development of the electroweak theory and decisive in
the development of the quark-parton model and quantum
chromodynamics. 
Recently, the discovery of neutrino oscillations, which implies that
neutrinos have mass and that the flavour eigenstates mix, has led to
the realisation that the SM is incomplete.

Measurements of the parameters that govern neutrino oscillations will
have a profound impact on our understanding of particle physics,
astrophysics and cosmology.
Such a breadth of impact justifies a far-reaching experimental
programme that exploits terrestrial and astrophysical sources of
neutrino.
Accelerator-based measurements of neutrino oscillations are an
essential component of the programme since they are the only means by
which each of the possible appearance channels can be studied with
sufficient precision.

In 2013 the International Committee for Future Accelerators
(ICFA)~\cite{ICFA:WWWSite} established a Neutrino
Panel~\cite{ICFA:nuPanelWWWSite} with the mandate to ``\textit{promote
international cooperation in the development of the accelerator-based
neutrino-oscillation program and to promote international
collaboration in the development of a neutrino factory as a future
intense source of neutrinos for particle physics
experiments}~\cite{ICFAnuPanel:Mandate:2013}. 
In its Initial Report the Panel outlined an ambitious programme by
which the discovery potential of the accelerator-based neutrino
programme could be optimised \cite{Cao:2014zra}.
Greater international cooperation was identified as being key to the
successful delivery of this programme.
The establishment of the Deep Underground Neutrino Experiment (DUNE)
\cite{DUNE:2015,Acciarri:2015uup}, the Short Baseline Programme (SBN)
\cite{SBN:2014} and the Long Baseline Neutrino Facility (LBNF) at the
Fermi National Accelerator Laboratory (FNAL) \cite{LBNF:2014} and the
CERN Neutrino Platform \cite{CENF:2015} as international facilities
for the advancement of the field represents  substantial progress in
the necessary internationalisation of the programme.
The Panel notes the recent developments in the consideration of the
international Hyper-K 
programme~\cite{Abe:2011ts,Abe:2015zbg,Hyper-Kamiokande:2016dsw}.
Together, the complementary long-baseline experiments DUNE and
Hyper-K \cite{Cao:2015ita}, the SBN programme and the CERN Neutrino
Platform will provide the basis for a robust discovery programme.

To complete our understanding of neutrino oscillations it will be
necessary to determine \cite{Cao:2014zra}:
\begin{itemize}
  \item Whether mixing among the three neutrino flavours violates the
    matter-antimatter (CP) symmetry.
    Such leptonic CP-invariance violation (CPiV) would be something
    new and might have cosmological consequences;
  \item The ordering of the three neutrino mass eigenstates is
    and what is the absolute neutrino-mass scale;
  \item Whether empirical relationships between neutrino-mixing
    parameters, or between neutrino- and quark-mixing parameters, can
    be established and whether the neutrino is its own antiparticle;
    and
  \item Whether the few measurements of neutrino oscillations that are
    not readily accommodated within the elegant framework of
    three-neutrino mixing are statistical fluctuations, systematic
    effects or indications that there is even more to discover.
\end{itemize}
By addressing these issues it may be possible to develop a theory that
can explain why neutrino masses are so tiny, at least a million times
smaller than any other known matter particle, and why the strength of
mixing among the neutrino flavours is so much stronger than the mixing
among the quarks.

The purpose of this document is to present concisely the elements of
the global accelerator-based neutrino programme in such a way that
branch or decision points can be identified.
The accelerator-based experiments that are in operation or that are
being planned will make substantial contributions to this programme
and have the potential to discover leptonic CP-invariance violation,
determine the mass hierarchy and, perhaps, find evidence for sterile
neutrinos.
Improvements in accelerator and detector techniques will be required
to take the programme forward, for example, to seek to establish
empirical relationships between neutrino- and quark-mixing
parameters.
Therefore, the accelerator and detector R\&D programmes are considered
alongside the neutrino-experiment programme.
Consideration of the accelerator-based programme as a whole will allow
choices to be made that exploit regional strengths and ambitions to
optimise the discovery potential.
In this way the impact of each contribution on the global programme
will be maximised.

The non-accelerator-based neutrino-physics programme is vibrant and
plays a crucial role in determining the properties of the neutrino.
The Neutrino Panel's terms of reference~\cite{ICFAnuPanel:ToR:2013}
restrict the scope of its recommendations to the accelerator-based
programme.

\subsection{The roadmap}
\label{SubSect:Intro:Roadmap}

A vibrant programme that is able to attract the interest of
researchers and the support of funding agencies and laboratories must:
\begin{itemize}
  \item Have both discovery potential and deliver critical
    measurements in the short term ($< 5$\,years) and in the medium
    term (between 5 and 15 years); and
  \item Develop the capabilities required to build on and go beyond the
    performance of the near- and medium-term experiments through
    appropriately-resourced detector and accelerator R\&D programmes.
\end{itemize}

By preparing this roadmap, the Panel seeks to identify an
accelerator-based programme that satisfies these imperatives.
In the short term (less than five years), experiments such as
T2K \cite{T2K:WWW}, NO$\nu$A \cite{NOvA:WWW}, MicroBooNE 
\cite{MiCroBooNE:WWW} and MINER$\nu$A \cite{MINERvA:WWW} will provide
a steady stream of results.
Over the five- to fifteen-year timescale, the medium-term programme
will seek evidence for CPiV by exploiting the Deep Underground
Neutrino Experiment (DUNE) in the USA and the Hyper-K experiment in
Japan.
The branch or decision points identified in the analysis of the
roadmap are intended to facilitate the discussions necessary to
maximise the scientific output of, and technological benefit from, the
global investment in the accelerator-based neutrino programme.

The categories into which the neutrino programme has been broken down
are:
\begin{enumerate}
  \item The study of neutrino oscillations; 
  \item Searches for sterile neutrinos; 
  \item The supporting programme that includes: 
    \begin{enumerate}
      \item The study of hadro-production and neutrino-nucleus
        scattering necessary to allow the neutrino flux and
        interaction rates to be estimated with the requisite
        precision; and 
      \item The detector, accelerator and software R\&D that supports
        the current programme and builds capability for the discovery
        programme in the medium to long term;
    \end{enumerate}
  \item Experiments that use neutrinos produced by nuclear reactors;
  \item Experiments that exploit non-terrestrial sources 
    and radio-active sources; and 
  \item The non-oscillation programme. 
\end{enumerate}
Categories 1, 2 and 3 constitute the accelerator-based neutrino
discovery and measurement programmes.
Categories 4 and 5 are included as the results from these programmes
may impact the accelerator-based programme.
The needs of the discovery and measurement programmes were considered
when analysing the timetable for the programme of supporting
measurements and R\&D (category 3).
The objectives and approved or proposed timetable is reported for each
of the projects considered together with an indication of the number
of scientists engaged in the execution or development of the activity.

The Panel's terms of reference \cite{ICFAnuPanel:ToR:2013} relate to
the accelerator-based neutrino programme.
In line with these terms of reference, the Panel provides conclusions
and recommendations on: the accelerator-based long-baseline
neutrino-oscillation programme; sterile-neutrino searches with
accelerator-based experiments; and the supporting programme.
For completeness, sections \ref{Sect:ReactorSource}
to \ref{Sect:NonOsc} contain brief summaries of the
non-accelerator-based programme.
The Panel's conclusions and recommendations are summarised in
section~\ref{App:Conc-n-rec}.

\subsection{The international accelerator-based neutrino-physics
  community}
\label{SubSect:Intro:Survey}

It is of interest to consider the strength of the neutrino-physics
community that exploits accelerator-generated neutrino beams.
Surveys of the particle-physics community have been carried out on a
national or regional basis 
\cite{RECFA:Survey:2010,STFC:PPAP:RoadMap:2015}.
No consistent survey of the international accelerator-based neutrino
community exists, making it necessary to draw information from a
number of sources to give an indication of the size of the community.

The European Committee on Future Accelerators (ECFA) \cite{ECFA:WWW}
carries out a regular survey of the particle-physics community in
Europe.
The most recent \cite{RECFA:Survey:2010}, published in 2010, was based
on data collected in 2009.
The ECFA survey counted personnel, rather than ``full-time
equivalents''; only persons spending 20\% or more of their time on
particle-physics activities were counted, each such person was counted
with a weight of one.
The fraction of research time allocated by a particular individual to a
particular project was then recorded.
Personnel were classified as permanent staff, time-limited
researchers, PhD students and engineers with a University degree.
With these definitions, $\sim 1000$ members (approximately 10\%) of
the European particle-physics community were engaged in neutrino
physics.
Of these, $\sim 40\%$ were involved in the accelerator-based
programme.
The 2009 survey is now somewhat outdated and ECFA is considering
repeating the survey in preparation for the European Strategy update
that is planned for 2019/20.

\begin{figure}
  \begin{center}                           
    \includegraphics[width=0.70\textwidth]{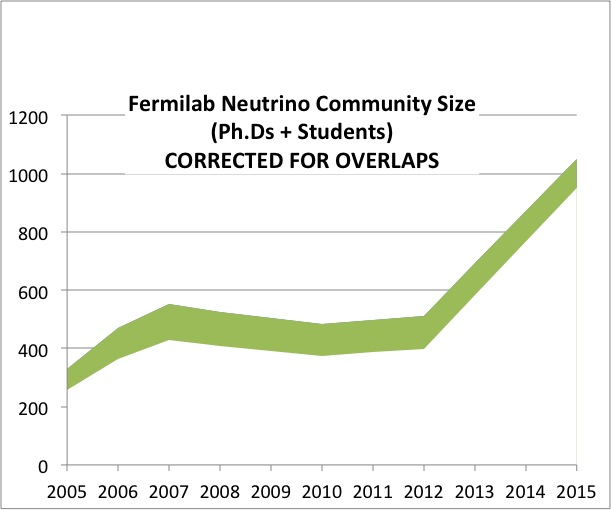}
  \end{center}
  \caption{
    The number of PhD-holding researchers and graduate students
    engaged in neutrino experiments based at Fermi National
    Accelerator Laboratory.
    The data has been corrected for ``overlaps''; cases in which an
    individual is listed on more than one experiment.
    The uncertainty on the overlap-correction is indicated by the
    shaded band.
    The data were taken from the archives of the Fermi National
    Accelerator Laboratory.
  }
  \label{Fig:Intro:FNALnuComm}
\end{figure}

The number of researchers involved in neutrino experiments based at
FNAL is shown in figure \ref{Fig:Intro:FNALnuComm}. 
Researchers holding a PhD and graduate students are included in the
count; all such researchers who are engaged in the programme are
included in the count, regardless of their affiliation.
The data has been drawn from the archives held at FNAL.
The raw number of researchers and graduate students has been corrected
for the effect of ``overlaps''; cases in which an individual is
recorded against more than one experiment.
The size of this correction is $~20\%$.
An estimate of the uncertainty arising from the overlap correction is
also shown.
The data indicate that the FNAL-based neutrino community grew by 
$\sim 55\%$ from 2005 to 2007.
The community was stable during the period when the Tevatron
$p\bar{p}$ collider was in operation (2007--2012).
The change of emphasis of the Laboratory in recent years is clearly
seen in the steady growth in the accelerator-based neutrino community
supported by FNAL.

In Canada, a survey of particle physics activities was conducted in
the context of a long-range planning exercise \cite{Geesaman:2015fha}
held in 2015.
According to this survey, approximately 35 scientists and graduate
students participate in accelerator-based long-baseline neutrino
physics through the T2K and Hyper-K collaborations.
An additional 70 scientists and graduate students are members of other
neutrino efforts, such as EXO~\cite{EXO:WWW},
IceCube~\cite{IceCube:WWW}, HALO~\cite{Zuber:2015ita} and
SNO+~\cite{SNOplus:WWW}.  

\begin{figure}
 \begin{center}
   \includegraphics[width=0.95\textwidth]{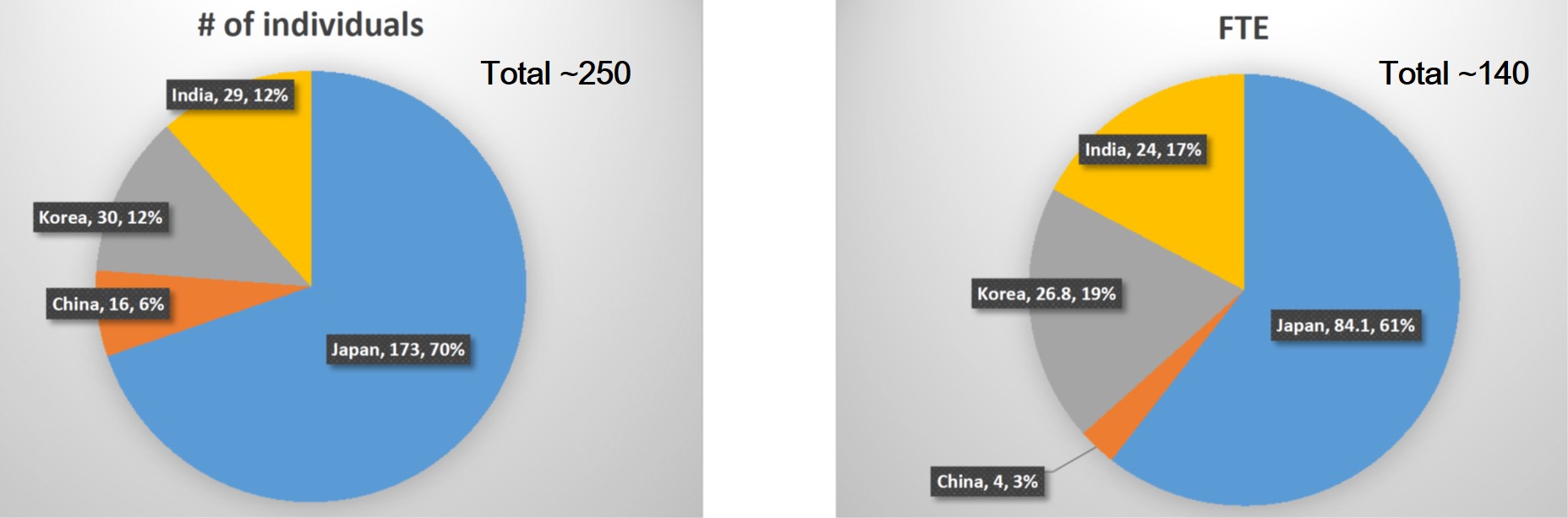}
 \end{center}
 \caption{
   The number of Asian researchers (staff, post-doc and students)
   engaged in accelerator-based neutrino experiments around the world.
   (a) number of individuals counted with a weight of one if the
   individual is working on an accelerator-based experiment with any
   fraction of their time.
   (b) full time equivalent (FTE) number of persons working on
   accelerator-based neutrino experiments.
 }
 \label{Fig:Intro:AsiaComm}
\end{figure}

For Asia, there had been no survey of the size of community for
accelerator-based neutrino experiments.
As a first trial, Asian members of the Panel compiled the number of
researchers (staff members, post-docs, and students) employed by Asian
institutes who are working on accelerator-based neutrino experiments
around the world; experiments such as T2K, Hyper-K,
MINOS \cite{MINOS:WWW}, DUNE, etc. were included in the survey.
Two statistics were prepared: (a) individuals were counted with weight
one if they were working on an accelerator-based experiment; the count
did not take into account the fraction of time spent by an individual
on an accelerator-based neutrino experiment; and (b) the full-time
equivalent effort invested in accelerator-based neutrino experiments.
The results of the survey are $\sim 250$ and $\sim 140$ for (a) and
(b) respectively.
The distribution of the number of individuals engaged in the
accelerator-based programme and the effort invested is shown for the
Asian countries taking part in the survey in figure 
\ref{Fig:Intro:AsiaComm}.

\subsection{Conclusions and recommendations}
\label{SubSect:Intro:Conc-n-rec}

\stepcounter{nuPanel-RM-Conc-Sect}

\noindent
\framebox[\textwidth][l]{
  \parbox[c]{0.98\linewidth}{
      \parbox[c]{0.98\linewidth}{
    \begin{description}
      \stepcounter{nuPanel-RM-Conc-Sect-Conc}
      \item[\arabic{nuPanel-RM-Conc-Sect}.\arabic{nuPanel-RM-Conc-Sect-Conc}:]
        The neutrino has a tiny mass, much smaller than any other
        fundamental fermion, and its type, or ``flavour'', changes as
        it propagates through space and time.
        These properties imply the existence of new phenomena not
        described by the Standard Model of particle physics and may
        have profound consequences for our understanding of the
        Universe. 
        The tiny neutrino mass seems likely to be related to phenomena
        that occur at very high energy scales, well beyond the reach
        of the present or proposed colliding-beam facilities.
        The study of the neutrino is therefore the study of physics
        beyond the Standard Model and a fundamentally important
        component of the particle-physics programme.
      \stepcounter{nuPanel-RM-Conc-Sect-Conc}
      \item[\arabic{nuPanel-RM-Conc-Sect}.\arabic{nuPanel-RM-Conc-Sect-Conc}:]
        The accelerator-based neutrino programme is global in
        scope, engagement and intellectual contribution.
        Continued and enhanced cooperation in a coherent global
        programme will maximise the impact of each individual
        contribution and of the programme as a whole.
        \begin{description}
          \stepcounter{nuPanel-RM-Conc-Sect-Rec}
          \item[\color{BlueViolet} Recommendation \arabic{nuPanel-RM-Conc-Sect}.\arabic{nuPanel-RM-Conc-Sect-Rec}:]
            \textbf{\color{BlueViolet} The present roadmap document
              should be revised and updated at appropriate intervals.
            }
        \end{description}
    \end{description}
  }

      \parbox[c]{0.98\linewidth}{
    \begin{description}
       \stepcounter{nuPanel-RM-Conc-Sect-Conc}
      \item[\arabic{nuPanel-RM-Conc-Sect}.\arabic{nuPanel-RM-Conc-Sect-Conc}:]
        By collating data from a number of sources the Panel has
        gained a partial understanding of the strength of the global
        accelerator-based neutrino community.
        Accurate, up-to-date, consistent and complete census data
        for the global accelerator-based neutrino community will be
        valuable in planning the development of the programme.
        \begin{description}
          \stepcounter{nuPanel-RM-Conc-Sect-Rec}
          \item[\color{BlueViolet} Recommendation \arabic{nuPanel-RM-Conc-Sect}.\arabic{nuPanel-RM-Conc-Sect-Rec}:]
            \textbf{\color{BlueViolet} ICFA should support the Panel
              in its efforts to work with the stakeholders to gather
              the necessary census data as part of the consultation
              process that will follow the completion of this roadmap
              discussion document. 
            }
        \end{description}
     \stepcounter{nuPanel-RM-Conc-Sect-Conc}
      \item[\arabic{nuPanel-RM-Conc-Sect}.\arabic{nuPanel-RM-Conc-Sect-Conc}:]
        Experiments that exploit extra-terrestrial sources of
        neutrinos, neutrinos produced by nuclear reactors and
        radio-active decays all have a fundamentally important role to
        play in the development of a complete understanding of
        neutrino physics. 
        The Panel's terms of reference restricted its considerations
        to the accelerator-based neutrino programme.
        During the period of consultation that followed the
        publication of the roadmap discussion document a compelling
        case was made that a future consultation or roadmapping
        process should encompass the field of neutrino physics as a
        whole.
        This case was made by members of the neutrino-physics
        community and by stakeholders in the programme.
        The Panel agrees that a holistic approach should be developed
        and welcomes the positive response it has received in initial
        discussions of the way forward.
    \end{description}
  }

  }
}

\cleardoublepage
\graphicspath{{02-Acc-nuOsc/Figures/}}

\section{Accelerator-based long-baseline neutrino-oscillation programme}
\label{Sect:AccBasedOsc}

Neutrino oscillations, in which a neutrino created in an eigenstate of
flavour $\alpha$ is detected in flavour state $\beta$, may readily
be described in terms of the mixing of three mass eigenstates,
$\nu_i$, $i=1,2,3$ \cite{Pontecorvo:1957qd,Maki:1962mu}.
The probability for the transition $\nu_\alpha \rightarrow \nu_\beta$
in vacuum is given by \cite{Agashe:2014kda}:
\begin{equation}
  P(\nu_\alpha \rightarrow \nu_\beta) = 
  \sum_{i,j} U_{\alpha i} U^*_{\beta i} U^*_{\alpha j} U_{\beta j}
  \exp \left[
       -i \frac{\Delta m_{ji}^2}{2} \frac{L}{E}
       \right] \; ;
\end{equation}
where $E$ is the neutrino energy, $L$ is the distance between source
and detector and $\Delta m_{ji}^2 = m^2_j - m^2_i$.
The unitary matrix, $U$, may be parameterised in terms of three mixing
angles, $\theta_{ij}$ and one phase parameter $\delta_{\rm CP}$:
\begin{equation}
  U = 
    \left(
      \begin{array}{c c c}
        1 &  0     & 0      \\
        0 &  c_{23} & s_{23} \\
        0 & -s_{23} & c_{23}
      \end{array}
    \right)
    \left(
      \begin{array}{c c c}
         c_{13}             &  0 & s_{13} e^{-i\delta_{\rm CP}} \\
         0                 &  1 & 0                  \\
        -s_{13} e^{i\delta_{\rm CP}} & 0 & c_{13}
      \end{array}
    \right)
    \left(
      \begin{array}{c c c}
        c_{12} & s_{12} & 0 \\
       -s_{12} & c_{12} & 0 \\
        0     & 0     & 1
      \end{array}
    \right)
  \label{Eq:SnuM}
\end{equation}
where $c_{ij}=\cos\theta_{ij}$ and $s_{ij}=\sin\theta_{ij}$.
The values of the various parameters that define the ``Standard
Neutrino Model'' (S$\nu$M) obtained in fits to the present data
obtained using terrestrial and non-terrestrial sources are given in
table \ref{Tab:LBL:Param} \cite{Agashe:2014kda}.
Possible Majorana phases can not be measured in neutrino-oscillation
experiments and are omitted from equation \ref{Eq:SnuM}.
Values for all of the mixing angles and $\Delta m^2_{21}$ have been
determined.
The magnitude of $\Delta m^2_{32}$ is also known.
The value of the CP-invariance violating phase, $\delta_{\rm CP}$, and the sign
of $\Delta m^2_{32}$ are unknown.

The goals of the future neutrino-oscillation programme are to:
\begin{itemize}
  \item Complete the S$\nu$M:
    \begin{itemize}
      \item Determine the mass hierarchy; and
      \item Search for (and discover?) leptonic CP-invariance
        violation;
    \end{itemize}
  \item Establish the S$\nu$M as correct description of nature:
    \begin{itemize}
      \item Determine precisely $\theta_{23}$ and the degree to which
        it differs from $\pi/4$;
      \item Determine $\theta_{13}$ precisely; and
      \item Determine $\theta_{12}$ precisely;
    \end{itemize}
  \item Search for deviations from the S$\nu$M:
    \begin{itemize}
      \item Provide redundant measurements of sufficient precision to
        test the unitarity of the neutrino-mixing matrix; and
      \item Search for sterile neutrinos and non-standard neutrino
        interactions.
    \end{itemize}
  \item Seek relationships between the parameters of the S$\nu$M or
    between neutrinos and quarks.
\end{itemize}
\begin{table}[t]
  \caption{
    Summary of neutrino-oscillation parameters~\cite{Agashe:2014kda}. 
    The best-fit values and $3\sigma$ allowed ranges of the 3-neutrino
    oscillation parameters, derived from a global fit to the current
    terrestrial and non-terrestrial oscillation data. 
    The definition of $\Delta m^2$ used is: $\Delta m^2 = m^2_{3} -
    (m^2_{2}+m^2_{1})/2$. 
    Thus, $\Delta m^2 > 0$, if $m_1 < m_2 < m_3$, and $\Delta m^2 < 0$
    for $m_3 < m_1 < m_2$. 
  } 
  \label{Tab:LBL:Param}
  \begin{center}
    \begin{tabular}{|c|c|c|}
      \hline
      Parameter & Value ($\pm 1\sigma$) & ($3\sigma$ range) \\
      \hline
      $\sin^2\theta_{12}$ & $0.308 \pm 0.017$ & 0.259 - 0.359\\
      $\sin^2\theta_{23}, \Delta m^2 > 0 $ & $0.437^{+0.033}_{-0.023}$ & 0.374 - 0.628 \\
      $\sin^2\theta_{23},\Delta m^2 < 0  $ & $0.455^{+0.039}_{-0.031}$ & 0.380 - 0.641 \\
      $\sin^2\theta_{13},\Delta m^2 > 0  $ & $0.0234^{+0.0020}_{-0.0019}$ & 0.0176 - 0.0295 \\
      $\sin^2\theta_{13},\Delta m^2 < 0  $ & $0.0240^{+0.0019}_{-0.0022}$ & 0.0178 - 0.0298 \\
      $\Delta m_{21}^2$ & $(7.54^{+0.26}_{-0.22}) \times 10^{-5}$~$\rm eV^2$ & $(6.99 - 8.18)\times 10^{-5}$~$\rm eV^2$   \\
      $|\Delta m^2|, \Delta m^2 > 0$ & $(2.43 \pm 0.06) \times 10^{-3}$~$\rm eV^2$ & $(2.23 - 2.61)\times 10^{-3}$~$\rm eV^2$   \\
      $|\Delta m^2|, \Delta m^2 < 0$ & $(2.38 \pm 0.06) \times 10^{-3}$~$\rm eV^2$ & $(2.19 - 2.56)\times 10^{-3}$~$\rm eV^2$   \\
      sign of $\Delta m_{32}^2$ & unknown &  \\
      $\delta_{\rm CP}$ & unknown &  \\
      \hline
    \end{tabular}
  \end{center}
\end{table}

The accelerator-based neutrino-oscillation experiments presently in
operation are presented in section \ref{Sect:AccBasedOsc:Present} and
the planned next-generation of experiments is presented in
section \ref{Sect:AccBasedOsc:Next}.
Projects that seek to go beyond the performance of the present or
planned experiments are presented in section
\ref{Sect:AccBasedOsc:Future}.

\subsection{Present accelerator-based neutrino-oscillation programme}
\label{Sect:AccBasedOsc:Present}

\subsubsection{T2K}
\label{Sect:AccBasedOsc:Present:T2K}

\begin{figure}
  \begin{center}
    \includegraphics[width=0.46\textwidth]{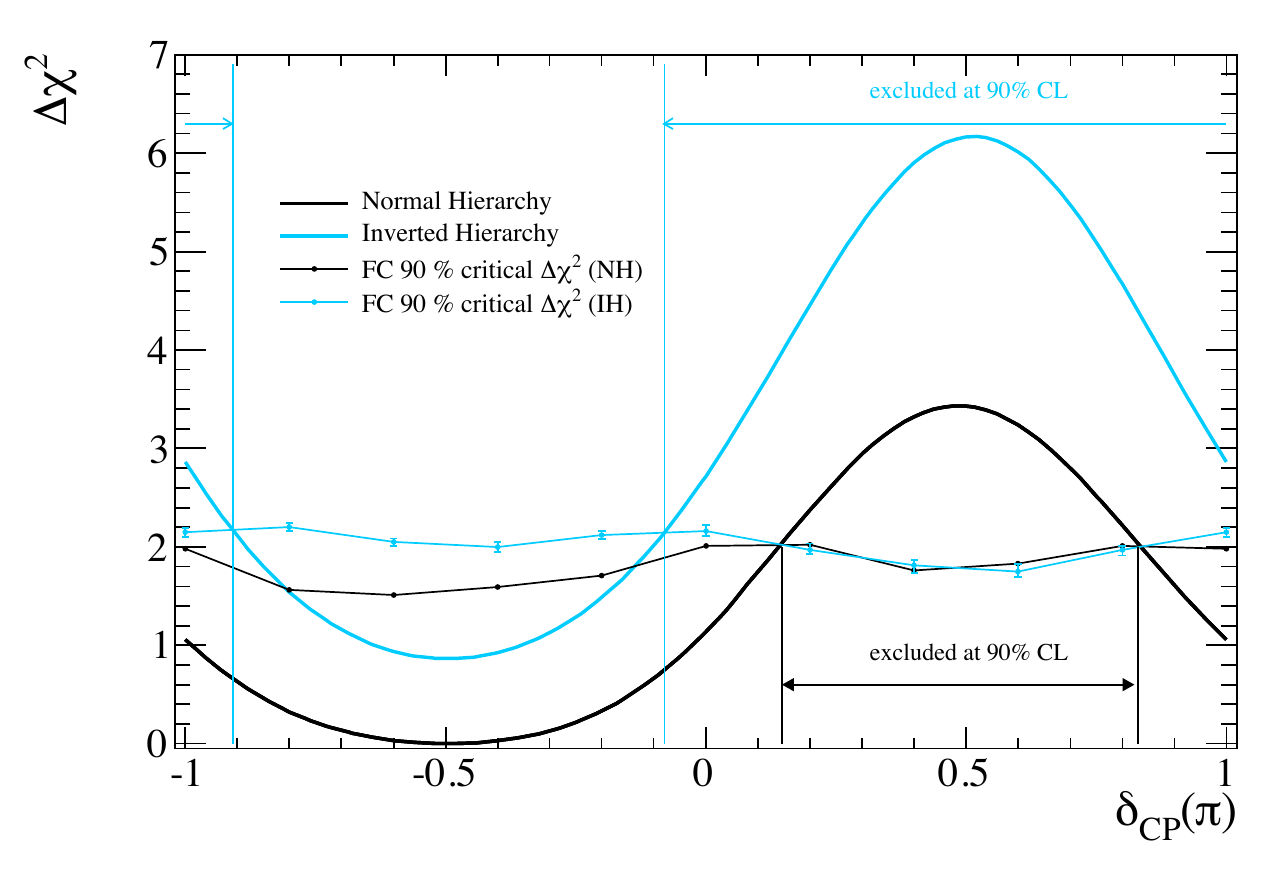}
    \quad \quad
    \includegraphics[width=0.44\textwidth]{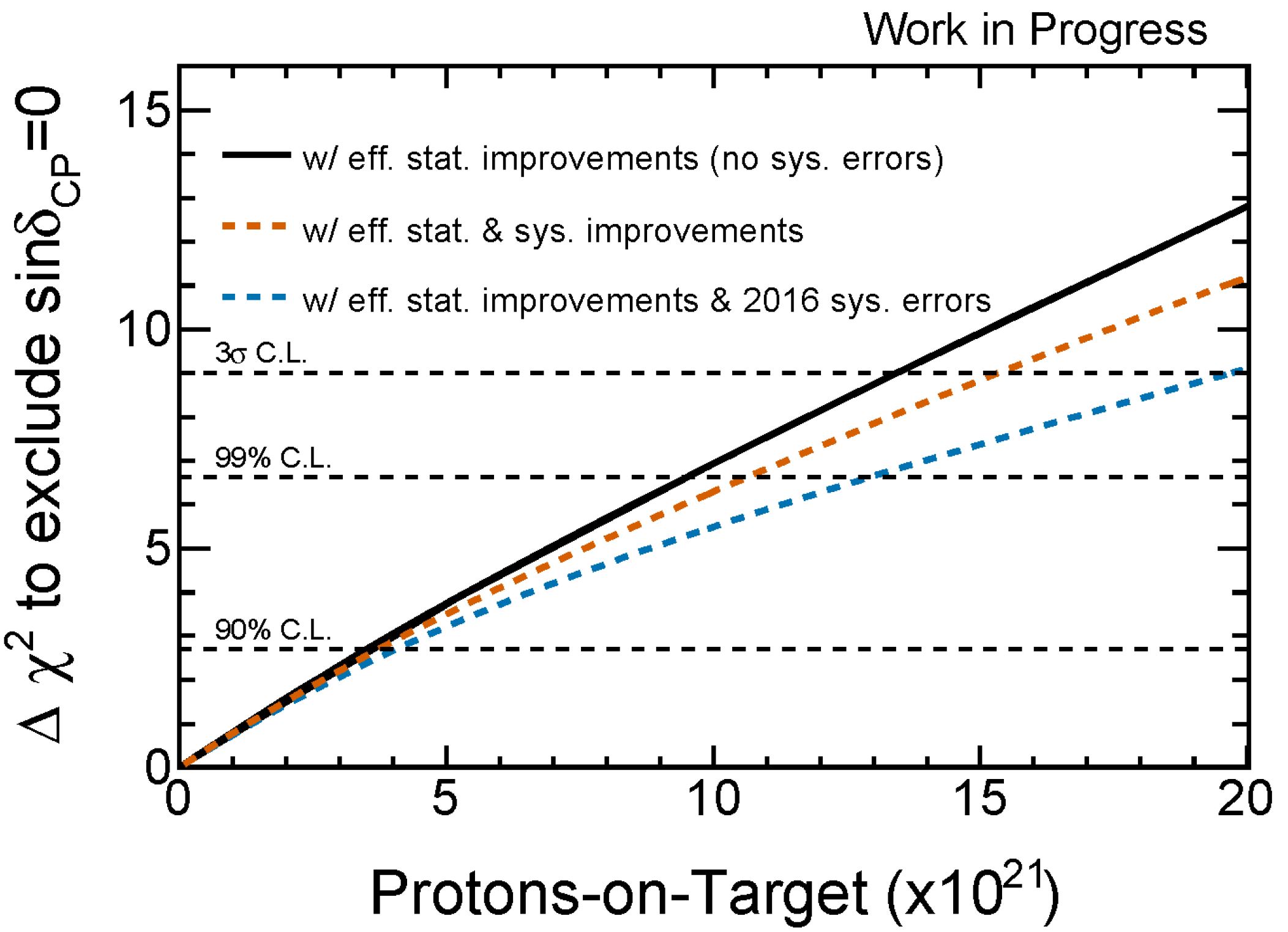}
  \end{center}
  \caption{
    Left panel: Profiled $\Delta \chi^2$ as a function of the CP phase
    ($\delta_{\rm CP}$) for a fit to T2K $\nu_e$ appearance data
    corresponding to $6.6 \times 10^{20}$\,POT combined with the
    reactor measurement of $\theta_{13}$ (taken from 
    \cite{Abe:2015awa}).
    Right panel: Sensitivity to CP violation as a function of POT with
    a 50\% improvement in the effective statistics, assuming the true
    the normal mass hierarchy and the true value of 
    $\delta_{\rm CP} = -\pi/2$ and 
    $\sin^2\theta_{23} = 0.50$ for different assumptions of the
    systematic uncertainties \cite{T2K2-EOI}.
  }
  \label{Fig:T2K:Sens}
\end{figure}

\noindent\textbf{Physics goals} \\
\noindent 
The collaboration uses the off-axis J-PARC neutrino beam, for which
the neutrino-energy distribution peaks at 0.6\,GeV, to collect large
samples in the near (ND280) and far (Super-K) detectors
to~\cite{Abe:2011ks}:
\begin{itemize}
  \item Search for CP-invariance violation using $\parenbar{\nu}_e$
    appearance;
  \item Measure $\theta_{23}$ with high precision using
    $\parenbar{\nu}_\mu$ disappearance;
  \item Make a variety of measurements of neutrino-nucleus
    interactions, to improve neutrino oscillation measurements;
  \item Contribute to the neutrino mass-hierarchy determination; and
  \item Search for non-standard interactions and exotic phenomena.
\end{itemize}
The T2K programme was originally approved for an exposure of
$7.8\times10^{21}$ protons on target (POT).
A data set corresponding to $1.5\times10^{21}$\,POT has been
accumulated.
T2K has published results based on $6.6 \times 10^{20}$\,POT
which, in conjunction with the reactor constraint on $\theta_{13}$,
disfavour values of $\delta_{\rm CP}$ around $\frac{\pi}{2}$ (see
figure \ref{Fig:T2K:Sens}).
A total exposure of $2\times10^{21}$\,POT is projected by the end of
Japanese financial year 2016.
The projected sensitivity to CPiV is shown in figure
\ref{Fig:T2K:Sens} \cite{Abe:2015awa}.
 \\

\noindent\textbf{Institutes 61; collaborators 500} \\
\noindent 
TRIUMF, University of British Columbia, University of Regina, University of Toronto, 
  University of Victoria, University of Winnipeg, York University (Canada);
Institute of Nuclear Physics of Lyon (IPNL), 
  Institute of Research into the Fundamental Laws of the Universe, CEA Saclay, 
  Laboratoire Leprince-Riguet, Ecole Polytechnique (IN2P3), 
  LPNHE, UPMC, Paris, (France); 
RWTH Aachen University (Germany);
INFN Bari and University of Bari, INFN Napoli and Napoli University, 
  INFN Padova and Padova University, INFN Roma and University of Roma ``La Sapienza'' (Italy);
Institute for Cosmic Ray Research (ICRR), Kamioka Observatory, University of Tokyo,
  ICRR, Research Center for Cosmic Neutrino (RCCN), University of Tokyo, 
  Kavli Institute for the Physics and Mathematics of the Universe, University of Tokyo, 
  High Energy Accelerator Research Organization (KEK), Kobe University, Kyoto University, 
  Miyagi University of Education, Okayama University, Osaka City University, 
  Tokyo Metropolitan University, University of Tokyo, Yokohama
  National University (Japan);
Institute for Nuclear Research (Cracow), National Centre for Nuclear Research (Warsaw), 
  University of Silesia (Katowice), Warsaw University of Technology, 
  University of Warsaw, Wroc{\l}aw University, (Poland);
INR (Russia), 
IFAE, Barcelona, UAM Madrid, IFIC , Valencia (Spain), 
ETH Zurich, University of Bern, University of Geneva (Switzerland);
Imperial College London, Oxford University, Queen Mary, University of London,
  Royal Holloway University of London, STFC Daresbury Laboratory, STFC Rutherford Appleton Laboratory, 
  University of Lancaster, University of Liverpool, University of Sheffield, 
  University of Warwick (United Kingdom);
Boston University, Colorado State University, Duke University,
  Louisiana State University, Michigan State University, Stony Brook University, 
  University of California, Irvine, University of Colorado, University of Pittsburgh, 
  University of Rochester, University of Washington (United States). \\

\noindent\textbf{Future programme}

\noindent 
Following the J-PARC Main Ring upgrade, the beam power will be 700\,kW
in 2018, 800\,kW in 2019 and 900\,kW in 2020.
Assuming these beam-power projections and ``five-cycle'' operation
(one ``cycle'' is 22--23 days of operation in a month), T2K will
achieve the approved number of POT goal by around 2021.
The far detector (Super-Kamiokande) will be loaded with gadolinium to
enhance the neutrino-tagging capability at a time to be determined.
Data-taking must be suspended during the detector work needed to
prepared for the gadolinium-loading procedure.
An upgraded near detector is being studied.
A second phase of the experiment, aiming at $>3\sigma$ evidence of
CP-invariance violation when CP is maximally violated by accumulating
$\sim2\times10^{22}$\,POT by around 2026 with the beam power reaching 
1.3\,MW, has been given Stage\,1 Status by the J-PARC Directorate.

\subsubsection{NO\boldmath{$\nu$}A}
\label{Sect:AccBasedOsc:Present:NOvA}

\begin{figure}
  \begin{center}
    \includegraphics[width=0.42\textwidth]{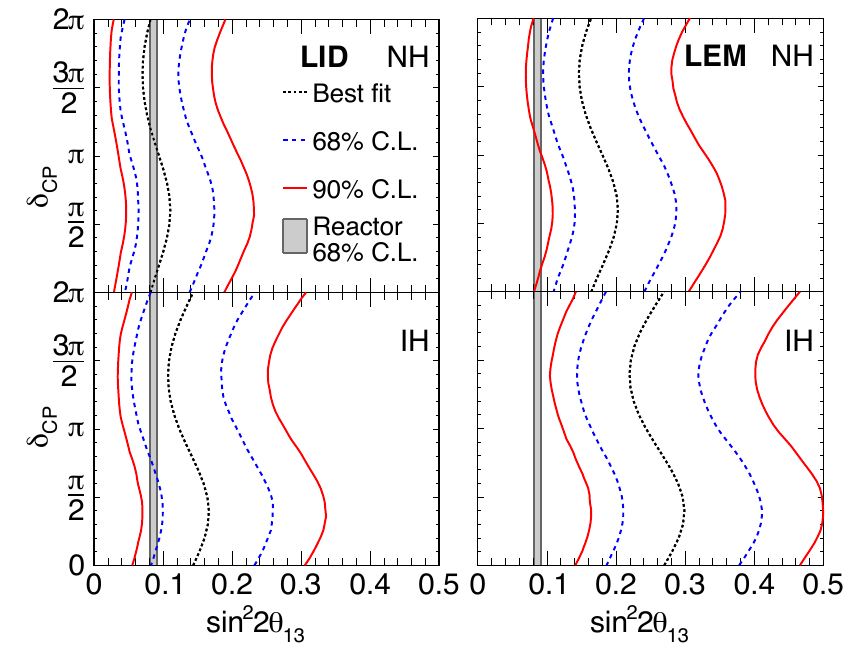}
    \includegraphics[width=0.54\textwidth]{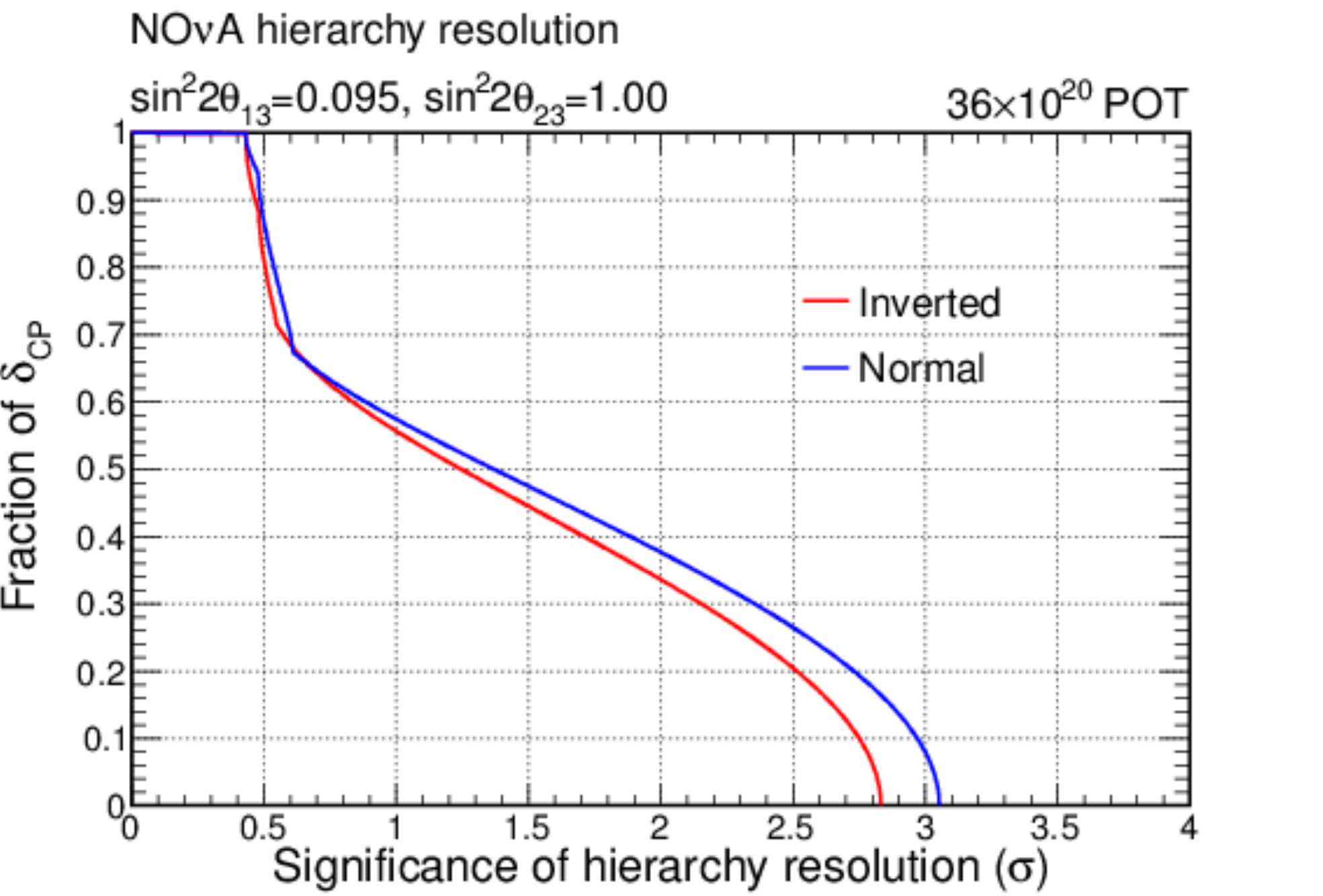}
  \end{center}
  \caption{
    Left panel: Allowed regions in the $\sin^22\theta_{13}, \delta_{\rm CP}$ 
    plane determined using the NO$\nu$A detector and an exposure
    corresponding to $2.74 \times 10^{20}$\,POT;
    14\,kt-equivalent with $\sin^2\theta_{23}$ fixed at
    0.5 \cite{Adamson:2016tbq}.
    Two event-selection algorithms are used: a likelihood-based
    selector, ``LID''; and library event matching, ``LEM''.
    Right panel: NO$\nu$A sensitivity to the mass hierarchy.
    The fraction of all values of $\delta_{\rm CP}$ for which the mass
    hierarchy can be resolved as a function of sensitivity with an
    exposure of $\sim 36 \times 10^{20}$\,POT \cite{Shanahan:LP15}.
  }
  \label{Fig:NOvA:Sens}
\end{figure}

\noindent\textbf{Physics goals} \\
\noindent
The NO$\nu$A detector is located 810\,km from the source of the Main
Injector neutrino beam \cite{Ayres:2004js}.
The off-axis angle of 14\,mrad results in a neutrino-energy spectrum
peaked at $\sim 2$\,GeV.
The collaboration will use the near and far detectors to:
\begin{itemize}
  \item Maximise sensitivity to the neutrino mass hierarchy;
  \item Constrain the value of $\delta_{\rm CP}$;
  \item Resolve the octant of $\theta_{23}$ at better than $1.5\sigma$
    for $80\%$ of all values of $\delta_{\rm CP}$ for either mass
    hierarchy;
  \item Achieve world-leading precision on $\Delta m^2_{32}$ and 
    $\sin^2\theta_{23}$; and
  \item Search for oscillations associated with sterile neutrinos.
\end{itemize}
Using an exposure of $2.74 \times 10^{20}$\,POT the NO$\nu$A
collaboration has isolated a $\nu_e$ appearance signal in the far
detector \cite{Adamson:2016xxw}.
These events have been used to determine allowed regions in the 
$\sin^22\theta_{13}, \delta_{\rm CP}$ plane (see figure 
\ref{Fig:NOvA:Sens}).
A first measurement of muon-neutrino disappearance has also been
made \cite{Adamson:2016tbq}.
The sensitivity to the mass hierarchy with the proposed exposure of
$36 \times 10^{20}$\,POT is shown in figure \ref{Fig:NOvA:Sens}.
 \\

\noindent\textbf{Institutes 41; collaborators 190} \\
\noindent 
Argonne National Laboratory,
Banaras Hindu University, India,
California Institute of Technology, 
Cochin University of Science and Technology, India, 
Institute of Physics of the Academy of Sciences of the Czech Republic,
Charles University in Prague,
University of Cincinnati,
Colorado State University,
Czech Technical University in Prague,
University of Delhi,
Joint Institute For Nuclear Research (Dubna),
Fermi National Accelerator Laboratory,
Universidade Federal de Goias, Brazil,
Indian Institute of Technology (Guwahati),
Harvard University,
Indian Institute of Technology (Hyderabad),
University of Hyderabad,
Indiana University,
Iowa State University,
University of Jammu, India,
Lebedev Physical Institute (Moscow),
Michigan State University,
University of Minnesota (Duluth),
University of Minnesota (Twin Cities),
The Institute for Nuclear Research (Moscow),
Panjab University,
University of South Carolina (Columbia),
South Dakota School of Mines and Technology,
Southern Methodist University (Dallas),
Stanford University,
Universidad del Atl\'antico, Barranqilla, Colombia,
University College London,
University of Sussex,
University of Tennessee (Knoxville),
University of Texas (Austin),
Tufts University,
University of Virginia (Charlottesville),
Wichita State University,
Winona State University,
The College of William \& Mary. \\

\noindent\textbf{Future programme} \\
\noindent
Results from a data set approximately twice that presented
in~\cite{Adamson:2016tbq,Adamson:2016xxw} was available in the summer
of 2016.
The power delivered by the Main Injector proton beam to the neutrino
target will increase to 700\,kW during operations in 2016.
The exact timetable for increasing the proton-beam power will be
determined by the level of losses in the Recycler.
It is likely that NO$\nu$A will switch to anti-neutrino running in
2017; the precise timetable for the change in polarity will be
determined by the evolution of the proton-beam power.
An exposure of $36\times 10^{20}$\,POT is expected to be delivered by
US fiscal year 2023.

\subsection{Next generation accelerator-based neutrino-oscillation
  programme} 
\label{Sect:AccBasedOsc:Next}

Efforts to deliver the large, high-precision data sets required to
observe CP-invariance violation, measure $\delta_{\rm CP}$, determine
the mass hierarchy and measure the neutrino-mixing parameters with a
precision that significantly exceeds that which will be achieved by
T2K and NO$\nu$A have coalesced around the Deep Underground Neutrino
Experiment (DUNE) served by the FNAL Long Baseline Neutrino Facility
(LBNF) and the Hyper-K experiment served by the J-PARC neutrino beam.
The physics program at DUNE and Hyper-K also includes neutrino
astrophysics and the search for proton decay.

The primary physics goals and projected timescales for the two
experiments are similar.
The complementarity of the two experiments \cite{Cao:2015ita} rests on
key differences in their specification:
\begin{description}
  \item{{\it Baseline}:} Hyper-K will be sited 295\,km from J-PARC,
    while DUNE will be located 1300\,km from FNAL.
    With these baselines, the energy at which the first oscillation
    maximum occurs is different; $\sim 600$\,MeV for Hyper-K and 
    $\sim 3$\,GeV for DUNE; and
  \item{{\it Neutrino energy spectrum}:} Hyper-K will be located at an
    off-axis angle of $2.5^\circ$, yielding a narrow neutrino-energy
    spectrum peaked at $\sim 600$\,MeV with a high
    signal-to-background ratio in the critical
    $\parenbar{\nu}_\mu\to\parenbar{\nu}_e$ channel.
    DUNE will be located on axis so that the beam with which it will
    be illuminated will have a broad energy spectrum, peaked at $\sim
    3$\,GeV, which will allow the second oscillation maximum to be
    studied.
\end{description}
Matter effects in the long-baseline programme at Hyper-K will be
small and neutrino-oscillation effects such as asymmetries in the
neutrino and anti-neutrino oscillations will be dominated by ``vacuum''
effects such as CP-invariance violation. 
Matter effects will be significant for DUNE, allowing a detailed study
of related phenomena and the resolution of the mass hierarchy. 
The deep underground location of both experiments permits detailed
studies of atmospheric neutrinos to be made over a large range of
energies and baselines.
The study of the atmospheric-neutrino sample is a complementary probe
of the oscillation physics.

The Hyper-K and DUNE detectors are each designed to give optimal
performance given the beam with which they will be illuminated.
Hyper-K will use a water Cherenkov detector since the technique is
proven, scaleable and cost effective for the detection of neutrino
interactions at $\sim 1$ GeV where low multiplicity channels such as
quasi-elastic and resonant single-pion production dominate. 
The high granularity and fine tracking capabilities of the
liquid-argon time-projection chamber (LAr-TPC) technology used at DUNE
will allow the reconstruction of the more complex events resulting
from neutrino interactions $\gsim 2$\,GeV. 

A review of experiments that exploit non-terrestrial sources of
neutrinos is given in section \ref{Sect:NonTerre} and a summary of
non-oscillation physics related to, or carried out using detectors
illuminated by, accelerator-based neutrino sources is given in
section \ref{Sect:NonOsc}.
Here it is interesting to note that the DUNE and Hyper-K detectors are
also complementary in their sensitivity to proton decay and
supernova-burst neutrinos.
For example, in proton decay, the best sensitivity for many modes is
achieved by a large water Cherenkov detector, however, for some key
modes such as $p\to K^++\bar{\nu}_\mu$, a LAr-TPC has the potential to
deliver nearly background-free samples due to its ability to detect
the complete final state.
Likewise, in the study of supernova-burst neutrinos, a LAr-TPC will
single out $\nu_e$ through the process 
$\nu_e+\;^{40}\mbox{Ar} \to e^- + \;^{40}\mbox{K}^*$, while a water
Cherenkov detector will primarily observe $\bar{\nu}_e$ through the
inverse $\beta$ decay process on the free protons in the water.
A megaton-scale water Cherenkov detector such as Hyper-K also has the
potential to extend the reach for observing supernov\ae~to the Mpc
scale, thereby including supernov\ae~in nearby galaxies such as M31 
(Andromeda). 

\subsubsection{Deep Underground Neutrino Experiment}
\label{Sect:AccBasedOsc:Next:DUNE}

\begin{figure}
\begin{center}
\includegraphics[width=0.40\textwidth]{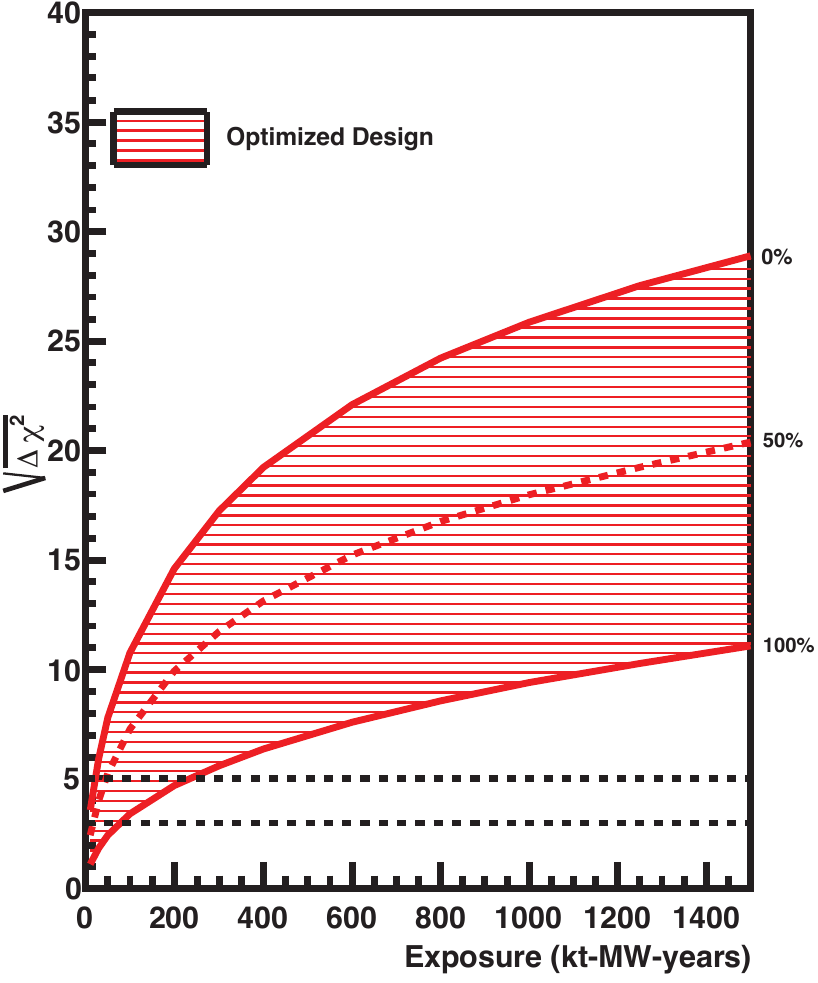}
\quad \quad \quad \quad
\includegraphics[width=0.388\textwidth]{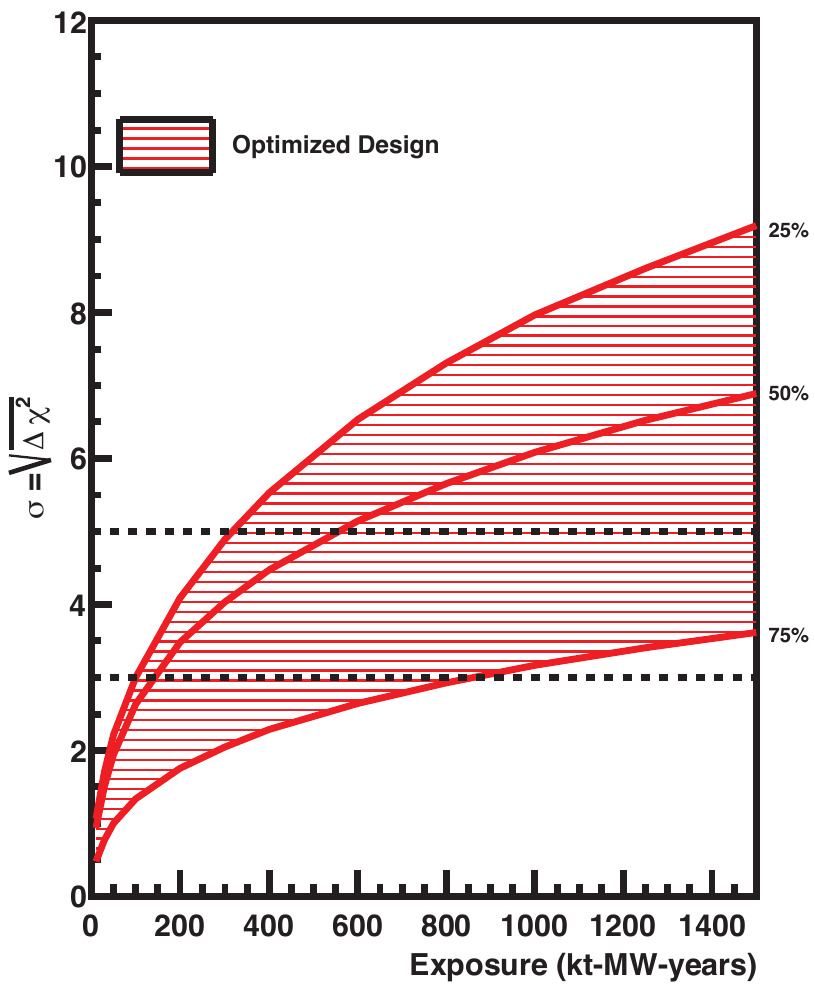}
\end{center}
\caption{
Left panel: DUNE sensitivity to the mass hierarchy: the minimum
significance with which the mass hierarchy can be determined is
plotted as a function of exposure for all values of the CP phase
($\delta_{\rm CP}$ (100\%), 50\% and in the most optimistic
scenario of maximum CPiV (0\%).
Right panel: DUNE sensitivity to CPiV: assuming the normal mass
hierarchy, the minimum significance with which CP violation can be
determined for 25\%, 50\% and 75\% of all values of
$\delta_{\rm CP}$ is plotted as a function of exposure.
In each plot, the results projected for the optimised design of
the LBNF neutrino beam are presented.
Taken from \cite{Acciarri:2015uup}.
}
\label{Fig:DUNE:Sens}
\end{figure}

\noindent\textbf{Physics goals/measurement programme} \\
\noindent 
The Deep Underground Neutrino Experiment (DUNE) has successfully
completed the ``CD1 refresh'' required by the DOE ``Critical
Decision'' process. Following a CD3a review in 2015, LBNF
has received the authorisation to start construction in the
US President Budget for FY2017.
The headlines of the collaboration's physics programme are
\cite{Acciarri:2015uup}:
\begin{itemize}
  \item CP-invariance violation in neutrino oscillations:
    \begin{itemize}
      \item Evidence for CPiV with a significance of $\geq 3 \sigma$
        for 75\% of all values of $\delta_{\rm CP}$ using an exposure
        of  
        850\,$\mbox{kt} \cdot \mbox{MW} \cdot \mbox{year}$;
      \item Observe CPiV with a significance of $\geq 5 \sigma$ for
        50\% of all values of $\delta_{\rm CP}$ using an exposure of 
        550\,$\mbox{kt} \cdot \mbox{MW} \cdot \mbox{year}$;
      \item Measure $\delta_{\rm CP}$ with a precision better than
        $10^\circ$. 
    \end{itemize}
\item Determination of the neutrino mass hierarchy:
\begin{itemize}
\item Determine the neutrino mass hierarchy 
with $\ge 5 \sigma$ significance for all values of
$\delta_{\rm CP}$ with an exposure of
230\,$\mbox{kt} \cdot \mbox{MW} \cdot\mbox{year}$.
\end{itemize}
\item Precision measurements and tests of the S$\nu$M:
\begin{itemize}
\item Measure $\sin^2\theta_{23}$ with a precision of
$\sim 0.005$, assuming $\sin^2\theta_{23}=0.45$;
\item Determine the $\theta_{23}$ octant with a significance of
3$\sigma$ if $\theta_{23}>48^\circ$ or $\theta_{23}<43^\circ$;
\item Measure $\sin^22\theta_{13}$ with a precision of
$\sim 0.003$ assuming $\sin^2 2\theta_{13}=0.085$; and $\sin^2\theta_{23}=0.45$;
\item Measure $\Delta m^2_{31}$ with a precision of $\sim
0.5\times10^{-5}$\,$\mbox{eV}^2$.
\end{itemize}
\item Search for proton decay with zero backgrounds:
\begin{itemize}
\item Provide 90\% CL sensitivities to $p\to K^+ +\bar{\nu}$ and $n\to K^+ +e^-$
for a nucleon lifetime of $\sim 6\times10^{34}$ years in 10 years.
\item Provide 90\% CL sensitivities to $p\to K^0 +e^+$, $p\to K^0 +\mu^+$ and $n\to \pi^- +e^+$ 
for a nucleon lifetime of $\sim 3\times10^{34}$ years in 10 years.
\end{itemize}
\item Detection of neutrinos from core-collapse supernov\ae:
\begin{itemize}
\item Observe 3--4 thousand neutrino interactions from an
intragalactic core-collapse supernova; and
\item Deliver unique sensitivity to $\nu_e$s from the
core-collapse process, particularly from the neutronisation
process.
\end{itemize}
\end{itemize}
Figure \ref{Fig:DUNE:Sens} shows the projected sensitivity for the
DUNE experiment as a function of exposure \cite{Acciarri:2015uup}.
An exposure of 288\,kt\,MW\,years will be achieved after seven years
running, a 40\,kt detector and a proton beam-power of 1.2\,MW. 
 \\

 \noindent\textbf{Institutes 149; collaborators 848} \\
 \noindent
 ABC Federal University, Brazil,
 APC-Paris,
 University of Alabama (Tuscaloosa),
 Federal University of Alfenas, Brasil,
 Aligarh Muslim University, India,
 Antananarivo University, Madagascar
 Argonne National Laboratory,
 University of Athens,
 Universidad del Atlantico, Kolumbia,
 Banaras Hindu University,
 Bhabha Atomic Research Center,
 University of Bern,
 Boston University,
 Brookhaven National Laboratory
 Institute of Physics CAS,
 CERN,
 Centro de Investigaciones Energeticas, Medioambientales y Tecnologicas (CIEMAT),
 Center for Nanotechnology Innovation (Pisa),
 University of California (Berkeley),
 University of California (Davis),
 University of California (Irvine),
 University of California (Los Angeles),
 California Institute of Technology,
 University of Cambridge,
 University de Campinas,
 University di Catania,
 Charles University,
 University of Chicago,
 University of Cincinnati,
 CINVESTAV, Mexico,
 Universidad de Colima,
 University of Colorado (Boulder)
 Colorado State University,
 Columbia University,
 Cornell University,
 Czech Technical University (Prague),
 Dakota State University,
 University of Delhi,
 Drexel University,
 Duke University,
 University of Durham,
 University Estadual de Feira de Santana, Brasil,
 Fermi National Accelerator Laboratory,
 University Federal de Goias,
 Gran Sasso Science Institute,
 Universidad de Guanajuato,
 Harish-Chandra Research Institute,
 University of Hawaii,
 University of Houston,
 Horia Hulubei National Institute of Physiscs and Nuclear Engineering,
 University of Hyderabad,
 IFAE, Barcelona,
 Illinois Institute of Technology,
 IIT Bombay,
 IIT Gowahani,
 IIT Hyderabad,
 Institute for Nuclear Research,
 Institute for Research in Fundamental Sciences (IPM),
 Idaho State University,
 Imperial College of Science Tech. \& Medicine,
 Indiana University,
 Iowa State University,
 Institute of Nuclear Physics of Lyon (IPNL),
 University of Jammu, India,
 University of Jyvaskyla,
 KTH Royal Institute of Technology,
 Kansas State University,
 Kavli Institute for the Physics and Mathematics of the Universe, University of Tokyo,
 High Energy Accelerator Research Organization (KEK),
 Kyiv National University,
 Koneru Lakshmaiah Education Foundation, India,
 Jagiellonian University (Cracow),
 Annecy-le-Vieux Particle Physics Laboratory,
 Laboratori Nazionali del Gran Sasso,
 Lancaster University,
 Lawrence Berkeley National Laboratory,
 University of Liege,
 University of Liverpool,
 University College London,
 Los Alamos National Laboratory,
 Louisiana State University,
 University of Lucknow, India,
 Madrid Autonoma University,
 University of Manchester,
 University of Maryland,
 Massachusetts Institute of Technology,
 University of Puerto Rico,
 Michigan State University,
 University di Milano, INFN Milano,
 INFN Milano Bicocca,
 University of Minnesota (Twin Cities),
 University of Minnesota (Duluth),
 NIKHEF,
 INFN Napoli,
 National Centre for Nuclear Research,
 Jawaharlal Nehru University,
 University of New Mexico,
 Northern Illinois University,
 Northwestern University,
 University of Notre Dame,
 Observatorio Nacional,
 Ohio State University,
 Oregon State Univ,
 University of Oxford,
 Pacific Northwest National Laboratory,
 Pennsylvania State Univ,
 PUCP,
 INFN Padova,
 Panjab University,
 University of Pavia, INFN Pavia,
 University of Pennsylvania,
 University di Pisa,
 University of Pittsburgh,
 Princeton University,
 Punjab Agricultural University,
 University of Rochester,
 SLAC National Accelerator Laboratory,
 STFC Rutherford Appleton Laboratory,
 CEA/Saclay, IPhT,
 Institute de Physique Theorique,
 University of Sheffield,
 University of Sofia,
 University of South Carolina,
 University of South Dakota,
 South Dakota School of Mines and Technology,
 South Dakota Science and Technology Authority,
 South Dakota State University,
 Southern Methodist University (Dallas),
 Stony Brook University,
 University of Sussex,
 Syracuse University,
 University of Tennessee,
 University of Texas (Arlington),
 University of Texas (Austin),
 TUBITAK Space Technologies Research Institute,
 Tufts University,
 Variable Energy Cyclotron Center,
 Instituto de Fisica Corpuscular (IFIC, Valencia),
 Virginia Tech,
 University of Warsaw,
 University of Warwick,
 Wichita State University,
 College of William \& Mary,
 University of Wisconsin,
 Wroc\l aw University,
 Yale University,
 Yerevan Institute for Theoretical Physics and Modeling,
 York University,
 ETH Z\"urich\\

\noindent\textbf{Future programme} \\
\noindent
Following the successful ``CD1 refresh'', the principal milestones of
the project are:
\begin{itemize}
\item Construction and operation of ProtoDUNE detectors at CERN (2016-2018);
\item Excavation of detector caverns at the Sanford Underground
Research facility (SURF) (2017-2021);
\item Construction of first two 10\,kt far detector modules (2022-2025);
\item Fill and commission first two far detector modules (2024-2026);
\item Install two more 10 kt far detector modules (2024-2027);
\item Fill and commission additional far detector modules (2026-2027);
\item Beam line complete (2026);
\item Near detector complete (2026); and
\item Project early completion in 2027; DOE CD4 (with 40 months contingency) in 2030.
\end{itemize}

\subsubsection{Hyper-Kamiokande} 
\label{Sect:AccBasedOsc:Next:Hyper-K}

\noindent\textbf{Physics goals/measurement programme} \\
An international proto-collaboration has been formed to carry out the
Hyper-K experiment.
The Institute for Cosmic Ray Research of the University of Tokyo
(UTokyo-ICRR) and the Institute of Particle and Nuclear Studies of the
High Energy Accelerator Research Organization (KEK-IPNS) have signed
an MoU affirming cooperation in the Hyper-K project to review and
develop the program in its comprehensive aspects.
The Japanese High-Energy-Physics community sets Hyper-K and ILC
as its top two projects.  The Japanese Cosmic Ray Community (CRC) also
sets Hyper-K as one of top priority projects and recognises 
Hyper-K as the next CRC very large project after the current KAGRA 
construction completes.  It has also been selected as one of 27 high priority
projects among 207 large-scale projects by Science Council of Japan
in ``Master plan 2014.''  

Hyper-K is a large underground water Cherenkov detector. 
The group has succeeded in developing new 50\,cm PMTs with good
single-photon-sensitivity and has re-optimised the detector
configuration.
Hyper-K is built in 2 cylindrical tanks that are 60\,m in depth and
74\,m in diameter, each yielding a fiducial mass of 187\,kt; i.e. the
total fiducial mass will be 374\,kt.
Approximately 80,000 50-cm PMTs are required to give 40\% photo
cathode coverage.
The proto-collaboration aims to realise the two tanks by staging the
construction such that the second tank begins operation six years
after the first.

\begin{figure}
  \begin{center}                           
    \includegraphics[width=0.45\textwidth]{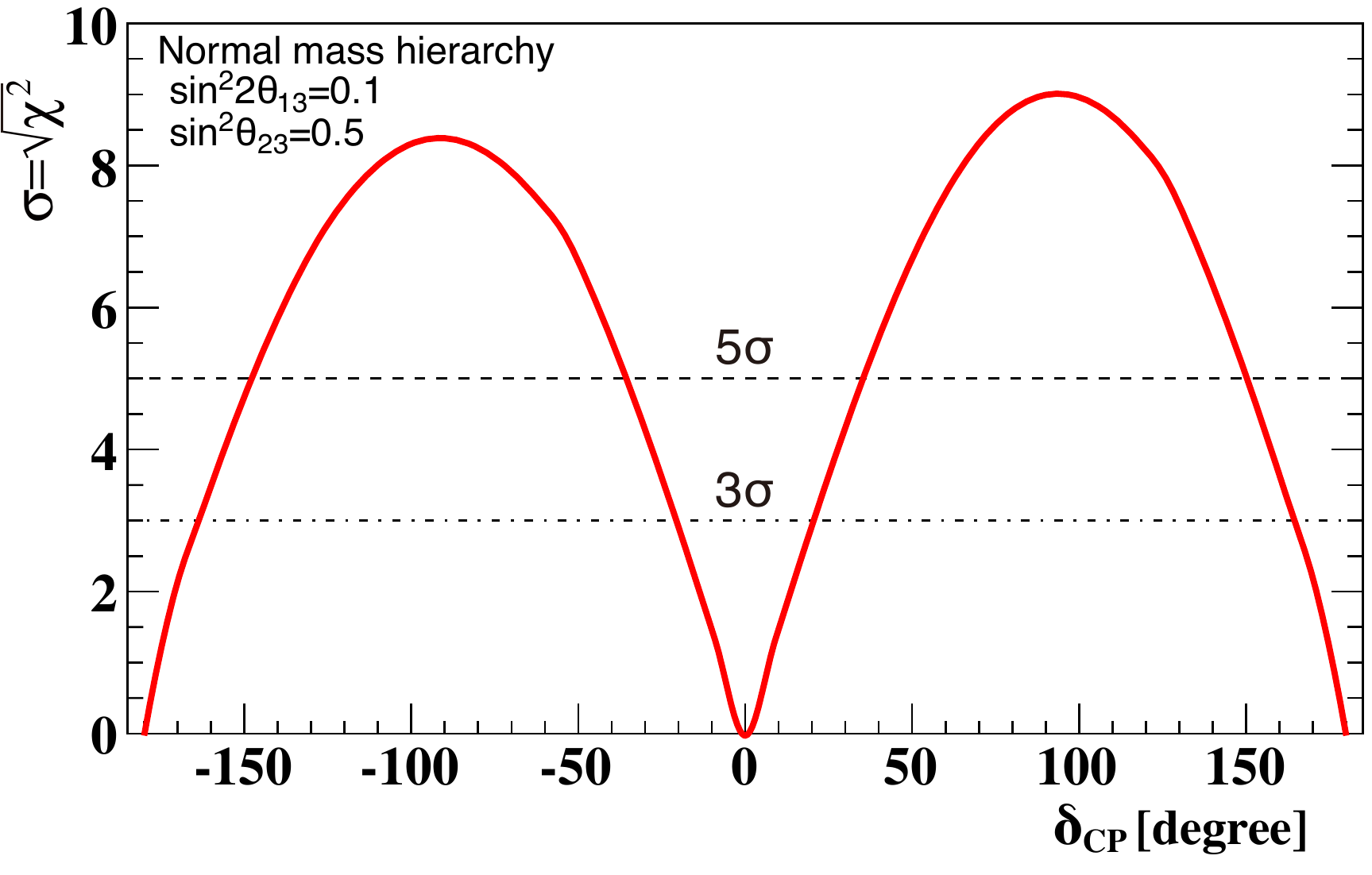}
    \includegraphics[width=0.47\textwidth]{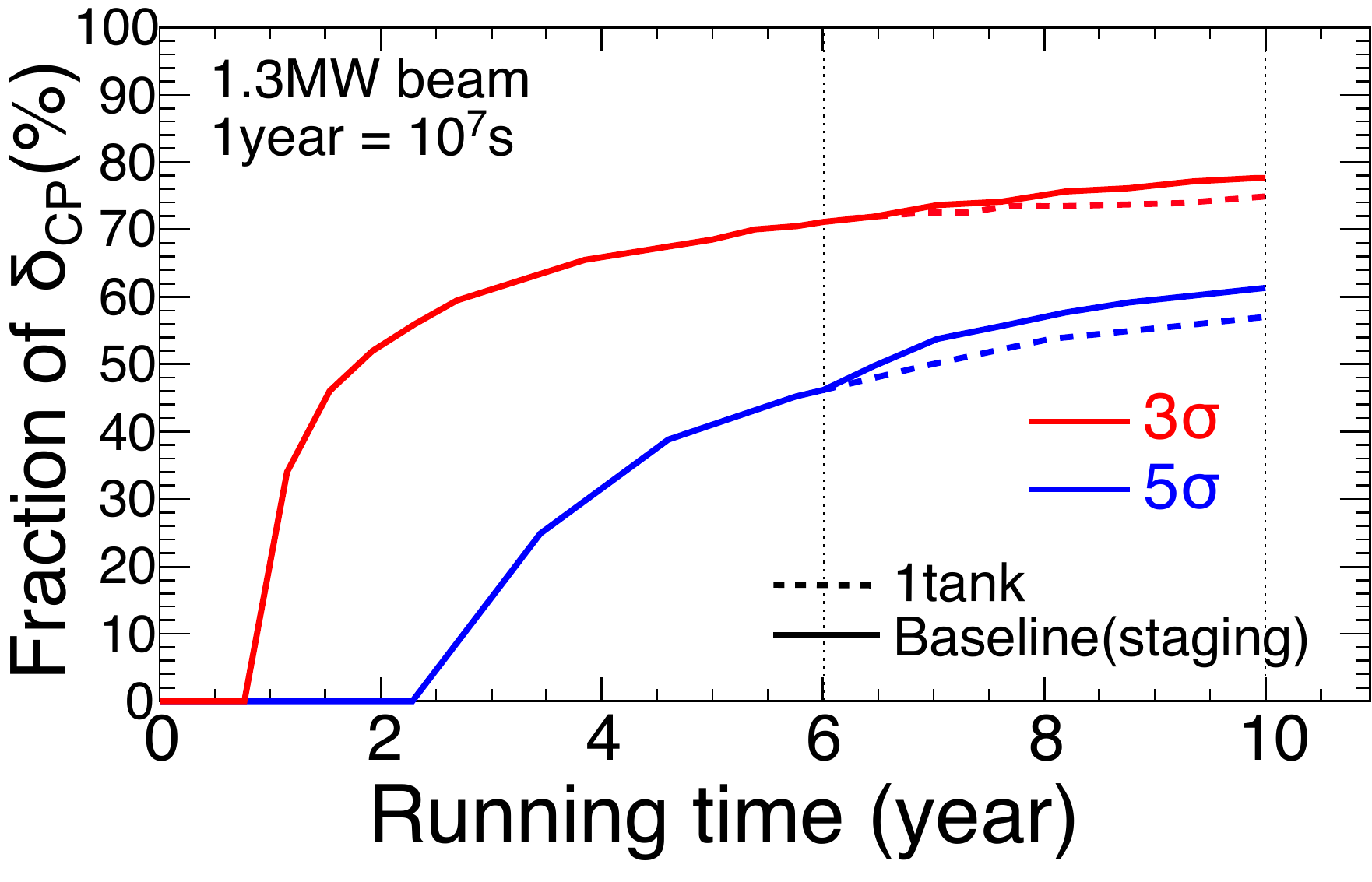}
  \end{center}
  \caption{
    Left panel: Expected significance to exclude $\sin\delta_{\rm CP}=0$ for
    the case of the normal mass hierarchy and assuming 10 years
    running.
    Right panel: fraction of all values of the CPiV phase, $\delta_{\rm
    CP}$, for which $\delta_{\rm CP}=0, \pi$ can be excluded at
    $3\sigma$ (red) or $5\sigma$ (blue).
    An exposure of 13\,MW\,$\times$\,$10^7$\,s is expected to be
    achieved after 10 years of operation.
  }
  \label{Fig:NBB:HKCPV}
\end{figure}

The headlines of the Hyper-K physics programme
are~\cite{Hyper-Kamiokande:2016dsw}:
\begin{itemize}
  \item CP violation in neutrino oscillations: 
    \begin{itemize}
      \item Evidence for CPiV with a significance of $\geq 3 \sigma$
        for 78\% of all values of $\delta_{\rm CP}$ in 10 years
        corresponding to an exposure of $1.3\,\mbox{MW}  \times 
        (0.187\,\mbox{Mt} \times 6 \cdot 10^7\,\mbox{sec} 
        + 0.374\,\mbox{Mt} \times 4 \cdot 10^7\,\mbox{sec})$; 
      \item Observe CPiV with a significance of $\geq 5 \sigma$ for
        62\% of all values of $\delta_{\rm CP}$ by 10 years;
        and
      \item Measure $\delta_{\rm CP}$ with a precision of $7^\circ$
        ($21^\circ$) precision for $\delta_{\rm CP}=0^\circ$
        ($90^\circ$).
    \end{itemize}
  \item Determination of the neutrino mass hierarchy:
    \begin{itemize}
      \item Determine the neutrino mass hierarchy with $>3 \sigma$
          significance after a few years by combination of 
          atmospheric neutrinos and beam data.
    \end{itemize}
  \item Precision measurements and tests of the S$\nu$M:
    \begin{itemize}
      \item Determine the $\theta_{23}$ octant with a significance of
        >3$\sigma$ if $\theta_{23}>49^\circ$ or $\theta_{23}<41^\circ$;
      \item Measure $\sin^2\theta_{23}$ with a precision of
        $\pm0.015$ assuming that $\sin^2 \theta_{23}=0.5$; and
      \item Measure $\Delta m^2_{32}$ with a precision of up to
        $\sim 1.4\times 10^{-5}\,\mbox{eV}^2$. 
    \end{itemize}
  \item Search for proton decay in various decay modes, including:
    \begin{itemize}
      \item Provide 90\% CL sensitivity to $p\to e^+ + \pi^0$, a
        dominant channel in a number of modern grand unified theories,
        for a proton lifetime of $1\times10^{35}$ years; and
      \item Provide 90\% CL sensitivity to $p\to K^+ + \bar{\nu}$, a
        key channel in grand unifies theories that include
        supersymmetry, for a proton lifetime of up to $3\times10^{34}$
        years.
    \end{itemize}
  \item Detection of Solar neutrinos:
    \begin{itemize}
      \item Record a data set corresponding to $>$250 $^8$B $\nu$s
        per day with two tanks;
      \item Measurement of the day/night flux difference with a
        precision of 1\% with one year of data taking;
      \item Make a first measurement of hep neutrinos; and
      \item Provide unique, high-precision measurements of
        $\Delta m^2_{21}$ and $\sin^2\theta_{12}$ using neutrinos
        (rather than anti-neutrinos). 
    \end{itemize}
  \item Detection of neutrinos from core-collapse supernov\ae:
    \begin{itemize}
      \item Observe $\sim130,000$ ($\sim26$) neutrinos with 2 tanks from an
        intragalactic (Andromeda) core-collapse supernova; and
      \item Observe 100 supernova relic neutrinos in 10 years
        with non-zero significance of 4.8$\sigma$.
    \end{itemize}
\end{itemize}
Figure \ref{Fig:NBB:HKCPV} shows the projected sensitivity for the
Hyper-K experiment which is updated from \cite{Abe:2015zbg}.
An exposure of 13\,MW\,$\times$\,$10^7$\,s will be achieved after
ten years assuming 
1:3 ratio between neutrino and anti-neutrino running time, 
one or two 187\,kt detector(s) and a proton beam-power of 1.3\,MW at
30\,GeV. \\

\noindent\textbf{Institutes 70; collaborators 277 (including interested members)} \\
\noindent 
  Boston University, 
  University of British Columbia, 
  University of California (Davis),
  University of California (Irvine), 
  California State University,
  Institute of Research into the Fundamental Laws of the Universe, 
  CEA Saclay, 
  Chonnam National University, 
  Dongshin University, 
  Duke University, 
  Ecole Polytechnique, 
  IN2P3-CNRS,
  University of Edinburgh, 
  University of Geneva, 
  University of Hawaii, 
  Imperial College London, 
  INFN Bari, Bari University,  Politecnico di Bari,
  INFN Napoli, Napoli University,
  INFN Padova, Padova University,
  INFN Roma,
  Institute for Nuclear Research of the Russian Academy of Sciences,
  Iowa State University, 
  High Energy Accelerator Research Organization (KEK),
  Kobe University, Kyoto University, Laboratori Nazionali di Frascati,
  Lancaster University, University of Liverpool, Los Alamos National Laboratory,
  Louisiana State University, University Autonoma Madrid, Michigan State University,
  Miyagi University of Education, Nagoya University, 
  Nagoya University, Kobayashi-Maskawa Institute for the Origin of Particles and the Universe, 
  Nagoya University, Institute for Space-Earth Environmental Research,
  National Centre for Nuclear Research, 
  Okayama University, Osaka City University, Oxford University,
  University of Pittsburgh, University of Regina, 
  Pontificia Universidade Catolica do Rio de Janeiro,
  University of Rochester, Queen Mary University of London,
  Royal Holloway University of London, 
  Universidade de Sao Paulo, University of Sheffield, 
  Seoul National University, Seoyeong University,
Stony Brook University,
  STFC Rutherford Appleton Laboratory, 
  Sungkyunkwan University, South Korea, 
  Research Center for Neutrino Science, 
  Tohoku University,
  University of Tokyo, 
  Earthquake Research Institute,
  University of Tokyo, 
  Institute for Cosmic Ray Research, 
  Kamioka Observatory,
  University of Tokyo, Institute for Cosmic Ray Research,
  Pennsylvania State University,
  Research Center for Cosmic Neutrinos, 
  University of Tokyo, 
  Kavli Institute for the Physics and Mathematics of the Universe (WPI), The University of Tokyo Institutes for Advanced Study,
  Tokyo Institute of Technology, 
  TRIUMF, 
  University of Toronto,
  University of Warsaw, 
  University of Warwick, 
  University of Washington,
  University of Winnipeg, 
  Virginia Tech, 
  Wroc\l aw University,
  Wroc\l aw University of Technology,
  Yerevan Physics Institute,
  York University\\
  
\noindent\textbf{Next steps} \\
\noindent
In 2017, a sub-committee of the Science Council of Japan
selected Hyper-K as one of the important large-scale
projects to be considered for funding.
The next steps in the Hyper-K approval process is a review by the
Japanese Ministry of Education, Culture, Sports, Science and Technology
(MEXT) in 2017.
Assuming this review to be successful, the principal project
milestones include:
\begin{itemize}
  \item Construction of far detector Hyper-Kamiokande (2018-2025); and
  \item Start data taking (2026).
\end{itemize}

\paragraph*{Addendum to the configuration of Hyper-Kamiokande}
An alternative configuration in which one of the large water Cherenkov
counters is located in Korea at a baseline of $\sim$1100\,km to
enhance the sensitivity of the leptonic CP-violation search and to
enhance sensitivity to the neutrino-mass ordering is seriously being
considered~\cite{Abe:2016ero}. 
Studies show that the matter effect combined with sensitivity to the
second oscillation maximum enhances the performance of the
configuration in which one detector is constructed in Korea over that
which would be obtained with both detectors at Kamioka.
The two identical detectors at different baselines allow a ratio
measurement to be made in which correlated uncertainties will cancel.
There are several candidate sites in Korea with substantial ($\sim
1$\,km) rock overburdens at off-axis angles of between $1^\circ$ and
$3^\circ$. 

\subsection{Future opportunities}
\label{Sect:AccBasedOsc:Future}

Substantial advances will be made by the programme described in
sections \ref{Sect:AccBasedOsc:Present} and
\ref{Sect:AccBasedOsc:Next}.
Individually and in combination, these experiments have the potential
to discover leptonic CP-invariance violation, determine the mass
hierarchy and improve the precision with which the parameters of the
S$\nu$M are known.
While much will have been achieved, it seems unlikely that the
precision that will be delivered will be sufficient to over-constrain
the S$\nu$M such that its self-consistency can be tested and the
degree to which three-flavour mixing gives the correct description of
nature established.
Further, precision beyond that which the programme described above can
deliver is likely to be required to establish empirical relationships
between neutrino- and quark-mixing parameters.

A number of experiments, or experimental facilities, have been
proposed to take the programme beyond the precision and sensitivity of
the experiments described in section \ref{Sect:AccBasedOsc:Next}.
These experiments are described below.
In each case, developments in accelerator and/or detector techniques
are required before the experiment or facility can be constructed.
The experiments or facilities will not be considered for construction
before the next-generation experiments are underway and the R\&D is
sufficiently advanced.
The R\&D programme, discussed in sections \ref{SubSect:SptPrg:DetDev}
and \ref{SubSect:SptPrg:AccDev}, should therefore be planned such that
the experiments described below can be considered for implementation
in the second half of the next decade, i.e. around 2025.

\subsubsection{Neutrino Factory}
\label{Sect:AccBasedOsc:Future:NF}

\noindent\textbf{Physics goals} \\
\noindent
The Neutrino Factory, in which a broad-band neutrino beam is produced
from the decay of muons confined within a storage ring, has been shown
to offer unparalleled performance in terms of its sensitivity to CPiV
and the precision with which it can be used to measure parameters of
the S$\nu$M 
\cite{Bandyopadhyay:2007kx,Choubey:2011zzq,Delahaye:2013jla,Bogomilov:2014koa}.
This exquisite performance results from the large flux of
$\parenbar{\nu}_e$ combined with the precise knowledge of the
neutrino-beam flux and energy spectrum.
Once sufficient muons have been transported to the muon-acceleration
system, the stored-muon energy can be tuned to optimise the physics
reach and in response to constraints imposed by detector technology
and source-detector distance.

The scientific potential of the facility may be summarised as
follows \cite{Bandyopadhyay:2007kx,Choubey:2011zzq,Delahaye:2013jla,Bogomilov:2014koa}: 
\begin{itemize}
  \item CP violation in neutrino oscillations: 
    \begin{itemize}
      \item Observe CPiV with $\gsim 5 \sigma$ significance for 80\%
        of all $\delta_{\rm CP}$ values with a ten-year exposure of a
        100\,kt magnetised iron detector and a flux corresponding to
        $10^{21}$ muon decays per year; and 
      \item Measure $\delta_{\rm CP}$ with $\lsim 5^\circ$ precision.
    \end{itemize}
  \item Precision measurements and tests of the three-flavour
    neutrino oscillation paradigm:
    \begin{itemize}
      \item Resolve the $\theta_{23}$ octant with a sensitivity of
        2$\sigma$ if $\theta_{23}>48^\circ$ or $\theta_{23}<43^\circ$;
      \item Measure $\theta_{13}$ with precision of 1.5\% assuming
        $\theta_{13}=9^\circ$; and 
      \item Measure $\Delta m^2_{31}$ with a precision of 0.5\%. \\
    \end{itemize}
\end{itemize}
In addition, the large electron- and muon-neutrino flux at the near
detector, the flux of which may be determined precisely by the
instrumentation of the storage ring, may be used to carry out a
definitive neutrino-nucleus scattering programme and search for
physics beyond the S$\nu$M. \\

\noindent\textbf{Institutes 46} \\
\noindent 
  Harish-Chandra Research Institute, 
  Brookhaven National Laboratory,
  Brunel University (London),
  CERN,
  Institute of Mathematical Sciences (Tamil Nadu), India,
  Institute for Particle Physics Phenomenology (Durham),
  Fermi National Accelerator Laboratory, 
  University of Geneva,
  University of Glasgow,
Max Planck Institute for Nuclear Physics (Heidelberg), 
  Imperial College London,
  Thomas Jefferson National Laboratory, 
  Saha Institute of Nuclear Physics (Kolkata),
  Kyoto University, 
  Research Reactor Institute (Osaka),
  Lawrence Berkeley National Laboratory, 
  University of California (Los Angeles),
  Universidad Autonoma de Madrid,  
  Michigan State University, 
INFN Milano Bicocca,
  University of Mississippi, 
  Institute for Nuclear Research of Russian Academy of Sciences,
  Tata Institute of Fundamental Research,
  Max Planck Institute for Physics (Munich), 
  Muons Inc., 
Napoli University,  
  Northwestern University, 
  Illinois Institute of Technology, 
  Oak Ridge National Laboratory, 
  Osaka University, 
  Oxford University,
Padova University, INFN Padova, 
  Princeton University,
  University of California (Riverside), 
INFN Roma Tre,
  STFC Rutherford Appleton Laboratory, 
  University of Sheffield, 
  University of Sofia, 
  Stony Brook University,
  University of South Carolina, 
  IPHC, 
University of Strasbourg,
  Toyko Metropolitan University,
  Instituto de Fisica Corpuscular (IFIC, Valencia), 
  Centro Mixto CSIC-UVEG,
  Virginia Tech, 
  University of Warwick,
  University of W\"{u}rzburg \\

\noindent\textbf{Next steps} \\
A number of feasibility studies of the Neutrino Factory have been
completed
\cite{:1900cvd,Finley:2000cn,Kuno:2001tb,Ozaki:2001,Apollonio:2008aa,Choubey:2011zzq,Delahaye:2013jla,Bogomilov:2014koa}.
The staged implementation of the Neutrino Factory has been studied by
the International Design Study for the Neutrino Factory
(IDS-NF)~\cite{Choubey:2011zzq} and in the Muon Accelerator Staging
Study (MASS) that was carried out within the US Muon Accelerator
Program (MAP)~\cite{Delahaye:2015yxa}. 
Work on the necessary high-power, pulsed proton beams is underway at
CERN, ESS, FNAL, J-PARC and RAL.
The MERIT experiment \cite{Efthymiopoulos:2008zz} at CERN demonstrated
the principle of the mercury-jet target, while the EMMA
accelerator \cite{Machida:2012zz,Owen:2012zzb} at the Daresbury 
Laboratory demonstrated the principle of the non-scaling Fixed Field
Alternating Gradient (FFAG) technique.
The remaining system demonstration is the proof-of-the principle of the
ionization cooling technique.
This is being carried out by the international Muon Ionization Cooling
Experiment (MICE) collaboration \cite{MICE-WWW} at the STFC Rutherford
Appleton Laboratory; further details are provided in
section~\ref{SubSect:SptPrg:AccDev}. 

The Neutrino Factory offers the potential to deliver sensitivity and
precision beyond those offered by the next generation of experiments
(DUNE and Hyper-K).
The successful demonstration of ionization cooling with MICE will
complete the proof-of-principal, system-level R\&D for the accelerator
facility.
The construction of a muon storage ring to serve a
neutrino-cross-section-measurement and, perhaps, a
sterile-neutrino-search, programme would demonstrate that muon
beams can serve a front-rank neutrino programme, allow experience to
be gained in the operation of such facilities and build the user
community necessary to mount the Neutrino Factory programme.

\subsubsection{DAE\boldmath{$\delta$}ALUS}
\label{Sect:AccBasedOsc:Future:Deadalus}

\noindent\textbf{Physics goals/measurement programme} \\
\noindent
DAE$\delta$ALUS 
\cite{Calabretta:2011nr,Alonso:2010fy,Alonso:2010fs,Scholberg:2012zz,Toups:2013dxa,Toups:2015pma}
will use three high-power proton cyclotrons to
generate neutrino beams from pion decay-at-rest.
It is proposed to place the cyclotrons at baselines of 1.5\,km, 8\,km
and 20\,km from a large liquid-scintillator or gadolinium-doped water
Cherenkov. 
The staged implementation of the programme has been considered.
The goals of the programme are to:
\begin{itemize}
  \item Measure $\delta_{\rm CP}$ with a precision of
    $5^\circ$--$10^\circ$ using the spectrum of $\bar{\nu}_e$ events
    generated by the transition $\bar{\nu}_\mu\to\bar{\nu}_e$; and 
  \item At an intermediate (IsoDAR \cite{Adelmann:2012kq}, see section
    \ref{SubSect:Sterile:SBL:Future:IsoDAR}) stage, search for sterile
    neutrino oscillations using $\nu_e$ from $^8\mbox{Li}$ decays at
    rest and a large scintillator detector $\sim 16$ meters away. \\
  \end{itemize}

\noindent\textbf{Institutes 36} \\
Amherst College,
  Argonne National Laboratory,
  Bartoszek Engineering,
  Black Hills State University,
  University of California,
  Sicilian Center of Nuclear Physics and Physics of Matter,
  The Cockcroft Institute for Accelerator Science \& the University of Manchester,
  Columbia University,
  Duke University,
  Harvard University,
  IBA Research,
  Imperial College London,
  Instituto De Fisica Nucleare,
  Iowa State University,
  Kavli Institute for the Physics and Mathematics of the Universe,
  Lawrence Livermore National Laboratory,
  Los Alamos National Laboratory,
  Massachusetts Institute of Technology,
  Michigan State University,
  New Mexico State University,
  North Carolina State University,
  Northwestern University,
  Paul Scherrer Institut,
  RIKEN, Japan, 
  University of South Carolina,
  Tahoku University,
Texas Agricultural and Mechanical University, 
  University of California (Los Angeles),
  University of Chicago,
  University of Huddersfield,
  University of Maryland,
  University of Wisconsin (Madison)
  University of Tennessee,
  University of Texas,
  University of Tokyo,
  Wellesley College. \\
  
\noindent\textbf{Next steps} \\
\noindent
The next steps in the development of the DAE$\delta$ALUS programme are
the:
\begin{itemize}
  \item Development of an ion source using $\mbox{H}_2^+$ ions to
    reduce space charge effects;
  \item Development of a $^8\mbox{Li}$ source using $\mbox{H}_2^+$ on
    a $^9\mbox{Be}$ target surrounded by a $^7\mbox{Li}$ sleeve. 
    This will be used to produce a low energy, $\beta$ decay-at-rest
    beam to study sterile neutrinos with a large liquid
    scintillator-detector; and
  \item Construction of a superconducting 800 MeV cyclotron and target
    and beam dump systems at a near site, followed by two more
    cyclotrons at medium and long baseline.
\end{itemize}

\subsubsection{ESSnuSB}
\label{Sect:AccBasedOsc:Future:ESSnuSB}

\noindent\textbf{Physics goals} \\
\noindent
It is proposed to upgrade the 2\,GeV ESS linac to deliver an average
power of 10\,MW to be shared between neutrino and neutron production
\cite{Baussan:2013zcy}. 
The neutrino beam of mean energy 0.4\,GeV would illuminate a
megaton-scale underground water-Cherenkov detector located 540\,km
from the ESS at the second oscillation maximum where the effect of
CPiV is approximately three times larger than at the first oscillation
maximum.
The low neutrino beam energy reduces the background from inelastic
scattering.
Assuming a ten-year exposure with two-years running time in neutrino
mode and eight-years in anti-neutrino mode, the goals of the programme
are to: 
\begin{itemize}
  \item Search for CPiV with a significance of $5\sigma$ over more
    than 60\% of all values of $\delta_{\rm CP}$ values; and  
  \item Perform a non-accelerator programme including, for example,
    the study of neutrinos from supernov\ae, the search for proton
    decay etc. \\
\end{itemize}

\noindent\textbf{Institutes 14} \\
  IHEP (Beijing),
  Virginia Tech,
  STFC Rutherford Appleton Laboratory,
  CERN,
  Cukurova University,
  University of Science and Technology (Cracow),
  Lund University,
  European Spallation Source,
  Universidad Autonoma de Madrid,
  INFN Padova,
  Ohridski University of Sofia, Bulgaria, 
  KTH Royal Instittue of Technology (Stockholm), 
  Universit\'e de Strasbourg,
  Uppsala University. \\

\noindent\textbf{Next steps} \\
\noindent
The next steps in the development of the ESSnuSB concept are the:
\begin{itemize}
  \item Design of an accumulator ring to reduce bunch lengths
    from 2.86\,ms to 1.5\,$\mu$sec to support horn focusing and
    the design of modifications of the LINAC to upgrade its average
    power form 5\,MW to 10\,MW by increasing the repetition frequency
    from 14\,Hz to 28\,Hz and to support $\mbox{H}^-$-ion
    acceleration, interleaved with the ESS proton beam, in order to
    facilitate injection of protons in the presence of protons already
    circulating in the ring;
  \item Development of a four target/horn system to spread the full
    beam power such that 1.25\,MW of power impinges on each; 
  \item Design of a system for the extraction of muons produced
    alongside the neutrinos in pion decay to serve nuSTORM, a Neutrino
    Factory or a Muon Collider; and  
  \item Identification and evaluation of a suitable underground site
    and the design of a cavern to house a large water Cherenkov
    detector.
    Surveys at the Garpenberg mine 540\,km from the ESS are under way. 
\end{itemize}

\subsection{Conclusions and recommendations}
\label{Sect:AccBasedOsc:ConcRec}

\stepcounter{nuPanel-RM-Conc-Sect}

\noindent
\framebox[\textwidth][l]{
  \parbox[c]{0.98\linewidth}{
        \begin{description}
      \stepcounter{nuPanel-RM-Conc-Sect-Conc}
      \item[\arabic{nuPanel-RM-Conc-Sect}.\arabic{nuPanel-RM-Conc-Sect-Conc}:]
        The present accelerator-based long-baseline
        neutrino-oscillation programme is vibrant and has substantial
        discovery potential. 
        \begin{description}
          \stepcounter{nuPanel-RM-Conc-Sect-Rec}
          \item[\color{BlueViolet} Recommendation \arabic{nuPanel-RM-Conc-Sect}.\arabic{nuPanel-RM-Conc-Sect-Rec}:]
            \textbf{\color{BlueViolet} Full exploitation of the present generation of
              experiments should continue thereby maximising their
              discovery potential and the scientific return on
              historical investment.
            }
        \end{description}
    \end{description}

        \begin{description}
      \stepcounter{nuPanel-RM-Conc-Sect-Conc}
      \item[\arabic{nuPanel-RM-Conc-Sect}.\arabic{nuPanel-RM-Conc-Sect-Conc}:]
        The measurements that will be made by the DUNE and the Hyper-K
        collaborations are complementary and the combination of the
        data from the two experiments offers the potential for
        insights beyond those that either experiment can provide on
        its own. 
        \begin{description}
          \stepcounter{nuPanel-RM-Conc-Sect-Rec}
          \item[\color{BlueViolet} Recommendation \arabic{nuPanel-RM-Conc-Sect}.\arabic{nuPanel-RM-Conc-Sect-Rec}:]
            \textbf{\color{BlueViolet} Both the DUNE and the Hyper-K programmes should be
              pursued through the established approval and funding
              processes.
            }
        \end{description}
    \end{description}

        \begin{description}
      \stepcounter{nuPanel-RM-Conc-Sect-Conc}
      \item[\arabic{nuPanel-RM-Conc-Sect}.\arabic{nuPanel-RM-Conc-Sect-Conc}:]
        The sensitivity of long-baseline neutrino-oscillation
        experiments may be enhanced by including events at the second
        oscillation maximum. 
        DUNE may exploit the second oscillation maximum since the LBNF
        broad-band beam will provide an interesting flux at the relevant
        energy.
        The baseline of Hyper-K places the relevant energy
        significantly below the peak of the J-PARC narrow-band 
        beam making it harder for Hyper-K to profit from the second
        oscillation maximum. 
        ESSnuSB has been optimised for the study of the second
        oscillation maximum thereby maximising its sensitivity to
        CP-invariance violation.  
        \begin{description}
          \stepcounter{nuPanel-RM-Conc-Sect-Rec}
          \item[\color{BlueViolet} Recommendation \arabic{nuPanel-RM-Conc-Sect}.\arabic{nuPanel-RM-Conc-Sect-Rec}:]
            \textbf{\color{BlueViolet} The complementarity and
              timeliness of the ESSnuSB proposal should be considered
              in the light of the likely timescales and performance of
              the DUNE programme and, if approved, the Hyper-K programme.
           }
        \end{description}
    \end{description}

        \begin{description}
      \stepcounter{nuPanel-RM-Conc-Sect-Conc}
      \item[\arabic{nuPanel-RM-Conc-Sect}.\arabic{nuPanel-RM-Conc-Sect-Conc}:]
        The focus of the long-baseline neutrino community has recently
        been on establishing DUNE and proposing Hyper-K. 
        If the science demands a further program with a performance
        that substantially exceeds that of the ambitious DUNE and
        Hyper-K experiments, new accelerator and/or detector
        technologies will be required.
        An R\&D program will be needed to deliver feasible options at
        the appropriate time. 
        This R\&D is likely to take many years and needs to be well
        justified and carefully planned. 
        \begin{description}
          \stepcounter{nuPanel-RM-Conc-Sect-Rec}
          \item[\color{BlueViolet} Recommendation \arabic{nuPanel-RM-Conc-Sect}.\arabic{nuPanel-RM-Conc-Sect-Rec}:]
            \textbf{\color{BlueViolet} ICFA should encourage a process, informed by the
            neutrino community, to assess the scientific case for a
            long-term accelerator and/or detector R\&D programme aimed
            at the post-DUNE/Hyper-K era as a first step in defining
            the R\&D programme that is required.
            Assessment of the scientific case will require sustained
            activity in neutrino theory and phenomenology including 
            significant developments in the understanding of
            neutrino-nucleus scattering.
           }
          \stepcounter{nuPanel-RM-Conc-Sect-Rec}
          \item[\color{BlueViolet} Recommendation \arabic{nuPanel-RM-Conc-Sect}.\arabic{nuPanel-RM-Conc-Sect-Rec}:]
            \textbf{\color{BlueViolet} The forum provided by the series of International
              Meetings for Large Neutrino Infrastructures is
              invaluable to ensure the coherent development of the 
              global programme and should be continued with a strong
              accelerator-based component. 
            }
        \end{description}
    \end{description}

  }
}

\cleardoublepage
\graphicspath{{03-Sterile/Figures/}}

\section{Sterile neutrino searches at accelerators}
\label{Sect:Sterile}

Three-flavour mixing provides a simple and elegant framework for
understanding the phenomenology of neutrino oscillations.
Most of the current oscillation measurements can be described within
this framework, enabling a self-consistent set of oscillation
parameters to be extracted from global fits to the data. 
However, the measurements are not yet sufficient to provide a
stringent and comprehensive test of the framework. 
Indeed, there are a number of measurements that, if interpreted as
resulting from neutrino oscillations, cannot be accommodated within 
the three-flavour framework. 
The resolution of these ``neutrino anomalies'' is an important part of
the future neutrino programme, one that will test the
three-flavour-mixing framework with reasonable precision.
Possible extensions to the framework involve the addition of
non-standard interactions and/or the addition of neutrino states
beyond the three known flavours. 
The intermediate vector bosons do not decay into these states,
therefore, if the additional states are light, they must also be
``sterile'', in the sense that they have no electroweak couplings. 
The search for sterile neutrinos is, perhaps, the primary way of
testing the three-flavour-mixing framework and of attempting to
resolve neutrino anomalies. 

Deviations from three-flavour mixing (neutrino anomalies) have been
reported from accelerator-based, reactor-based, and
radioactive-source-based measurements \cite{Declais:1994su,Aguilar:2001ty,AguilarArevalo:2008rc,Abdurashitov:2009tn,Kaether:2010ag,Mention:2011rk,Aguilar-Arevalo:2013pmq}. 
Individually, these tensions with three-flavour mixing, which are at
the level of two- to four-standard-deviations, do not provide
definitive evidence of new physics. 
Some, or all, of the anomalies may be due to statistical fluctuations
and/or systematic effects that have not been taken into account. 
Taken together, the anomalies suggest new neutrino-flavour transitions
with a frequency characterised by $L/E \sim 1$\,m/MeV, but the
evidence is inconclusive. 
This $L/E$ corresponds to a $\Delta m^2 \sim 1$\,eV$^2$ which is much
larger than the two $\Delta m^2$s determined from the
three-flavour-mixing measurements.
The anomalies are intriguing and persistent enough to warrant
definitive investigation.

The accelerator-based neutrino anomalies come from the
LSND \cite{Aguilar:2001ty} and MiniBooNE 
\cite{AguilarArevalo:2008rc,Aguilar-Arevalo:2013pmq} experiments,
which provide evidence for the appearance of 
$\nu_e (\overline{\nu}_e)$ in a $\nu_\mu (\overline{\nu}_\mu)$ beam
with $L/E \sim 1$\,m/MeV. 
These experiments, however, could not distinguish between electrons
and photons and therefore could not establish definitively the
phenomenon as due to neutrino-flavour transitions.

The next steps in the accelerator-based sterile-neutrino-search
program are well under way. 
MINOS+ completed its data taking at FNAL in 2016 and will measure the
time-dependent oscillation probabilities over a long-baseline at
energies that correspond to oscillation times away from the
three-flavour-mixing oscillation maxima~\cite{Tzanankos:2011zz}.
This enables the MINOS+ collaboration to search for small deviations
from expectations that could be attributed to additional oscillations
characterised by as yet unobserved values of $\Delta m^2$.
In addition, a new liquid-argon TPC, MicroBooNE, has just begun
taking data at FNAL \cite{Chen:2007ae}.
Within a few years, MicroBooNE will be capable of confirming (or
otherwise) the MiniBooNE anomaly and distinguishing between a
$\nu_e$-appearance or photon-appearance interpretation of the
phenomenon.
In 2018, MicroBooNE will be complemented with new near (SBND) and far
(ICARUS) liquid-argon TPCs to make the three-detector SBN program
setup \cite{Antonello:2015lea} that, after a few years of data taking,
is expected to provide definitive resolutions of the LSND and
MiniBooNE observations.   
These timelines suggest that around 2020 a decision could be made
about the need for further accelerator-based short-baseline
experiments beyond those in the presently foreseen program.

\subsection{Present accelerator-based sterile-neutrino searches}
\label{SubSect:Sterile:SBL:Present}

\subsubsection{MINOS+}
\label{SubSect:Sterile:SBL:Present:MINOS+}

\begin{figure}
  \begin{center}
    \includegraphics[width=0.40\textwidth]{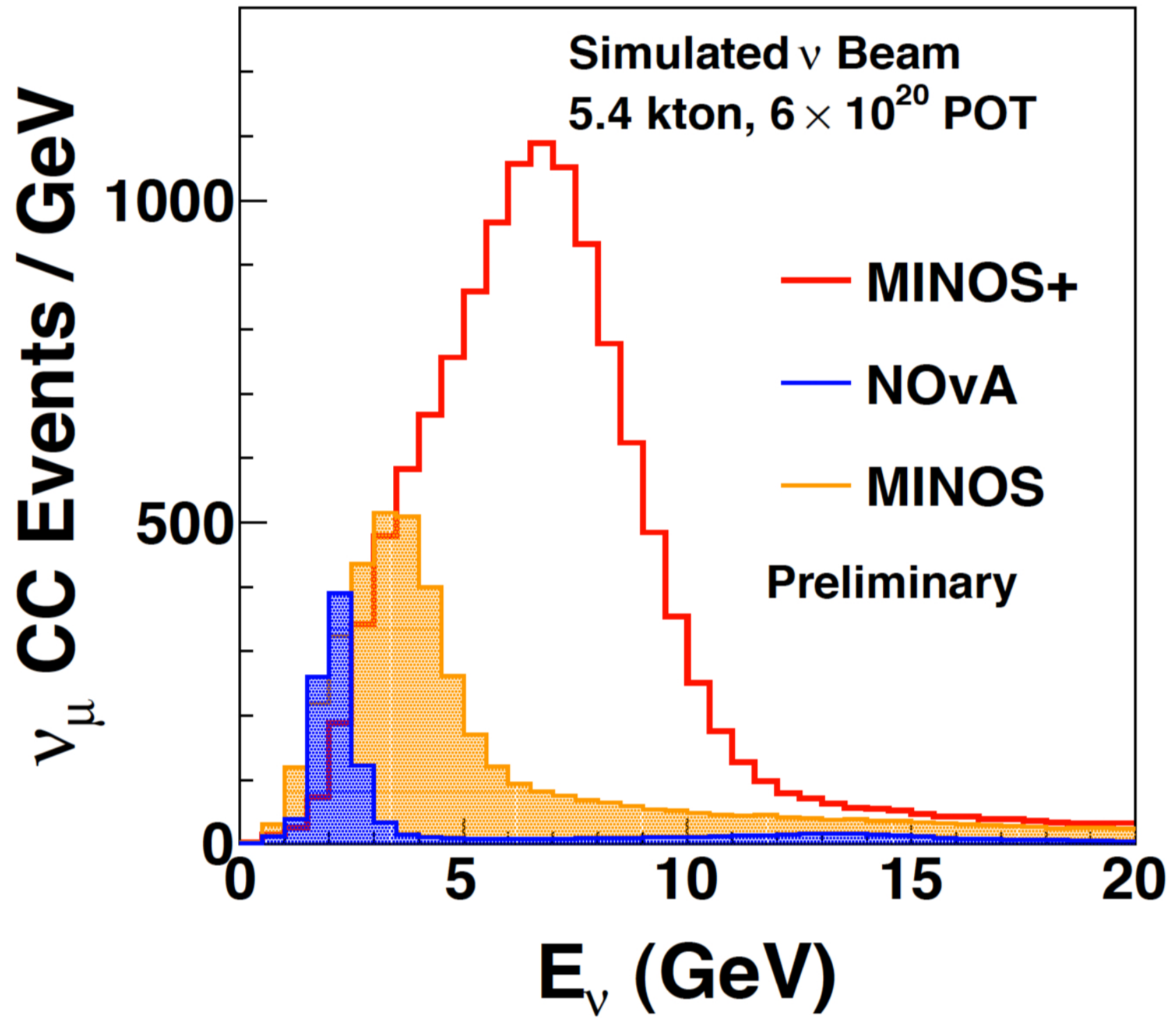}
    \quad \quad \quad \quad 
    \includegraphics[width=0.50\textwidth]{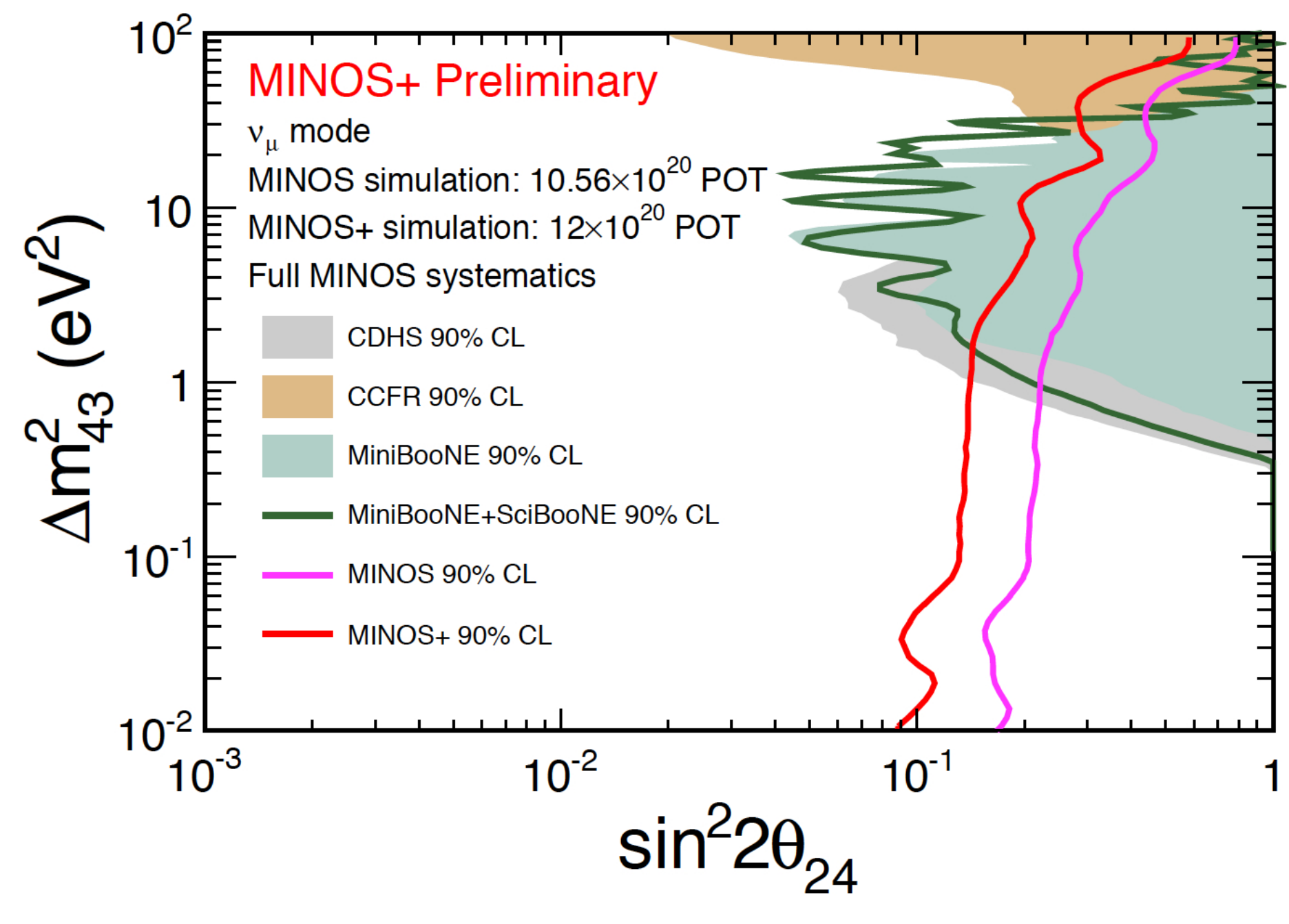}
  \end{center}
  \caption{
    Left panel: The NuMI neutrino energy spectrum for the MINOS+
    medium-energy tune, shown as the red solid line. 
    For comparison, the NO$\nu$A spectrum is shown as the blue
    shaded histogram and the low-energy tune used by MINOS is shown as
    the gold histogram.
    Right panel: MINOS+ reach in probing $\nu_\mu$ disappearance into
    sterile neutrinos for NuMI neutrino running (red solid line).
    The MINOS+ sensitivity is compared to that of CDHS (grey shaded
    region), CCFR (orange shaded region) and MiniBooNE (green shaded
    region).
    A recent, preliminary limit provided by MINOS is shown as the
    solid purple line.
  }
  \label{Fig:MINOSplus}
\end{figure}

\noindent\textbf{Physics goals} \\
\noindent
The ``NuMI'' neutrino beam from the FNAL Main Injector has been tuned
to deliver a wide-band beam on axis to the MINOS near and far
detectors.
The neutrino energy spectrum (see figure \ref{Fig:MINOSplus}) spans
the range 2\,GeV to 15\,GeV and an exposure corresponding to 
$\sim 10^{21}$\,POT was delivered by the end of 2016.
The goals of the MINOS+ programme are to \cite{Tzanankos:2011zz}:
\begin{itemize}
  \item Enhance understanding of the S$\nu$M scheme and constrain or
    discover new phenomena that go beyond it. 
  \item Search for sterile neutrinos using muon-neutrino disappearance
    and electron-neutrino appearance; and 
  \item Search for non-standard interactions and exotic phenomena such
    as large extra dimensions.
\end{itemize}
Figure \ref{Fig:MINOSplus} (right panel) shows the projected
sensitivity of the experiment to a single sterile neutrino with an
effective mixing angle $\theta_{24}$ and mass-squared splitting
$\Delta m^2_{43}$.
 \\

\noindent\textbf{Institutes 30; collaborators 50} \\
BNL, 
    Caltech,
    University of Cambridge,
    University of Cincinnati,
    Federal University of Goias,
    Fermi National Accelerator Laboratory,
    Harvard University,
    University of Houston,
    Iowa State,
    Lancaster University,
    Los Alamos National Laboratory,
    University College London, 
    University of Manchester,
    University of Minnesota (Twin-Cities),
    University of Minnesota (Duluth),
    Otterbein University,
    University of Oxford,
    University of Pittsburgh,
    University of South Carolina,
    Stanford University,
    University of Sussex,
    University of Texas at Austin,
    Tufts University,
    University of Campinas,
    University of Sao Paulo,
    University of Warsaw,
    College of William \& Mary \\

\noindent\textbf{Future programme} \\
MINOS+ ran to the end of 2016 and it is anticipated that all data
analyses will be completed within a year of the end of data taking.

\subsubsection{MicroBooNE}
\label{SubSect:Sterile:SBL:Present:MicrobooNE}

\noindent\textbf{Physics goals} \\
\noindent
MicroBooNE is a 170\,t liquid-argon TPC located 470\,m from the source
of the FNAL Booster Neutrino Beam \cite{Chen:2007ae}.
The detector will be operated for three years on its own and then
continue as the intermediate detector in the Short Baseline Neutrino
programme (see below) \cite{Antonello:2015lea}.
In standalone operation, MicroBooNE will integrate an exposure
corresponding to $\approx 6.6 \times 10^{20}$\,POT.
The principal goals of the collaboration are to: 
\begin{itemize}
  \item Investigate the currently-unexplained excess of low-energy
    electromagnetic events observed by MiniBooNE;
  \item Measure neutrino cross sections on argon for multiple
    reaction channels; and 
  \item Advance the development of the liquid-argon TPC technology
    towards the realisation of the future short- and long-baseline
    neutrino programs. \\
\end{itemize}

\noindent\textbf{Institutes 25; collaborators 161} \\
University of Bern,
  Brookhaven National Laboratory,
  University of Cambridge,
  University of Chicago,
  University of Cincinnati,
  Columbia University,
  Fermi National Accelerator Laboratory,
  Illinois Institute of Technology,
  Kansas State University,
  Lancaster University,
  Los Alamos National Laboratory,
  University of Manchester,
  Massachusetts Institute of Technology,
  University of Michigan (Ann Arbor),
  New Mexico State University,
  Oregon State University,
  Otterbein University,
  University of Oxford,
  Pacific Northwest National Laboratory,
  University of Pittsburgh,
  Princeton University,
  Saint Mary's University of Minnesota,
  SLAC,
  Syracuse University,
  Virginia Tech,
  Yale University \\

\noindent\textbf{Future programme}

\noindent
The cross-section measurements that MicroBooNE will make are reported
in section \ref{SubSect:SptPrg:nuFT}.
The principal goal of the sterile-neutrino search programme is to:
\begin{itemize}
  \item Determine the source of the MiniBooNE low-energy excess
    with a statistical significance $>5\sigma$ if it arises due to
    a source of electrons and with a statistical significance
    $>4\sigma$ if it arises due to a source of photons.
    These sensitivities assume $6.6\times10^{20}$ protons on
    target.
\end{itemize}
The achievement of the exploitation milestones depends on the rate of
beam delivery to the Booster Neutrino Beam.
To complete the low-energy analysis will require the full POT request,
$6.6\times10^{20}$ protons on target.
It is anticipated that the full data set needed for this analysis will
be delivered by mid-to-late 2018, allowing the low-energy-analysis
results to be presented in 2019.

First neutrino cross-section results will be prepared using the first
year's data set ($\sim 2 \times 10^{20}$\,POT were delivered by summer
2016).
In particular, the collaboration is actively working on
charged current (CC) inclusive, CC $0\pi$, and neutral current $\pi^0$
analyses.
By summer 2016, the statistics of the MicroBooNE data set exceeded
that of ArgoNeuT.

\subsection{Next-generation accelerator-based sterile-neutrino searches}
\label{SubSect:Sterile:SBL:Next}

\subsubsection{Short Baseline Neutrino (SBN) programme}
\label{SubSect:Sterile:SBL:Next:SBN}

\begin{figure}
  \begin{center}
    \includegraphics[width=0.48\textwidth]{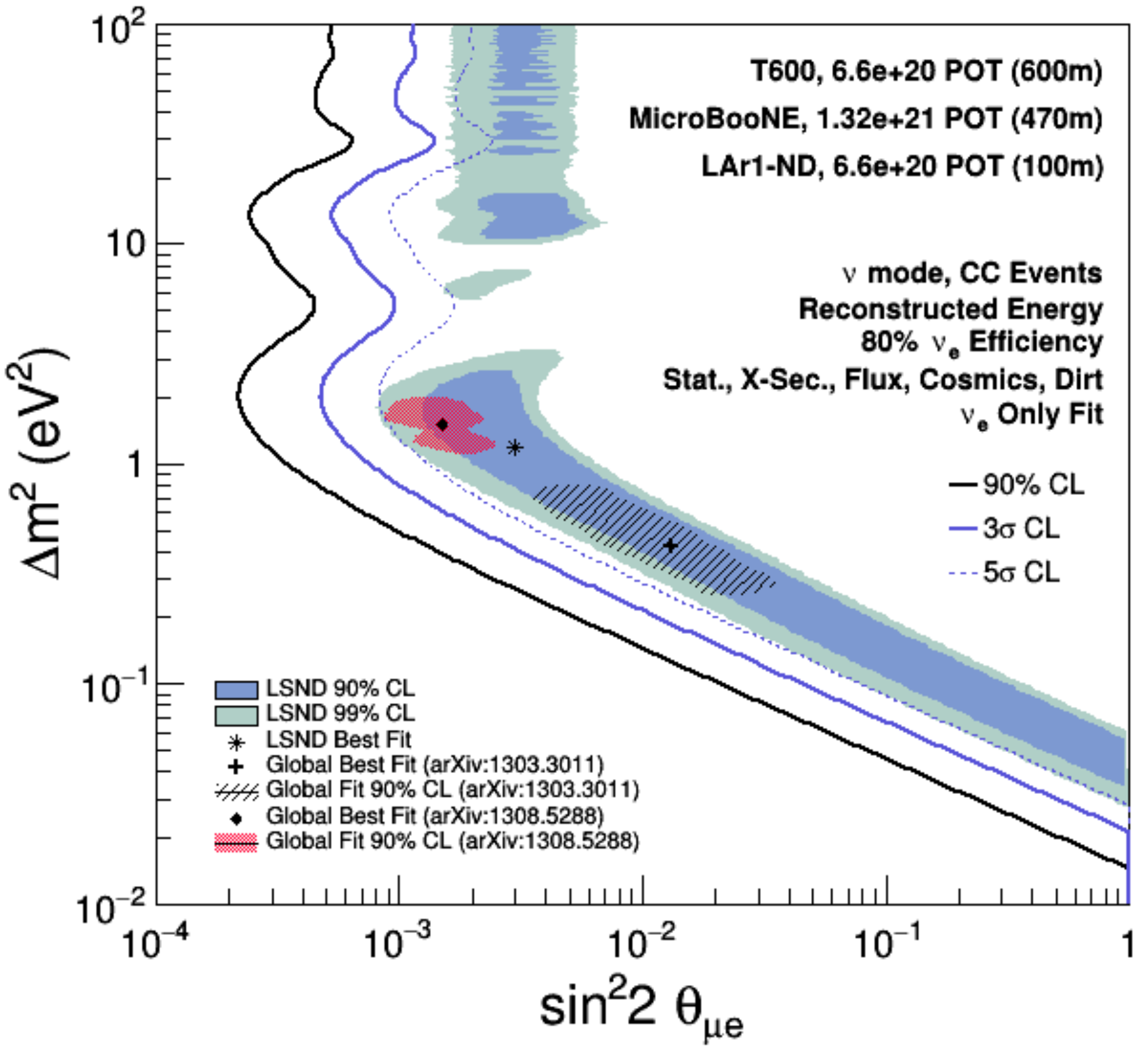}
    \quad \quad 
    \includegraphics[width=0.44\textwidth]{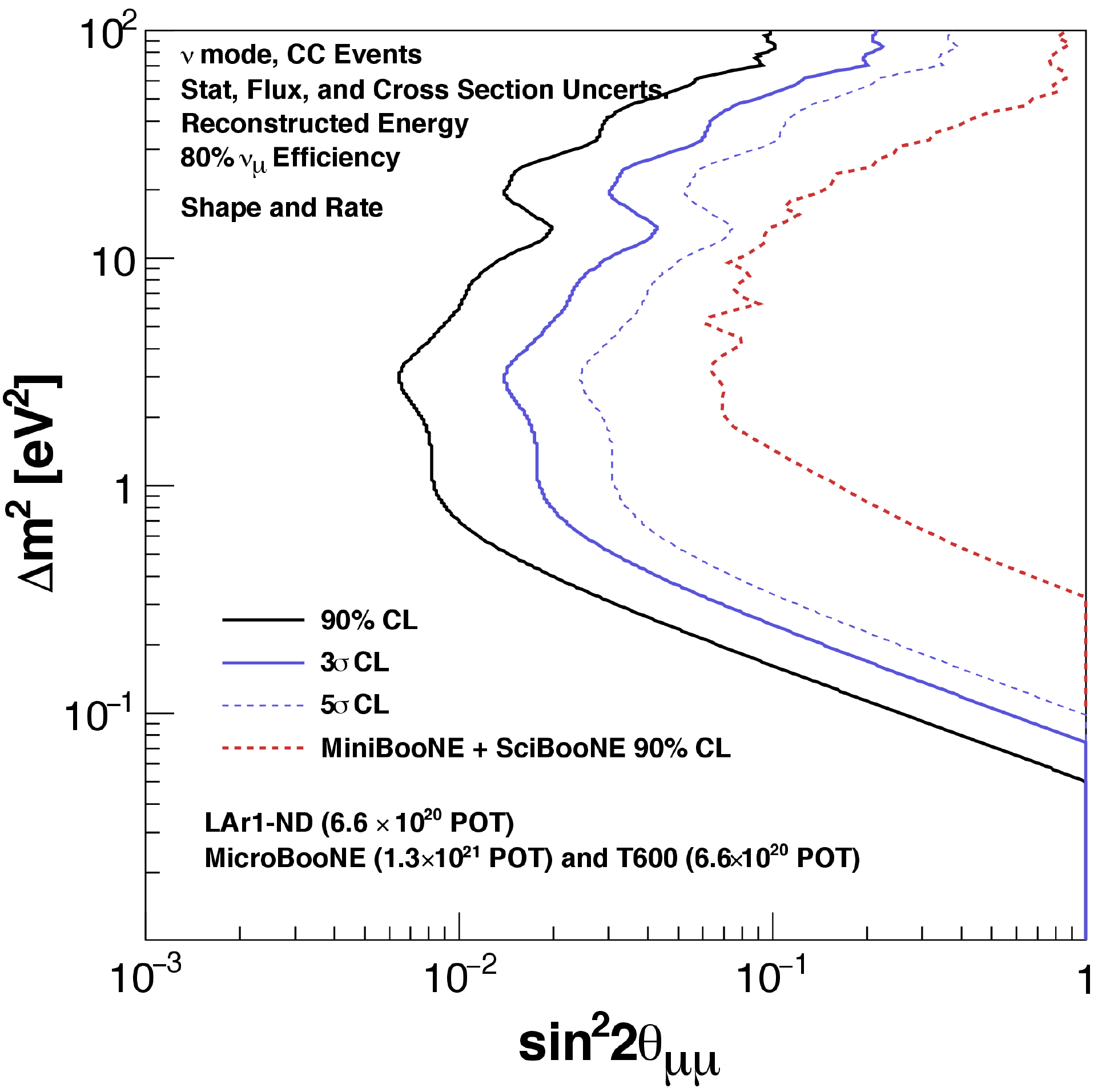}
  \end{center}
  \caption{
    Left panel: SBN Program to sensitivity to
    $\nu_\mu \rightarrow \nu_e$ oscillations.
    The regions of parameter space allowed by LSND and two global fits
    are also shown as the shaded bands.
    Right panel: SBN Program sensitivity to $\nu_\mu$ disappearance.
    The exclusion limit derived from the MiniBooNE and SciBooNE data
    are also shown.
    Figures taken from \cite{Antonello:2015lea}.
  }
  \label{Fig:SBNsens}
\end{figure}

\noindent\textbf{Physics goals} \\
The Short Baseline Neutrino (SBN) Program \cite{Antonello:2015lea}
will exploit the Booster Neutrino Beam (BNB) at FNAL by continuing
MicroBooNE running beyond its initial 3 years, and complement the
170\,t MicroBooNE detector at a baseline of 470\,m by adding a new
220\,t LAr-TPC (Short Baseline Near Detector - SBND) at a baseline of
110\,m and installing a refurbished 760\,t ICARUS-T600 detector at a
baseline of 600\,m. 
Thus the SBN setup will have near, intermediate, and far detectors,
all of them based on liquid-argon TPCs.   
The SBN Program is designed to deliver a programme that includes the
resolution of the experimental anomalies in short-baseline neutrino
physics and the most sensitive search to date for sterile neutrinos at
the eV mass-scale through both appearance and disappearance
oscillation channels (see figure \ref{Fig:SBNsens}).
In addition the SBN Program includes the study of neutrino-argon cross
sections with millions of interactions using the well-characterised
neutrino fluxes of the BNB.
The SBN detectors will also record events from the off-axis flux of
the NuMI neutrino beam with its higher electron-neutrino content and
different energy spectrum.
 \\

\noindent\textbf{SBND institutes 30; collaborators 138} \\
\noindent
Argonne National Laboratory,
University of Bern,
Brookhaven National Laboratory,
University of Cambridge,
University of Campinas,
CERN,
University of Chicago,
Columbia University,
Federal University of ABC,
Federal University of Alfenas,
Fermi National Accelerator Laboratory,
Illinois Institute of Technology,
Indiana University (Bloomington),
Kansas State University,
Lancaster University,
University of Liverpool,
Los Alamos National Laboratory,
 University of Manchester,
University of Michigan,
Massachusetts Institute of Technology,
University of Oxford,
Pacific Northwest National Laboratory,
University of Pennsylvania,
University of Puerto Rico,
University of Sheffield,
Syracuse University,
University of Texas Arlington,
University College London,
Virginia Tech,
Yale University \\

\noindent\textbf{MicroBooNE institutes} \\
\noindent
As listed above. \\

\noindent\textbf{ICARUS institutes 20; collaborators 56} \\
\noindent
CERN, 
Catania University, INFN Catania,
Pavia University, INFN Pavia, 
Padova University, INFN Padova, 
Gran Sasso Science Institute, Italy, 
Laboratori Nazionali di Gran Sasso, Italy,
Institute of Nuclear Physics (Cracow), 
Laboratori Nazionali di Frascati (Roma),
Laboratori Nazionali del Gran Sasso, Italy,
INFN Milano Bicocca,
INFN Milano,
INFN Napoli, 
Institute for Nuclear Research of the Russian Academy of Sciences (Moscow),
University of Silesia (Katowice) Poland,
Wroc\l aw University, Poland,
Argonne National Laboratory,
Colorado State University, 
Los Alamos National Laboratory,
Fermi National Accelerator Laboratory,
University of Pittsburgh,
SLAC \\

\noindent\textbf{Future programme} \\
\noindent
Preparations for the SBN program are under way with civil construction
ongoing and the refurbishment of the ICARUS-T600 detector proceeding
along with the design and construction of the new SBND detector.  
The projected start of data taking in the SBN era is 2018. 
The Program is approved for an exposure of $6\times10^{20}$\,POT on
top of the $6\times10^{20}$\,POT to be accumulated by MicroBooNE in
the pre-SBN-era.
These data sets will enable a search for $\nu_\mu \rightarrow \nu_e$
appearance with a $5\sigma$ sensitivity covering the full LSND-allowed
region and the search for anomalies in the disappearance modes both
for $\parenbar{\nu}_\mu$ and $\parenbar{\nu}_e$.

\subsubsection{JSNS\boldmath{$^2$}}
\label{SubSect:Sterile:SBL:Next:JSNS2}

\noindent\textbf{Physics goals/measurement programme} \\
\noindent
The JSNS$^2$ experiment aims to search for the existence of neutrino
oscillations with $\Delta m^2$ near $1$~\,eV$^2$ at the J-PARC
Materials and Life Science Experimental Facility
(MLF)~\cite{Harada:2016rou}. 
An intense neutrino beam from muon decay at rest is available from the
1\,MW, 3\,GeV proton beam from the J-PARC Rapid Cycling Synchrotron
striking the spallation-neutron target.
The neutrinos come predominantly from $\mu^+$ decay, 
$\mu^+\rightarrow e^+ +\bar{\nu}_{\mu} +\nu_e$, which allows the
search for $\bar{\nu}_{\mu} \rightarrow \bar{\nu}_e$ transitions
through the detection of the inverse $\beta$-decay reaction, 
$\bar{\nu}_e+p \rightarrow e^++ n$, and the photons subsequently
produced through neutron capture on gadolinium in the 50\,t fiducial
mass of two liquid-scintillator detectors located 24\,m away from the
mercury target.
A cross-section measurement programme will also be carried out using
neutrinos with energies of a few times 10\,MeV. \\  

\noindent\textbf{Institutes 12; collaborators 31} \\
   University of Alabama,
   Brookhaven National Laboratory,
   Colorado State University,
   University of Florida,
   JAEA,
   High Energy Accelerator Research Organization (KEK),
   University of Kyoto,
   Los Alamos National Laboratory,
   University of Michigan,
   University of Osaka, Research Center for Nuclear Physics,
   University of Tohoku (RCNS) \\

\noindent\textbf{Future programme}

\noindent
The background rate and energy response have been measured using a
500\,kg prototype plastic-scintillator detector.
Stage 1 approval was granted by the J-PARC PAC in 2015 based on these
measurements. 
The next steps in the development of the programme are:
\begin{itemize}
  \item The completion of a Technical Design Report (TDR) that will be
    used to seek approval to start construction of the experiment from
    the J-PARC PAC.
    The TDR will be completed within the next year; and 
  \item Following approval, construct the detector within a further
    1.5 years and start the experiment.
\end{itemize}

\subsection{Future opportunities for accelerator-based
  sterile-neutrino searches}
\label{SubSect:Sterile:SBL:Future}

\subsubsection{IsoDAR} 
\label{SubSect:Sterile:SBL:Future:IsoDAR}

\noindent\textbf{Physics goals} \\
\noindent
IsoDAR is an isotope decay-at-rest experiment that had been proposed
as the first stage of a phased Decay-At-rest Experiment for
$\delta_{\rm CP}$ studies (DAE$\delta$ALUS, see section 
\ref{Sect:AccBasedOsc:Future:Deadalus}) that would ultimately search
for evidence of CP-invariance violation in the neutrino sector 
\cite{Alonso:2012zv,Toups:2013dxa,Shaevitz:2015uar}.
Using the injector cyclotron setup being devised for DAE$\delta$ALUS,
IsoDAR would be a short-baseline experiment using a $\sim 1$\,kt
scintillator-based detector.
With an electron-anti-neutrino flux of sufficient intensity from
$^8$Li decay at rest, the experiment would facilitate a conclusive
test of sterile-neutrino-oscillation models using the $L/E$ dependence
of inverse beta-decay.
In the report from the US High Energy Physics Advisory Panel (HEPAP)
Particle Physics Projects Prioritization Panel
(P5) \cite{HEPAP:P5:2014}, IsoDAR is one of the experiments the
development of which ``\ldots is not yet advanced to a point at which
it would be possible to consider recommendations to move forward.''
However ``\ldots the R\&D for these projects would fit as candidates
in the small projects portfolio, with a path to eventual
implementation presumably being among the evaluation criteria.''
IsoDAR would provide a definitive way to explore electron-antineutrino
disappearance and could therefore be a candidate to add to the
programme if the R\&D confirms its feasibility in intensity and cost.

\subsubsection{nuSTORM}
\label{SubSect:Sterile:SBL:Future:nuSTORM}

\noindent\textbf{Physics goals} \\
\noindent 
The Neutrinos from Stored Muons (nuSTORM) facility provides neutrino
beams from the decay of 3.8\,GeV muons confined within a storage ring  
\cite{Adey:2013afh,Adey:1537983,Kyberd:2013ii,Kyberd:2013ij,Adey:2013pio,Lackowski:2013ria,Adey:2014rfv}.
It is proposed that the muon beam be instrumented with a
magnetised-iron-scintillator sampling calorimeter at a distance of
$\sim 1.5$\,km from the end of the production straight and a
magnetised near detector at a distance of 50\,m from the end of the
straight.
Pions are injected into the neutrino-production straight.
The pion-muon transition in the first pass of the beam through the
production straight creates a flash of $\nu_\mu$ from pion decay.

\begin{figure}
  \begin{center}
    \includegraphics[width=\textwidth]{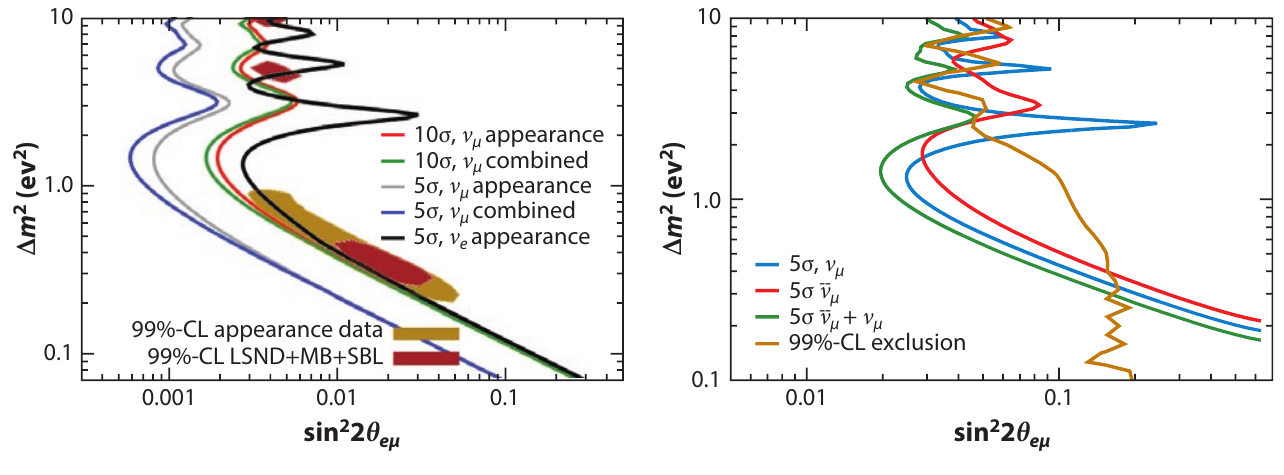}
  \end{center}
  \caption{
    nuSTORM sensitivity to sterile neutrinos in a 3+1 model.
    Left panel: The combination of $\nu_\mu$ appearance and
    disappearance experiments expressed in terms of the effective
    mixing angle $\theta_{e\mu}$.
    Right panel: Sensitivity of the $\nu_\mu$ disappearance experiment
    from the pion-decay source and the $\bar{\nu}_\mu$ disappearance
    experiment from the muon-decay source.
    The green line represents the sensitivity of the combination of
    the experiments.
    Taken from \cite{Adey:2015iha}. 
  }
  \label{Fig:nuSTORM:Sens}
\end{figure}

With an appropriately optimised detector and an exposure of $\sim
10^{18}$ muon decays, corresponding to 10\,years of operation, nuSTORM
can provide a 10$\sigma$ exclusion limit in the region of
sterile-neutrino parameter space indicated by the results of the LSND
and MiniBooNE experiments (see figure 
\ref{Fig:nuSTORM:Sens}) \cite{Adey:2015iha}. 
nuSTORM is also capable of supporting a definitive neutrino-nucleus
cross-section measurement programme (see section 
\ref{SubSect:SptPrg:nuFT:DS}) and the R\&D programme required for the
development of muon accelerators for particle physics.
 \\

\noindent\textbf{Institutes (36), collaborators (112)} \\
\noindent 
  Fermi National Accelerator Laboratory, 
  Institute of Physics, Orissa, India,
  Muons Inc., 
University of Geneva,
  University of Warwick,
  University of Strasbourg, 
  University of Glasgow,
  York University,
  University of Oxford,
  Thomas Jefferson National Accelerator Facility, 
  University of Sheffield,
Instituto de Fisica Corpuscular (IFIC, Valencia), 
  Imperial College London, 
Virginia Tech,
  University of California, Riverside, 
  STFC Rutherford Appleton Laboratory, 
  Northwestern University, 
University of Santiago de Compostela,
Max Planck Institute for Nuclear Physics (Heidelberg), 
  Lancaster University,
  Osaka University,
  Brunel University, 
  Kyoto University,
  TRIUMF,
  University of Toronto,
  University of Liverpool,
  Princeton University, 
  University of South Carolina, 
  University College London,
  Institute for Particle Physics Phenomenology, Durham, 
  Wroc\l aw University, Poland, 
  CERN,
  Illinois Institute of Technology, 
  INFN Padova, 
  The University of British Columbia,
University of W\"{u}rzburg \\

\noindent\textbf{Next steps} \\
\noindent 
nuSTORM allows the study of two-thirds of all possible appearance and
disappearance channels in which signals for sterile neutrinos may
appear.
In addition, the detailed knowledge of the neutrino flux allows the
neutral-current rate to be measured precisely.
These features make nuSTORM a candidate to succeed the present
generation of sterile-neutrino-search experiments.
Further, as described in section \ref{SubSect:SptPrg:nuFT:DS}, nuSTORM
is capable of delivering a $\nu N$-scattering programme that includes
the precise determination of $\parenbar{\nu}_e N$-scattering cross
sections.
The nuSTORM facility can also support the accelerator R\&D programme
required to deliver the capability necessary to mount a future
neutrino factory of muon collider \cite{Adey:2015iha}.

A proposal to the FNAL PAC \cite{Adey:2013pio} was well received, but
nuSTORM was not recommended by P5 \cite{HEPAP:P5:2014}.
An expression of interest to mount nuSTORM at CERN was well received
by the CERN SPSC \cite{Adey:2013afh,Adey:1537983}.
nuSTORM was discussed in Europe as part of a potential international
muon-accelerator programme 
\cite{EUmuonDiscussion:2015,EUmuonOppoDoc:2015} and will be studied
by the CERN's Physics Beyond Colliders study
group~\cite{CERN:PBC:WWW:2017}.

The SBN Programme at FNAL will reach its full sensitivity by 
$\sim 2020$. 
On this timescale the results from the SBN and other
sterile-neutrino-search experiments will be available; these results
may indicate that nuSTORM should be considered as part of the
sterile-neutrino programme.
Since, in addition, nuSTORM allows precise studies of
$\parenbar{\nu}_eN$ scattering to be carried out, the Panel believes
2020 to be a natural point at which to consider nuSTORM as part of the
sterile-neutrino programme.
\vfill

\subsection{Conclusions and recommendations}
\label{SubSect:Sterile:SBL:InterimConcRec}

\stepcounter{nuPanel-RM-Conc-Sect}

\noindent
\framebox[\textwidth][l]{
  \parbox[c]{0.98\linewidth}{
        \begin{description}
      \stepcounter{nuPanel-RM-Conc-Sect-Conc}
      \item[\arabic{nuPanel-RM-Conc-Sect}.\arabic{nuPanel-RM-Conc-Sect-Conc}:]
        Unambiguous evidence for sterile neutrinos would constitute
        a breakthrough of fundamental significance and would
        revolutionise the field while unambiguous confirmation of the
        short-baseline anomalies would warrant energetic
        investigation.
        \begin{description}
          \stepcounter{nuPanel-RM-Conc-Sect-Rec}
          \item[\color{BlueViolet} Recommendation \arabic{nuPanel-RM-Conc-Sect}.\arabic{nuPanel-RM-Conc-Sect-Rec}:]
            \textbf{\color{BlueViolet}
              The present generation of accelerator-based
              sterile-neutrino-search experiments, including those
              that constitute the SBN Program at FNAL, should be
              exploited so as to maximise their sensitivity.
            }
        \end{description}
      \stepcounter{nuPanel-RM-Conc-Sect-Conc}
      \item[\arabic{nuPanel-RM-Conc-Sect}.\arabic{nuPanel-RM-Conc-Sect-Conc}:]
        Results from the SBN Program and other
        sterile-neutrino-search experiments will be available by
        $\approx$2020.
        It will then be timely to decide on the future direction of
        the accelerator-based sterile-neutrino-search programme.    
        \begin{description}
          \stepcounter{nuPanel-RM-Conc-Sect-Dec}
          \item[\color{RedViolet} Decision point \arabic{nuPanel-RM-Conc-Sect}.\arabic{nuPanel-RM-Conc-Sect-Rec}:]
            \textbf{\color{RedViolet}
              \boldmath{$\approx$}2020: Decide on the future direction
              of the accelerator-based sterile-neutrino-search
              programme.
            }
        \end{description}
    \end{description}

  }
}
\clearpage

\noindent
\framebox[\textwidth][l]{
  \parbox[c]{0.98\linewidth}{
        \begin{description}
      \stepcounter{nuPanel-RM-Conc-Sect-Conc}
      \item[\arabic{nuPanel-RM-Conc-Sect}.\arabic{nuPanel-RM-Conc-Sect-Conc}:]
        Beyond the SBN Program, the way forward will depend on the
        strength of the evidence for sterile neutrinos.
        \begin{description}
          \stepcounter{nuPanel-RM-Conc-Sect-Rec}
          \item[\color{BlueViolet} Recommendation \arabic{nuPanel-RM-Conc-Sect}.\arabic{nuPanel-RM-Conc-Sect-Rec}:]
            \textbf{\color{BlueViolet}
              The sensitivity, cost, schedule and relative strengths
              of the proposed next-generation accelerator-based
              sterile-neutrino-search experiments (IsoDAR, nuSTORM)
              should be evaluated in preparation for a decision to be
              made on the future direction of the
              sterile-neutrino-search programme
              in \boldmath{$\approx$}2020.
              In the mean time, the R\&D programme necessary to
              establish the requisite capabilities should be carried
              out. 
            }
        \end{description}
    \end{description}

  }
}

\cleardoublepage
\graphicspath{{04-Supporting-programme/Figures/}}

\section{Supporting programme}
\label{Sect:SBL}

For the long- and short-baseline neutrino programmes to reach their
full sensitivity requires a programme of measurement the results of
which are to be used in the estimation of the neutrino flux and to
estimate the rate and the characteristics of neutrino interactions in
the near and far detectors.
Accelerator and detector R\&D programmes are required to deliver the
present generation of experiments and to create the capability to
mount the next generation.
Innovation and development of software tools is also required to meet
the needs of the programme.
The development of the programme required to support the
neutrino-oscillation and sterile-neutrino-search experiments is
discussed in the paragraphs that follow.

\subsection{Hadroproduction}
\label{SubSect:SptPrg:HadroProd}

Conventional neutrino beams produced from the decay in flight of pions
and kaons will serve the present and next generation accelerator-based
neutrino experiments.
The absolute flux normalisation, the neutrino energy spectrum and the
flavour-composition of the neutrino beam are determined by the
absolute yield and momentum distributions of the hadrons produced in
the proton-nucleus interactions in the target.
All present and planned long-baseline experiments exploit a near
detector to constrain the un-oscillated neutrino flux.
Extrapolation of the neutrino flux from the near to the far detector
is necessary; the secondary-particle distributions are a critical
input to this extrapolation.
Therefore, the precise knowledge of hadron-production cross sections
from the particle-production target is required for
neutrino-oscillation experiments to deliver to their specified
precision.

However, in general, the hadroproduction data that is ``tailor-made''
for a particular neutrino-oscillation experiment,
i.e. particle-production cross sections measured using protons of same
energy interacting with the same target material and in the same
configuration as is used in the oscillation experiment, is scarce.
Phenomenological parametrisations of the available data are therefore
used to estimate the production rates for a particular experiment.
This procedure introduces a source of systematic uncertainty.
Therefore, dedicated hadroproduction experiments that use the same
target material at the same proton energy as in the oscillation
experiment are an essential part of the accelerator-based neutrino
programme. 

Several experiments have been mounted to make the necessary
measurements.
The HARP experiment took place at CERN and made essential
contributions to the K2K and MiniBooNE experiments.
The MIPP experiment used a 120\,GeV proton beam to measure particle
production spectra for the experiments served by the Main Injector
neutrino beam. 
At present, the only running hadro-production experiment is NA61 at
CERN.

\subsubsection{Facilities and experiments}
\label{SubSubSect:SptPrg:HadroProd:Exp}

\paragraph*{NA61/SHINE}

\noindent\textbf{Physics goals} \\
\noindent
The principal goals of the NA61 experiment are to
\cite{Antoniou:2006mh}:
\begin{enumerate}
  \item Measure inclusive spectra and fluctuations in nucleus-nucleus
    collisions;
  \item Measure hadron production spectra of relevance to the
    prediction of the cosmic-ray flux; and 
  \item Provide hadron production spectra of relevance to the
    prediction of neutrino fluxes generated in conventional neutrino
    sources.
\end{enumerate}

To serve the T2K experiment, NA61 took data between 2007 and 2009 with
thin targets ($\sim 6$M triggers) and with a replica of the T2K
graphite target ($\sim 13$M triggers) 
\cite{Abgrall:2011ae,Abgrall:2011ts,Abgrall:2012pp,Abgrall:2013wda,Abgrall:2013qoa,Abgrall:2014xwa,Aduszkiewicz:2015jna,Abgrall:2015hmv}. 
Final results for the thin-target data are already incorporated into
the T2K physics analysis. 
Currently, the uncertainty of the neutrino-flux ($\Phi$) prediction at
the T2K near or far detector is $\delta\Phi~\sim 10$\%.
This uncertainty is dominated by the hadron-production uncertainty
of $\delta\Phi_{\mbox{H}}~\sim 9$\%. 
The uncertainty on the near-to-far extrapolation (far-to-near flux
ratio) now reaches the $\sim 3$\% level.
To reduce the systematic uncertainties will require the measurement of
the hadrons produced in secondary and tertiary interactions
inside/outside the particle-production target.
This will require measurements with a variety of beams (proton and
pion), at various momenta and with various target materials.
Such a programme is possible if requested.

There is a new initiative in NA61 to provide hadron production data
for US-based neutrino experiments \cite{Antoniou:2006mh}.
The goal of this programme is to measure identified-hadron spectra
with an uncertainty of 4--5\%.
Measurements of hadron production from thin carbon and aluminium
targets as well as replicas of the particle-production targets in use
at FNAL will be made.
Running with proton and pion beams with momenta in the range 30\,GeV/c
to 120\,GeV/c has been approved.
These measurements are expected to bring the total uncertainty of the
Main Injector neutrino-beam flux down to 5--6\%. \\
  
\noindent\textbf{Institutes 37; collaborators 145} \\
\noindent
National Nuclear Research Center, Baku, Azerbaijan,
University of Sofia,
Ru{\dj}er Bo\v{s}kovi\'c Institute, Zagreb, Croatia,
LPNHE, University of Paris VI and VII, Paris, France,
Karlsruhe Institute of Technology, Karlsruhe, Germany,
Fachhochschule Frankfurt, Frankfurt, Germany,
University of Frankfurt, Frankfurt, Germany,
University of Athens, Athens, Greece,
Wigner Research Centre for Physics of the Hungarian Academy of Sciences, Budapest, Hungary,
Institute for Particle and Nuclear Studies, Tsukuba, Japan,
University of Bergen, Bergen, Norway,
Jan Kochanowski University in Kielce, Poland,
National Centre for Nuclear Research, Warsaw, Poland,
Jagiellonian University, Cracow, Poland,
University of Silesia, Katowice, Poland,
University of Warsaw, Warsaw, Poland,
University of Wroc{\l}aw, Wroc{\l}aw, Poland,
Warsaw University of Technology, Warsaw, Poland,
Institute for Nuclear Research, Moscow, Russia,
Joint Institute for Nuclear Research, Dubna, Russia,
National Research Nuclear University (Moscow Engineering Physics Institute), Moscow, Russia,
St. Petersburg State University, St. Petersburg, Russia,
University of Belgrade, Belgrade, Serbia,
ETH Z\"urich, Z\"urich, Switzerland,
University of Bern, Bern, Switzerland,
University of Geneva, Geneva, Switzerland,
Fermilab, Batavia, USA,
Los Alamos National Laboratory, Los Alamos, USA,
University of Colorado, Boulder, USA,
University of Pittsburgh, Pittsburgh, USA. \\

\noindent\textbf{Future programme}

\noindent
NA61 started taking data in 2007.
NA61 is approved to run from until 2018 to accumulate the data
required to support the FNAL neutrino programme and complete
measurements of nucleus-nucleus scattering with xenon and lead beams.
The NA61 collaboration is preparing a proposal to extend the programme
of measurement beyond 2020.

\subsection{Neutrino interactions}
\label{SubSect:SptPrg:nuFT}

The increased precision of the measurements made by present and future
long- and short-baseline neutrino experiments places greater demands
on the precision with which neutrino-nucleus interactions must be
understood.
Neutrino-scattering uncertainties are a dominant source of systematic
uncertainty in current experiments such as T2K.
Studies of the sensitivity of future experiments show that it is very
important to reduce systematic uncertainties to the 1\% level.
At present, individual neutrino-nucleon/nucleus cross sections are
known with a precision of $\sim 20\%$; the level of uncertainty
depending on the reaction mechanism.
Recent neutrino-nucleus-scattering measurements reveal significant
deficiencies in our understanding of the physics of neutrino
scattering and point to the importance of previously neglected
interactions, such as multi-nucleon correlations.
At the same time, a coherent and unified theoretical view of
neutrino-nucleus interactions that encompasses all recent developments
does not exist and, in some cases, models with similar underlying
assumptions produce inconsistent results.
To advance our understanding to the level required to support the
future neutrino-oscillation program requires a coordinated campaign of
measurement and theoretical calculation. 
In the near term this program seeks to address the following
questions:
\begin{itemize}
  \item What is causing the apparent tensions between different data
    sets (e.g. single charged-pion production)?
    Are the comparisons valid? 
    If they are, do the discrepancies point to processes that are
    being neglected or to mis-modelling? 
    What additional measurements and/or theoretical calculations are
    needed to resolve remaining tensions?
  \item What is causing the large differences between theoretical
    calculations and/or event generator predictions?
    What additional measurements and/or theoretical calculations are
    needed to address these differences? 
  \item What additional nuclear-theory calculations are needed and
    what is the optimal path for getting these improved models
    consistently and completely incorporated into neutrino-event
    generators so that they can be used to interpret
    neutrino-oscillation measurements?
    Can these advanced calculations cover the necessary kinematic
    space? 
    Can they be extended reliably to nuclei as heavy as argon?
  \item Are there nuclear effects that impact the axial current and if
    so, how large are these effects and how uncertain are they?
  \item Nuclear physics aside, is there physics associated with the
    basic neutrino-nucleon interaction that requires further
    attention, e.g. the elastic axial-vector form factor, the
    nucleon-$\Delta$ axial form factors, and/or radiative corrections?
  \item Can the relation between different approaches (e.g. 
    local/global Fermi gas, spectral function, random phase
    approximation, relativistic mean field, super-scaling, ab initio 
    and Green's Function Monte Carlo to name a few) be understood?
  \item Are there specific ways in which the neutrino experimental and
    theoretical (nuclear and particle) communities can communicate
    more effectively in order to make progress more rapidly?
\end{itemize}

To address these points efficiently and to make rapid progress
requires an effective and structured means of communication between
both the nuclear and particle experimental and theoretical
communities.   
NuSTEC (Neutrino Scattering Theory Experiment Collaboration)
\cite{Morfin:nuSTEV:2015}, consisting of experimentalists from every
neutrino interaction experiment, theorists from major nuclear
theoretical initiatives and representatives of all major Monte Carlo
simulation tools has been formed explicitly to address this need. 

On the experimental side, we can expect new data in the coming few
years from MicroBooNE, MINER$\nu$A \cite{MINERvA:WWW}, NO$\nu$A, SBND
\cite{SBN:2014}, and T2K.
This data will span neutrino energies ranging from a few hundred MeV
up to a few GeV and will probe a variety of nuclear targets including
helium, carbon, oxygen, argon, iron and lead.
It is generally recognised that for the next-generation programme to
address the questions listed above it must:
\begin{itemize}
  \item Include measurements of both $\parenbar{\nu}_\mu N$ and
    $\parenbar{\nu}_e N$ scattering to support the future neutrino
    CP-invariance violation program; 
  \item Report cross sections in the form of physical observables so
    as to decrease the model dependence of the measurements;
  \item Cover a large range of energies, targets, and final-state
    topologies to constrain uncertainties in theoretical models and
    simulation tools;
  \item Maintain as large a muon (electron) angular acceptance as
    possible and examine hadron emission to probe the physics of
    multi-nucleon correlations in charged-current neutrino
    interactions;
  \item Include tests of how well we know electron- versus
    muon-neutrino scattering to support future
    $\parenbar{\nu}_e$-appearance searches; and  
  \item Rely on accurate neutrino-flux simulations to produce precise
    neutrino cross-section measurements.
\end{itemize}
The experiments listed above will carry the programme forward for the
next four to five years.

The measurement of neutral-current coherent elastic neutrino-nucleus
scattering is important for experiments searching for astrophysical
phenomena such as supernov\ae~and in searches for dark matter.
The reaction is predicted by the Standard Model but is very difficult
to measure because the signal comes from low-energy nuclear recoil
only. 
Coherence requires that the neutrino energy should not exceed a few
tens of MeV. 
The most promising experimental projects are the COHERENT
\cite{Akimov:2015nza} and $\nu$GeN \cite{Bednyakov:Ed:2014}
experiments. 
COHERENT proposes to set up three detectors with different target
materials (xenon, germanium, CsI[Na]) close to the target at the
Spallation Neutron Source (SNS) at Oak Ridge National Laboratory. 
$\nu$GeN will be located at the Kalininskaya nuclear plant in Russia
and use low-threshold germanium detectors developed at JINR in Dubna. 

While it is challenging to predict the exact requirements of the
programme in the next decade, it is likely that increased precision
will be required.
NuPRISM \cite{Bhadra:2014oma} and nuSTORM 
\cite{Adey:2013afh,Adey:1537983,Kyberd:2013ii,Kyberd:2013ij,Adey:2013pio,Lackowski:2013ria,Adey:2014rfv}
are examples of experiments that have been proposed to provide the
necessary increase in precision.
NuPRISM exploits the off-axis technique to detect neutrinos at a
variety of narrow energy bands.
nuSTORM produces beams with equal quantities of muon and electron
neutrinos with a well known energy spectrum from the decay of muons
confined within a storage ring.

Over the coming years, significant theoretical advances are also to be
anticipated.
For example, reliable ab-initio computations of nuclear response
functions are expected to be performed in the region of the
quasi-elastic peak for light nuclear targets (nuclei up to carbon).
A collaboration of theorists within NuSTEC has been formed with the
goal of expanding these computations to nuclei up to argon and into
the relativistic regions necessary for pion production.  
Together with the experimentalists and representatives of the
Monte Carlo developers of NuSTEC these improvements will be
incorporated in the major Monte Carlo simulation tools.  
These considerations indicate that a decision on the future direction
of the neutrino-scattering programme could be made around 2020.

Studies at present and future experiments that help to ``decouple''
nuclear effects will be of benefit in developing a detailed
understanding of neutrino interactions.
Such studies include measurements of exclusive electron- and
photon-nucleus scattering cross sections as a function of the number
of pions and protons in the final state. 
In addition, measurements of pion-nucleus scattering cross sections as
a function of pion kinetic energy will be valuable to benchmark the
cascade models used in Monte Carlo simulations. 

\subsubsection{Facilities and experiments}
\label{SubSect:nuFT:Exp}

\paragraph*{MINER\boldmath{$\nu$}A}

\noindent\textbf{Physics goals} \\
\noindent 
MINER$\nu$A is a fine-grained detector that uses the Main Injector
neutrino (NuMI) beam \cite{MINERvA:WWW}.
The experiment is approved to collect an exposure corresponding to 
$10 \times 10^{20}$\,POT in neutrino mode and $12 \times 10^{20}$\,POT
in anti-neutrino mode in the medium-energy beam in addition to the
samples that have been collected using the low-energy beam.
The principal goals of the approved programme are to:
\begin{itemize}
  \item Study both signal and background reactions relevant to
    oscillation experiments;
  \item Study nuclear effects in inclusive reactions and exclusive
    final states;
  \item Study neutrino-scattering processes as a function of neutrino
    energy;
  \item Study differences between neutrinos and anti-neutrinos; and
  \item Measure the ``EMC Effect'' using neutrino deep inelastic
    scattering on lead, iron, carbon and hydrocarbon targets. \\
\end{itemize}

\noindent\textbf{Institutes 20; collaborators 70} \\
\noindent 
  Brazilian Center for Research in Physics (CBPF),
  Fermi National Accelerator Laboratory,
  University of Florida,
  University of Geneva,
  University of Guanajuato, Mexico,
  Hampton University,
  Massachussetts College of Liberal Arts,
  Northwestern University,
  Oregon State University,
  Otterbein University,
  Oxford University.
  Pontifical Catholic University of Peru,
  University of Pittsburgh,
  University of Rochester,
  Rutgers University,
  Tufts University,
  University of Minnesota (Duluth),
  Universidad Nacional de Ingenier\'ia, Peru,
  Universidad T\'ecnica Federico Santa Mar\'ia, Chile,
  College of William \& Mary. \\

\noindent\textbf{Measurement programme} \\
\noindent
MINER$\nu$A completed the data taking using the low-energy beam, which
has a mean energy of 3.5\,GeV, with an exposure corresponding to
$4\times10^{20}$\,POT.
Running using the medium-energy beam, with mean energy 6\,GeV, has
started and an exposure corresponding to $9\times10^{20}$ POT in
neutrino mode has been accumulated.
Running continues with the objective of integrating the approved
exposure.

\paragraph*{CAPTAIN-MINER\boldmath{$\nu$}A}

\noindent\textbf{Physics goals}  \\
\noindent 
CAPTAIN-MINER$\nu$A \cite{MINERvA:WWW,Berns:2013usa} proposed the use
of a detector system consisting of the CAPTAIN LArTPC and MINER$\nu$A
fine-grained detector in the on-axis NuMI beam. 
The principal goals of the experiment were to:
\begin{itemize}
  \item Measure neutrino-argon cross sections in the energy region of
    the first oscillation maximum for DUNE;
  \item Measure cross-section ratios on argon and scintillator to test
    models of nuclear effects; and 
  \item Take $6\times10^{20}$ POT in both neutrino mode and
    antineutrino mode in the NuMi medium-energy beam.
\end{itemize}
Captain-MINER$\nu$A received Phase~1 approval from the Fermilab PAC in
June 2015 but was not approved for funding in the summer of 2016.
This entry in the roadmap is retained for completeness. \\

\noindent\textbf{Institutes 31} \\
\noindent
  University of Alabama,
  Argonne National Laboratory,
  Brookhaven National Laboratory,
  University of California (Davis),
  University of California (Irvine),
  University of California, (Los Angeles),
  University of California, (Davis),
  Brazilian Center for Research in Physics (CBPF),
  University of Florida,
  Fermi National Accelerator Laboratory,
University of Guanajuato, Mexico,
  Hampton University,
  University of Hawaii,
  University of Houston,
  Lawrence Berkeley National Laboratory,
  Los Alamos National Laboratory,
  Louisiana State University,
  Massachusetts College Liberal Arts,
  Massachusetts Institute of Technology,
  University of Minnesota (Twin Cities),
  University of Minnesota (Duluth),
  University of New Mexico,
  Oregon State University,
  University of Pennsylvania
  University of Pittsburgh,
Pontifical Catholic University of Peru,
  University of Rochester,
  Stony Brook University,
  Tufts University,
  Universidad Nacional de Ingenieria, Peru,
  Universidad Tecnica Federico Santa Maria, Chile,
  College of William and Mary. \\

\paragraph*{MicroBooNE}

\noindent\textbf{Physics goals} \\
\noindent
An outline of the MicroBooNE experiment was given in section
\ref{SubSect:Sterile:SBL:Present:MicrobooNE}.
In addition to the sterile-neutrino search programme, MicroBooNE will
study the properties of neutrino interactions in argon and will
measure the cross sections for a wide variety of inclusive and
exclusive channels. \\

\noindent\textbf{Institutes} \\
\noindent
The institutes that form the MicroBooNE collaboration are listed in
section \ref{SubSect:Sterile:SBL:Present:MicrobooNE}. \\

\noindent\textbf{Measurement programme} \\
\noindent
The MicroBooNE collaboration plans to measure charged-current (CC) and
neutral-current (NC) neutrino-interaction cross sections on argon for
neutrino energies in the range $\sim 0.2<E_\nu<2$\,GeV with a
precision of $\sim 10\%$. 
This uncertainty is dominated by neutrino-flux uncertainties and is at
the level with which the Booster Neutrino beam flux was determined for
MiniBooNE.
It may be possible to reduce this uncertainty after incorporating
recent results from the HARP experiment which measured hadrons
produced by a replica of the Booster Neutrino Beam (BNB) target.

Specific cross section measurements will include:
\begin{itemize}
  \item $\nu_\mu$ CC inclusive (muon kinematics, hadronic energy,
    neutrino energy);
  \item $\nu_\mu$ CC $0\pi$ (muon and proton kinematics, proton
    multiplicities); 
  \item $\nu_\mu$ CC $\pi^+$ (muon and pion kinematics, nucleon
    emission); 
  \item $\nu_\mu$ CC $\pi^0$ (muon and pion kinematics, nucleon
    emission);
  \item $\nu_\mu$ NC $\pi^0$ (pion kinematics, nucleon emission);
  \item $\nu_\mu$ NC $\pi^\pm$ (pion kinematics, nucleon emission);
  \item $\nu_\mu$-induced single photon production; and 
  \item $\nu_\mu$-induced kaon, $\eta$, $\rho$, multi-pion
    production.
\end{itemize}
$\nu_e$ and $\nu_\mu$ cross section measurements can also be made from
the far off-axis NuMI beam, although this beam is not expected to be
as well-known as the on-axis BNB.

\paragraph*{Short Baseline Near Detector (SBND)}

\noindent\textbf{Physics goals} \\
\noindent
The SBND will record $\sim 2$ million interactions per year 
($\sim 1.5$ million $\nu_\mu$ CC and $\sim 12$k $\nu_e$ CC) in the TPC
active volume. 
Measurements of $\nu_\mu$Ar and $\nu_e$Ar cross sections will be
made.
The well-characterised neutrino flux will be used to measure:
\begin{itemize}
  \item Rare processes such as $\nu$-e scattering, strange-particle
    production, and coherent scattering with an argon nucleus; and
  \item Differential cross sections for exclusive channels and studies
    of nuclear effects in $\nu$Ar interactions. \\
\end{itemize}

\noindent\textbf{Institutes} \\
\noindent 
The institutes that form the SBND collaboration are listed in
section \ref{SubSect:Sterile:SBL:Next:SBN}. \\

\noindent\textbf{Measurement programme} \\
\noindent
Sufficient statistics to produce cross section measurements on partial
data set will be acquired quickly.
It is anticipated that all cross-section analysis will be completed
within a few years of the end of the approved exposure
($6.6\times10^{20}$\,POT).

\paragraph*{NO\boldmath{$\nu$}A}

\noindent\textbf{Physics goals} \\
\noindent 
The NO$\nu$A near detector is placed off-axis in the NuMI beam to
measure neutrino interaction cross sections with a liquid-scintillator
detector as part of the NO$\nu$A neutrino-oscillation programme.
The principal goals for the near detector are to:
\begin{itemize}
  \item Advance the understanding of neutrino-nucleus interactions in
    an intermediate energy range; 
  \item Constrain the off-axis NuMI flux with measurements of
    better-known interactions; and 
  \item Search for non-standard interactions and exotic phenomena.
\end{itemize}
To achieve these goals data will be taken under the following
conditions:
\begin{itemize}
  \item Neutrino energy range between 1\,GeV and 20\,GeV, with the
    bulk of $\nu_\mu$ flux between 1\,GeV and 3\,GeV; and
  \item Baseline exposure: $36\times 10^{20}$\,POT over 6 years.
    The fraction of the total exposure to dedicated to anti-neutrino
    running remains to be determined. \\
  \end{itemize}

\noindent\textbf{Institutes} \\
\noindent 
The institutes that form the NO$\nu$A collaboration are listed in
section \ref{Sect:AccBasedOsc:Present:NOvA}. \\

\noindent\textbf{Measurement programme} \\
\noindent 
The NO$\nu$A neutrino-interaction programme will run through the duration
of the NO$\nu$A experiment.  
Approximately 850,000 muon-neutrino charged-current interactions,
before cuts, are expected annually in neutrino-mode running in a 10\,t
fiducial mass.
This sample will contain contributions from quasi-elastic (nominally
26\%), resonant (39\%), pion-continuum/DIS (34\%) and coherent (1\%)
channels.
A neutral-current sample of one-third the size will also be
collected.
The electron-neutrino content is about 0.6\% of the total flux.  
First results from the NO$\nu$A neutrino-interaction program are
anticipated in US fiscal year 2016 and regular updates from the
multiple ongoing analyses will be released as they become available.

\paragraph*{T2K (ND280)}

\noindent\textbf{Physics goals} \\
\noindent
The T2K ND280 detector is sited 280\,m from the source of the J-PARC
off-axis neutrino beam to allow large neutrino- and
anti-neutrino-scattering data sets, which can be distinguished
event-by-event using ND280's 0.2\,T magnetic field, to be collected.
The principal goal of the ND280 experiment is to:
\begin{itemize}
  \item Make a variety of measurements of neutrino-nucleus
    interactions to improve the precision of the T2K
    neutrino-oscillation measurements. \\
\end{itemize}

\noindent\textbf{Institutes} \\
\noindent 
The institutes that form the T2K collaboration are listed in
section \ref{Sect:AccBasedOsc:Present:T2K}. \\
  
\noindent\textbf{Measurement programme} \\
\noindent 
The experiment is approved to take data until 2021 by which time it is
expected that an exposure corresponding to $7.8 \times 10^{21}$\,POT
will have been delivered.
The plan is to collect an equal exposure (POT) in neutrino and
antineutrino modes.

To achieve this exposure it is planned that the power to the
J-PARC neutrino target will rise to 750\,kW in 2019.  
An upgrade to the near detector suite is being studied.
A second phase of the experiment, exceeding 750\,kW beam power, is
under discussion. 

\paragraph*{WAGASCI test experiment}

\noindent\textbf{Physics goals} \\
\noindent
The principal goals of the WAGASCI experiment are to measure the
charged-current neutrino cross section on water and the ratio of the
neutrino-water cross section to the neutrino-plastic and neutrino-iron
cross sections with high precision and large angular acceptance.
The detector exploits a 3D grid-like structure of scintillator bars to
detect charged particles with $4\pi$ acceptance and high efficiency.
The experiment will be exposed to the T2K off-axis neutrino beam at
J-PARC. 
By focusing positive particles, the beam will be used to measure the
neutrino-water cross section with a total uncertainty of 10\% and the
cross section ratio water-to-hydrocarbon with a total uncertainty of
3\% using an exposure corresponding to $1 \times 10^{20}$\,POT. 
By focusing negative particles, the anti-neutrino-water cross section
and the cross-section ratio will be measured with a statistical
uncertainty of 5\% using $2\times 10^{20}$\,POT. \\ 

\noindent\textbf{Institutes 8} \\
University of Tokyo, 
Institute for Nuclear Research of the Russian Academy of Sciences,
Laboratoire Leprince-Ringuet, Ecole Polytechnique,
Kyoto University,
Osaka City University,
Institute for Cosmic Ray Research, University of Tokyo, 
University of Geneva,
Kavli Institute for the Physics and Mathematics of the Universe,
University of Tokyo.  \\  

\noindent\textbf{Milestones} \\
\noindent
The WAGASCI test experiment was approved by the IPNS/KEK Director and
endorsed by the J-PARC PAC in 2014.
Construction began in the summer of 2015.
The principal milestones for the project are:
\begin{itemize}
  \item Autumn 2016: start of data taking with the WAGASCI prototype at
    on-axis in the J-PARC neutrino beam;
  \item Within calendar year 2017: start data taking with the full
    WAGASCI detector off-axis in the J-PARC neutrino beam.
\end{itemize}

\paragraph*{ANNIE}

\noindent\textbf{Physics goals} \\
\noindent
The ANNIE experiment has been proposed to measure the rate of neutron
production in neutrino-water interactions using the Booster Neutrino
Beam at Fermilab. 
The goal is to contribute to the fundamental understanding of neutrino
interactions in nuclei. 
The measurements will benefit neutrino and nucleon-decay studies that
exploit large water-Cherenkov and liquid-scintillator detectors. \\

\noindent\textbf{Institutes 16} \\
\noindent 
Argonne National Laboratory,
Fermi National Accelerator Laboratory,
Iowa State University,
Ohio State University,
Queen Mary University of London,
University of California at Berkeley,
University of California at Davis,
University of California at Irvine,
University of Chicago,
University of Sheffield. \\
  
\noindent\textbf{Programme} \\
\noindent
In 2015 ANNIE was approved for a Phase I run.
The principal milestones of the project are:
\begin{itemize}
  \item Phase I; technical development and background
    characterisation: installation of the necessary equipment began in
    the summer of 2015 with data taking before the shutdown of the
    Booster Neutrino Beam in 2016.
    Additional data taking with prototype LAPPDs and micro-channel
    plates (MCPs) being possible in the last quarter of calendar 2016
    and the first two quarters of 2017; 
  \item Phase IIa; ANNIE physics run I with a moderate photo-cathode
    coverage: installation will start in the summer of 2017 with data
    taking from the autumn of 2017 through to spring 2018; and
  \item Phase IIb; ANNIE physics run II with full photo-cathode
    coverage: installation in the summer of 2018 with running from the
    autumn of 2018 to the autumn of 2020.  
\end{itemize}

\subsubsection{Future opportunities}
\label{SubSect:SptPrg:nuFT:DS} 

\paragraph*{nuSTORM} 

\begin{figure}
  \begin{center}
    \includegraphics[width=0.4375\textwidth]{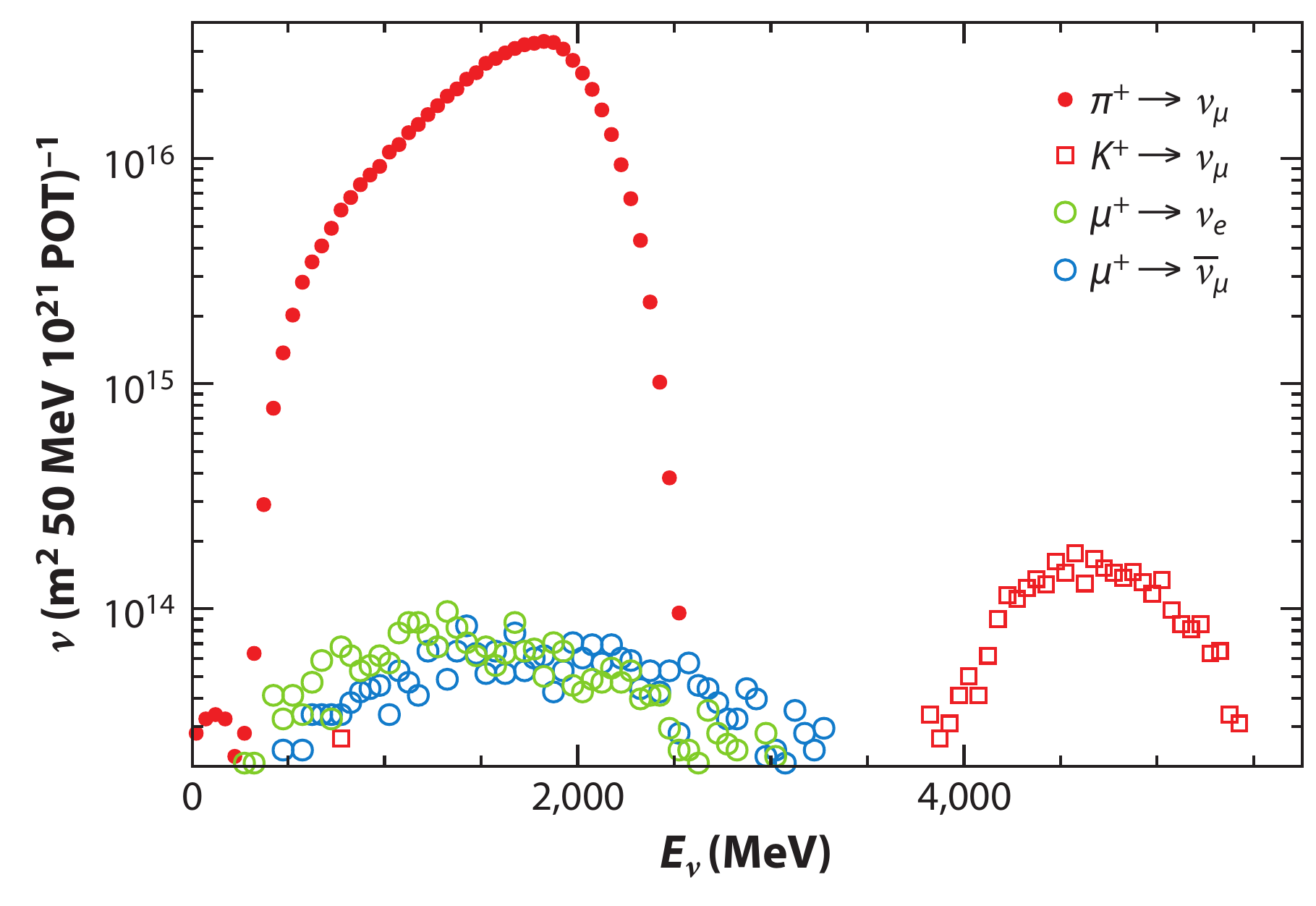} 
    \quad \quad \quad \quad 
    \includegraphics[width=0.4525\textwidth]{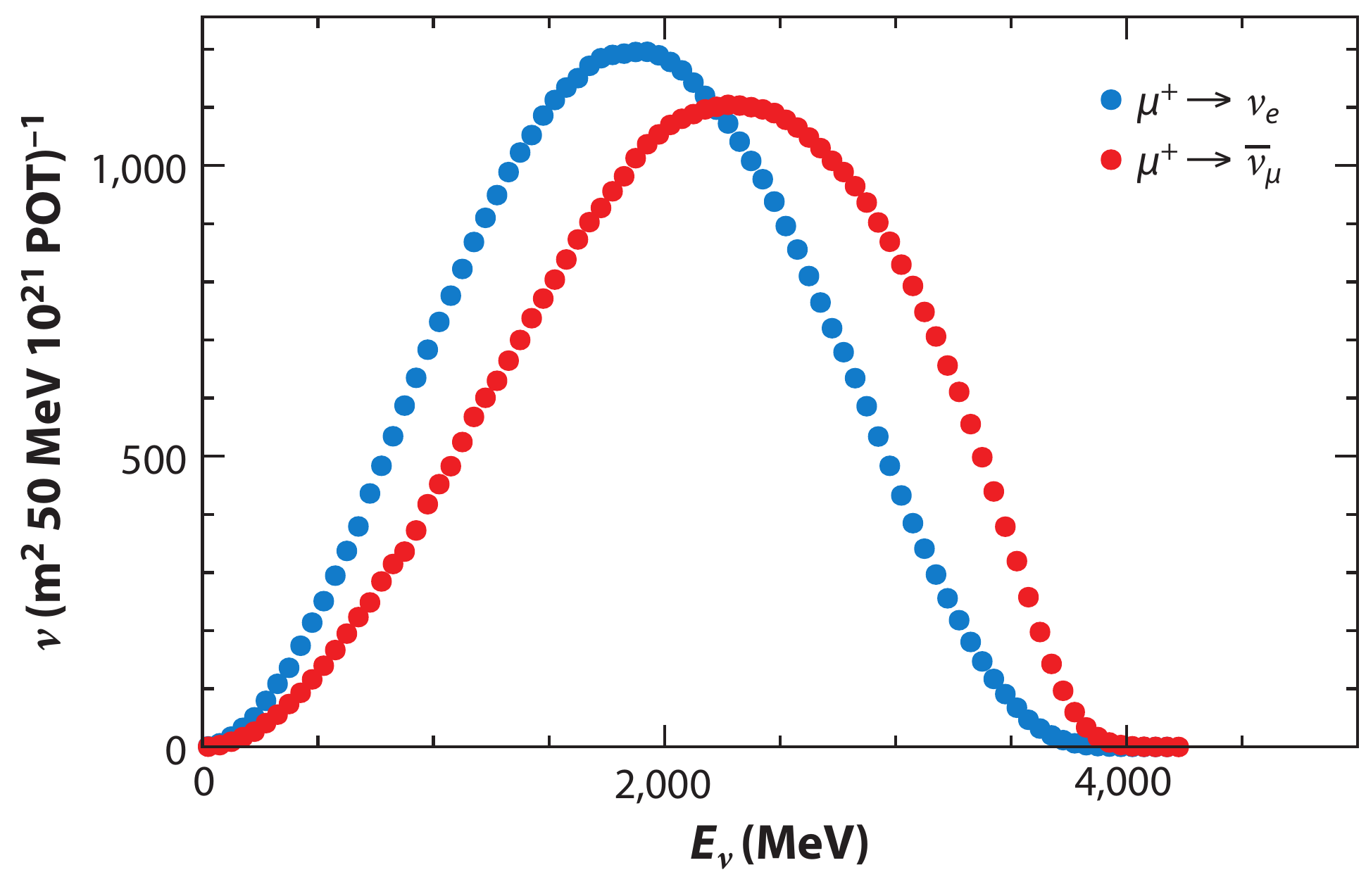} 
  \end{center}
    \caption{
      Left panel: simulated neutrino flux at the near detector
      location arising from pion decay during injection into the
      production straight.
      Right panel: simulated neutrino flux at the near detector
      location arising from muon decay.
    }
    \label{Fig:nuSTORM-flux}
  \end{figure}

\noindent\textbf{Physics goals/measurement programme} \\
\noindent
The nuSTORM concept was outlined in section
\ref{SubSect:Sterile:SBL:Future:nuSTORM}.
For the future neutrino-scattering programme nuSTORM offers the
following advantages:
\begin{itemize}
  \item The neutrino flux is precisely known; the normalisation may be
    determined to better than 1\% from the storage-ring
    instrumentation;
  \item The energy spectrum may be calculated precisely, exploiting
    constraints imposed by the storage ring and its instrumentation.
    The falling edge of the spectrum at high energy can be used to
    calibrate the detector response; and
  \item The muon-decay beam contains equal numbers of $\nu_\mu$ and
    $\bar{\nu}_e$ (or $\bar{\nu}_\mu$ and $\nu_e$).
    Since the $\pi \rightarrow \mu$ transition takes place in the
    neutrino-production straight, a bright ``flash'' of
    $\parenbar{\nu}_\mu$ is produced in the first pass of the beam
    through the production straight (see figure
    \ref{Fig:nuSTORM-flux}).
\end{itemize}
The simulated flux at the near detector 50\,m from the end of the
production straight is shown in figure \ref{Fig:nuSTORM-flux}.
The event rate in a 100\,T detector is large, making it possible to
measure single-differential cross sections at the few percent level.
The flux is sufficient for double-differential cross-section
measurements to be made.
 \\

\noindent\textbf{Institutes} \\
\noindent 
The list of institutions was given in the nuSTORM entry in  section
\ref{SubSect:Sterile:SBL:Future:nuSTORM}. \\

\noindent\textbf{Outlook} \\
\noindent 
The unique features of the nuSTORM beam (large $\parenbar{\nu}_e$
flux, precisely known energy spectrum) may be of substantial value in
determining cross sections for use in the analysis of the next
generation long-baseline oscillation experiments (DUNE and Hyper-K)
when the data samples are large.
Accounting for correlated and uncorrelated sources of systematic
uncertainty with a precision commensurate with the statistical
precision of the data samples will be critical in the combination of
results from a number of experiments.
The constraints that nuSTORM can provide are likely to be particularly
valuable in this combination.

Over the next four years or so, results from the present
neutrino-scattering programme will inform detailed studies of the
systematic uncertainties that will impact the search for CPiV.
Consideration of these studies may indicate that the precision which
nuSTORM can provide will be of benefit when the statistical weight of
the DUNE and Hyper-K data samples are large.
Since the DUNE and Hyper-K experiments seek to accumulate large data
sets towards the end of the next decade the Panel believes 2020 to
be a natural point at which to consider nuSTORM as part of the future
neutrino-nucleus cross-section measurement programme.

The possibility that nuSTORM can serve a neutrino-scattering programme
will be considered within the CERN Physics Beyond Colliders study
group~\cite{CERN:PBC:WWW:2017}.

\subsection{Detector development}
\label{SubSect:SptPrg:DetDev}

A substantial detector R\&D effort is underway to deliver the
capabilities necessary to realise the present and next generation of
accelerator-based neutrino-oscillation experiments.
The principal elements of this programme are reviewed in the
paragraphs which follow.

\subsubsection{ProtoDUNE-SP at the CERN Neutrino Platform}
\label{SubSubSect:SptPrg:DetDev:ProtoDUNE}

\noindent\textbf{Physics goals} \\
\noindent
ProtoDUNE-SP \cite{KutterPawloski:ProtoDUNE:2015} is a full scale
prototype single-phase liquid-argon TPC (LAr-TPC) that will be
operated at the CERN Neutrino Platform \cite{Nessi:CENF:2015}. 
The principal goal is to demonstrate the construction techniques that
will be used in the DUNE first 10\,kt single-phase detector module.
ProtoDUNE-SP will also allow experience to be gained with the
reconstruction and analysis of data from a large LAr-TPC.
Its contribution to the DUNE programme will be:
\begin{itemize}
  \item The measurement of the response of the detector to charged
    particles; and
  \item The validation of the performance of all the components, and
    the commissioning of the detector. \\
\end{itemize}

\noindent\textbf{Institutes} \\
Since protoDUNE-SP is part of the DUNE programme, the institute and
collaborator lists are those presented in section
\ref{Sect:AccBasedOsc:Next:DUNE}. \\

\noindent\textbf{Future programme} \\
Schedule is driven by the DUNE timescale: detector commissioning and
data run before end of 2018.

\subsubsection{WA104 at the CERN Neutrino Platform; ICARUS refurbishment}
\label{SubSubSect:SptPrg:DetDev:WA104}

\noindent\textbf{Physics goals} \\
The ICARUS-WA104 collaboration is refurbishing the ICARUS T600
liquid-argon TPC with new cryostats, new fast, and high-performance
electronics and with an upgraded light-detection system. 
The two TPCs were moved from the Gran Sasso Laboratory in Italy to
CERN in December 2014.
After the modules have been refurbished at CERN they will be shipped
to FNAL to be used as the far detector in the Short Baseline Neutrino
Programme described in section 
\ref{SubSect:Sterile:SBL:Next:SBN}. \\

\noindent\textbf{Institutes 10; collaborators 60} \\
\noindent
  CERN, 
  Catania University, INFN Catania, 
Pavia University, INFN Pavia, 
Padova University, INFN Padova, 
Gran Sasso Science Institute, Italy
Laboratori Nazionali di Frascati (Roma),
Laboratori Nazionali del Gran Sasso, Italy,
  INFN Milano Bicocca, 
  INFN Milano,
  INFN Napoli \\

\noindent\textbf{Future programme} \\
The refurbished detectors will be moved from CERN to Fermilab in 2017.
The modules will be installed at the 600\,m location on the Booster
Neutrino Beam for start of operation in 2018. 

\subsubsection{ProtoDUNE-DP (WA105) at the CERN Neutrino Platform;
  LBNO Dual-phase prototype}
\label{SubSubSect:SptPrg:DetDev:WA105}

\noindent\textbf{Physics goals} \\
\noindent
The WA105 collaboration is building a large liquid-argon TPC working
in double-phase (liquid and gas) mode
\cite{WA105:WWW:2014,Cantini:2015naq}. 
The TPC will be operated at the CERN Neutrino Platform
\cite{Nessi:CENF:2015}.
The TPC will contain a $6\times6\times6$\,m$^3$ volume of liquid
argon in a cryostat that is based on an industrial liquid-natural-gas
vessel.
Charge amplification and readout occurs in the gas phase allowing long
drift distances, a lower energy threshold and an improved
signal-to-noise ratio to be achieved.
The main goals are the development and test of the double-phase
technique, the validation of the technologies proposed to deliver
larger scale detectors and the measurement of the response of the
detector to beams of charged particles with energies ranging from
1\,GeV to 10\,GeV. Since December 2015, the project is embedded in the
DUNE experiment as ProtoDUNE-DP, one of the two prototype efforts
towards the deployment of four 10 kt FD detectors at SURF.

The WA105 technical programme includes:
\begin{itemize}
  \item The delivery of very high LAr purity in a non-evacuated tank
    (an oxygen contamination of $< 100$\,ppt is needed for an electron
    lifetime of 10\,ms);
  \item The construction and operation at very high voltage of a large
    field cage;
  \item Commissioning and operation of a large area micro-pattern charge
    readout system; 
  \item Implementation of the cold front-end charge readout
    electronics;
  \item Test of the long term stability of the WLS coating; and
  \item Integrated light readout and electronics.
\end{itemize}
The WA105 physics programme includes:
\begin{itemize}
  \item The reconstruction of $e$, $\pi^\pm$, $\pi^0$ and muons;
  \item Electron/$\pi^0$ separation;
  \item Calorimetry to reconstruct the full neutrino energy;
  \item Hadronic secondary interactions; and
  \item Measurements of pion/proton interaction cross sections on
    argon.  \\
\end{itemize}

\noindent\textbf{Institutes 20; collaborators 152} \\
Barcelona/IFAE (ESP),
Bucharest(RO), 
CERN, 
Centro de Investigaciones Energeticas, Medioambientales y Tecnologicas (CIEMAT), 
ETH Z\"{u}rich (CH), 
IFIN-HH (RO), IN2P3/APC, IN2P3/IPNL, IN2P3/LAPP, IN2P3/LPNHE, IN2P3/OMEGA,
Iwate University (Japan),
 CEA Saclay, Jyvaskyla (FI), 
Kyiv National University, National Centre for Nuclear Research (POL), 
National Institute of Technology, Kure College (Japan),
High Energy Accelerator Research Organization (KEK), 
University College London (UK), 
University Texas Arlington (USA)
\\

\noindent\textbf{Future programme} \\
\noindent
The principal steps in the development of the WA105 programme are:
\begin{itemize}
  \item Build the cryostat by April 2017;
  \item Assemble the detector in the cryostat in December 2017; and
  \item Be ready to take beam data in March 2018.
\end{itemize}

\subsubsection{Liquid Argon In A Testbeam (LArIAT)}
\label{SubSubSect:SptPrg:DetDev:LArIAT}

\noindent\textbf{Physics goals} \\
\noindent
LArIAT, the Liquid Argon In A Testbeam experiment, is located at the
Fermilab Test Beam Facility in a tertiary beam of pions, muons,
electrons, protons, and kaons, with momentum in the range of 
$\sim 200$\,MeV/c--1400\,MeV/c.
Operating a LArTPC in a well-understood beam of charged particles
enables detailed characterisation of the response of the detector to
each type of particle (corresponding to a range of deposited
ionization energy, $dE/dx$) over the full range of incoming particle
energies. 
The experimentally measured response characteristics will be used to
tune simulations for neutrino experiments in order to reproduce more
accurately the characteristics of particles traveling through the
argon. 
The data collected in LArIAT will also allow the collaboration to
evaluate how well the existing LArTPC particle identification
algorithms are functioning and to improve these algorithms based on
data rather than simulation.

During its two running periods, LArIAT will record more than 1 million
interactions of charged particles in liquid argon in the range of
energies that is relevant to current and future short- and
long-baseline neutrino experiments. 
The products of neutrino interactions seen in MicroBooNE, SBND, and
DUNE exactly match the type and energy range of particles collected in
LArIAT. 
The collaboration will use the data collected in this LArTPC to:
\begin{itemize}
  \item Measure total and exclusive interaction cross sections of
    charged pions, kaons, and protons;
  \item Experimentally evaluate the capability of LArTPCs to
    distinguish electrons from photons, which is key to understanding
    the primary background of neutrino oscillation experiments;
  \item Develop criteria for charge-sign determination in the absence
    of a magnetic field, based on topologies; 
  \item Study kaon identification, and its implications for
    proton-decay experiments that exploit LAr-TPCs;  
  \item Study stopping muons and decay electrons; and
  \item Extend the use of scintillation-light readout beyond its
    traditional function in neutrino detectors (for triggering
    purposes) to a better system that can aid in full calorimetric
    energy reconstruction.  \\
\end{itemize}

\noindent\textbf{Institutes (21), collaborators (85)} \\
 Universidade Federal do ABC,
 Universidade Federal de Alfenas,
 Boston University,
 Universidade Estadual de Campinas,
 University of Chicago,
 University of Cincinnati,
 Fermi National Accelerator Laboratory,
 Universidade Federal de Goias, 
 Istituto Nazionale di Fisica Nucleare,
 KEK,
 Lousiana State University,
 University of Manchester, 
 Michigan State University,
 University of Minnesota, Duluth,
 University of Pittsburgh,
 Syracuse University,
 University of Texas, Arlington,
 University of Texas, Austin,
 University College London,
 The College of William and Mary,
 Yale University. \\

\noindent\textbf{Future programme} \\
\noindent
Sufficient statistics to produce differential cross section
measurements for charged pions have been acquired and the additional
data being collected in Run-II will enable high-statistics cross
section measurements of particles that are less abundant in the beam. 
All analyses listed above will be completed within a few years of the
end of the approved Run-II period (ending July 2016).

Future plans under consideration for additional short runs include:
\begin{itemize}
  \item Direct comparison of 3-mm versus 5-mm wire pitch, under the
    reproducible experimental conditions afforded by the test-beam
    environment;
  \item Addition of LAPPDs (new, very fast large area photon
    detectors) to the beamline; 
  \item Novel scintillation light collection studies in the TPC using:
    LAPPDs, ARAPUCA (a new type of light wavelength converter and
    concentrator), acrylic bars with SiPM readout,
    capacitatively-coupled SiPM readout via TPC wire planes, and
    detection of the infrared component of LAr scintillation light; 
  \item Methane doping of the LAr for charge response linearization;
    and
  \item New cold electronics, including a fast ADC stage.
\end{itemize}

\subsubsection{CHIPS}
\label{SubSubSect:SptPrg:DetDev:CHIPS}

\noindent\textbf{Physics goals} \\
\noindent
The Cherenkov detectors In mine PitS (CHIPS) collaboration seeks to
develop a water containment and purification system for deployment in
lakes created by mining (``pits'') \cite{Adamson:2013xka}.
The water-containment system will be instrumented with a relatively
sparse (7\%--10\%) covering of 3-inch photo-multipliers which use time
and charge information to reconstruct electron- and muon-neutrino
events, thus forming a water Cherenkov detector that has a low cost
per kilo-ton.
The deepest pit illuminated by the Main Injector neutrino beam from
Fermilab has been identified (the ``Wentworth'' pit) and a 35\,t
prototype detector was deployed underwater for two seasons. 
The full detector is envisaged to be capable of measuring the CP
phase, $\delta_{\rm CP}$, with a precision of $\sim 30^\circ$
standalone and with a precision of $\sim 20^\circ$ when CHIPS is
combined with T2K and NO$\nu$A. 
Water purification has been demonstrated and light-attenuation lengths
greater than 50\,m at a wavelength of 405\,nm has been demonstrated.
In the present phase of the experiment the collaboration seeks to
demonstrate a water containment structure that would be applicable
to any water-Cherenkov detector.  \\

\noindent\textbf{Institutes (13), collaborators (36)} \\
University of Cincinnati,
Fermi National Accelerator Laboratory, 
Iowa State University (Ames), 
University College London,
University of Minnesota (Twin Cities),
University of Minnesota (Duluth),
University of Pittsburgh, 
Stanford University, 
University of Texas (Austin),
College of William \& Mary,
University of Wisconsin (Madison),
Large Lakes Observatory, University of Minnesota (Duluth),
Brookhaven National Laboratory\\

\noindent\textbf{Future programme} \\
\noindent
The principal steps in the proposed CHIPS development programme are:
\begin{itemize}
  \item Construction of 30\,m diameter endcaps on the surface at the
    Wentworth pit;
  \item Technical review followed by construction of 10\,kt structure;
    and
  \item Instrumentation of the  10\,kt detector.
\end{itemize}

\subsubsection{NuPRISM and TITUS}
\label{SubSubSect:SptPrg:DetDev:nuPRISMTITUS}

Currently, two ``intermediate'' baseline detectors, NuPRISM and TITUS,
to be situated between 1\,km and 2\,km from the target of the J-PARC
neutrino beam are under consideration.\\

\noindent {\bf Physics goals}\\
\noindent
NuPRISM exploits the trend for the peak of the energy spectrum of
conventional neutrino beams to move towards lower and narrower energy
distributions as one moves ``off'' the beam axis. 
By observing neutrinos produced at a large range of off-axis angles in
a large water Cherenkov detector, NuPRISM can measure
neutrino-interaction properties as a function of the incident neutrino
energy due to this variation. 
The method allows the direct measurement of critical quantities and
distributions, such as the outgoing lepton differential cross section
as a function of neutrino energy, that are inaccessible to
conventional techniques where only the flux-averaged distributions can
be measured in a model-independent way. 
Extensions of this technique also allow NuPRISM data to be used to
predict the lepton kinematics of the oscillated neutrinos at a far 
detector site (e.g. Super-Kamiokande in the case of T2K or
Hyper-Kamiokande) and to constrain cross section differences between
$\nu_{e,\mu}/\bar{\nu}_{e,\mu}$ that are important sources of
systematic uncertainty in a CP-invariance violation study.
As described below, gadolinium sulfate (Gd$_2$(SO$_4$)$_3$) can be
dissolved in the water to enhance the detection of neutrons emitted in
neutrino interactions.
The same energy variation also allows highly sensitive searches for
large $\Delta m^2$ neutrino oscillations that may result from the
presence of sterile neutrinos. 
With an exposure of $7.5 \times 10^{21}$ protons-on-target in neutrino
mode during the T2K second phase, NuPRISM has sensitivity to exclude
LSND allowed values of the sterile oscillation parameters with
$5\sigma$ significance for most values of $\Delta m^2$, and with
$3\sigma$ significance for all values of $\Delta m^2$. 

TITUS is a proposed water Cherenkov detector in the J-PARC neutrino
beam consisting of a cylindrical vessel oriented horizontally to
increase containment of high energy muons. 
Situated at the same off-axis angle as Hyper-Kamiokande, the
intermediate distance reduces near/far flux differences in the
unoscillated flux to $\lsim 1\%$. 
A magnetised muon range detector (MRD) allows sign selection for muons
which exit the water volume.
A key element of the detector design is Gd$_2$(SO$_4$)$_3$ loading;
the very high cross section and energetic photon emission from neutron
capture on gadolinium nuclei provide a clean signature by which to
identify neutrons emitted from neutrino interactions. 
The Gd$_2$(SO$_4$)$_3$-loading allows TITUS to perform detailed
studies of neutron emission in neutrino interactions with a powerful
capability to separate $\nu_\mu$ and $\bar{\nu}_\mu$ interactions
using the MRD. 
This may have important applications for the long-baseline
neutrino oscillation program at Hyper-K, where neutron
counting can allow statistical separation of neutrino and
anti-neutrino interactions, leading to improvements in sensitivity. 
Moreover, the physics programme of the TITUS detector goes beyond the
oscillation analysis extending to neutrino interaction measurements,
supernova detection, Standard Model measurements and dark matter
searches.  \\

\noindent {\bf Institutes}\\
\noindent
Currently, TITUS is formally part of the Hyper-Kamiokande
collaboration, but its unique benefits are also applicable to T2K and
its second phase.
NuPRISM is proposed as a separate experiment that can serve as an
intermediate detector for T2K or Hyper-Kamiokande while independently
pursuing other measurements such as neutrino interaction studies and
sterile neutrino searches. \\

\begin{description}
  \item{\underline{NuPRISM: institutes 27; collaborators 50}} \\
  University of British Columbia, 
University of California (Irvine),
University of Geneva, 
High Energy Accelerator Research Organization (KEK), 
IFAE, Barcelona,
Imperial College London, 
Institute for Nuclear Research of the Russian Academy of Sciences, 
Kavli Institute for the Physics and Mathematics of the Universe, 
Todai Institute for Advanced Study, 
University of Tokyo,
Kyoto University, 
Michigan State University, 
Stony Brook University, 
Osaka University,  Research Center for Nuclear Physics (RCNP), 
University of Regina, Canada, 
University of Rochester, 
University of Sheffield,
STFC Rutherford Appleton Laboratory, 
University of Tokyo, Institute for Cosmic Ray Research, Research Center for Cosmic Neutrinos, 
Tokyo Institute of Technology, 
University of Toronto, 
TRIUMF,  
Warsaw University of Technology, 
York University, Canada, 
Tokyo Metropolitan University\\
\end{description}

\noindent {\bf Next steps:} \\
Stage 1 approval was granted to NuPRISM in 2016.
The next steps are:
\begin{itemize}
  \item The construction of an intermediate detector facility outside
    of J-PARC to house detector; and
  \item Detector construction, with the goal of starting operations as
    soon as possible after the J-PARC Main Ring power supply upgrades
    that will increase the repetition cycle needed for $>$750 kW
    neutrino beams at J-PARC.
\end{itemize}
TITUS:
\begin{itemize}
  \item Submit proposal for running during T2K-II.
  \item Phased implementation.
\end{itemize}

\subsubsection{Deep sea multi-Megaton detectors}
\label{SubSubSect:SptPrg:DetDev:DeepSea}

\noindent\textbf{Physics goals} \\
\noindent
The concept for a detector that exploits a large volume of sea water
to produce a multi-megaton-scale detector illuminated by a powerful
neutrino beam has been described in \cite{Vallee:2016xde}. 
The KM3NeT collaboration is currently validating second generation,
compact, modular instrumentation that is suited for the detection in
sea water of neutrinos with energies of a few GeV
\cite{Brunner:2013lua,Adrian-Martinez:2016fdl}.
This technology would allow volumes of order tens of Megatons to be
instrumented.
Such a large detector located at a sufficiently long baseline 
($\ge 2500$\,km to yield a first oscillation maximum above $\sim
5$\,GeV) would have the potential to measure oscillation parameters
with high precision.
 
Opportunities for such programs would rely on the feasibility of a
neutrino beam from, for example, Fermilab to the existing
infrastructures NEPTUNE and OOI offshore of British Columbia, which
have been established as deep-sea observatories; this project is
referred to as the ``Pacific Neutrinos''
project~\cite{Vallee:2016xde}. 
Alternatively, a beam might be sent to the KM3NeT/ORCA detector that
is being developed offshore of Toulon (see 6.1.4).   
To assess the potential of such configurations, quantitative studies
must be performed with an optimised detector configuration.
An in-situ validation of the technology with the ORCA detector should
be carried out, prototypes should be deployed in NEPTUNE/OOI and an 
investigation of the characteristics of the deep-sea candidate sites
should be carried out.

\subsubsection{THEIA; advanced scintillation detector concept}
\label{SubSubSect:SptPrg:DetDev:THEIA}

\noindent\textbf{Physics goals} \\
\noindent
THEIA \cite{Gann:2015fba,WsNuPrgInJp:Svoboda:2015} is a proposed
large-scale experiment using water-based liquid-scintillator (WbLS)
technology.
The detector would be located underground at the LBNF far site at the
Homestake mine in South Dakota and would provide a complementary
program to that which the DUNE LAr-TPCs will carry out. 
THEIA would provide a broad program of physics, including
solar-neutrino measurements, the study of geoneutrinos, supernova
neutrinos, nucleon decay, the determination of the neutrino-mass
hierarchy, the search for evidence of leptonic CP-invariance violation
and the search for neutrinoless double-beta decay.
The collaboration plans to mount demonstrations of the required WbLS
and fast-timing technology by 2021. \\

\noindent\textbf{Institutes (24)} \\
\noindent
RWTH Aachen University (Germany),
University of Alberta,
Tsinghua University Beijing,
Brookhaven National Laboratory,
Brunel University,
University of California (Berkeley),
University of California (Davis),
University of California (Irvine),
University of Chicago,
Columbia University,
TU Dresden,
University of Hamburg,
FZ Juelich,
University of Hawaii at Manoa,
Hawaii Pacific University,
Iowa State University,
University of Jyvaskyla,
Laurentian University,
Lawrence Berkeley National Laboratory,
Lawrence Livermore National Laboratory,
Los Alamos National Laboratory,
Johannes Gutenberg-University Mainz,
University of Maryland,
MIT,
TUM Physik-Department,
University of Oulu,
University of Pennsylvania,
Princeton University,
Queens University,
Sandia National Laboratories,
University of Sheffield,
University of Toronto,
Virginia Polytechnic Inst. \& State University,
University of Washington

\subsection{Accelerator development}
\label{SubSect:SptPrg:AccDev}

Advances in accelerator technique are required to deliver the
next-generation of experiments.
In addition, R\&D is required to provide beams for which the
associated systematic uncertainties can be reduced substantially below
those of the conventional, pion-decay, beams.
Such beams will be necessary to take the programme beyond the next
generation of accelerator-based neutrino experiments.

\subsubsection{MICE}
\label{SubSubSect:SptPrg:AccDev:MICE}

\noindent\textbf{Physics goals} \\
\noindent
Muon beams of low emittance provide the basis for the intense,
well-characterised neutrino beams necessary to elucidate the physics
of flavour at the neutrino factory and to provide lepton-antilepton
collisions of up to several TeV at the muon collider.
To deliver beams with the properties necessary to meet the
specifications of these facilities requires that the volume of phase
space occupied by the tertiary muon beam be reduced (cooled).
Conventional beam-cooling techniques can not be used as the short
muon lifetime (2.2\,$\mu$s at rest) would lead to unacceptably large
loss of beam intensity.
Ionization cooling is the novel technique by which it is proposed to
cool the muon beam.
The international Muon Ionization Cooling Experiment (MICE) has been
approved to \cite{MICEproposal:2003,MICE:Note:53:2003}:
\begin{itemize}
  \item Design, build, commission and operate a realistic section of
    cooling channel; and
  \item Measure its performance in a variety of modes of operation
    and beam conditions.
\end{itemize}
The results will allow the optimisation of cooling-channel design for
use at the neutrino factory and muon collider. \\

\noindent\textbf{Institutes 28; collaborators 155} \\
\noindent
Department of Atomic Physics, St.~Kliment Ohridski University of Sofia,
Institute of High Energy Physics, Chinese Academy of Sciences, 
Sichuan University, 
Sezione INFN Milano Bicocca, 
Sezione INFN Napoli and Dipartimento di Fisica, 
Sezione INFN Pavia and Dipartimento di Fisica, 
Sezione INFN Roma Tre e Dipartimento di Fisica,
Osaka University, Graduate School of Science, 
High Energy Accelerator Research Organization (KEK), 
Nikhef, Amsterdam, 
CERN, 
DPNC, Section de Physique, Universit\'e de Gen\`eve, 
School Of Engineering and Design, Brunel University, 
STFC Daresbury Laboratory, 
School of Physics and Astronomy, Kelvin Building, The University of Glasgow, 
Department of Physics, Blackett Laboratory, Imperial College London, 
Department of Physics, University of Liverpool, 
Department of Physics, University of Oxford, 
STFC Rutherford Appleton Laboratory, 
Department of Physics and Astronomy, University of Sheffield, 
Department of Physics, University of Strathclyde, 
Department of Physics, University of Warwick, 
Brookhaven National Laboratory, 
Fermilab, 
Illinois Institute of Technology, Chicago, 
Department of Physics and Astronomy, University of Iowa, 
Lawrence Berkeley National Laboratory, 
University of Mississippi, 
University of California, Riverside. \\

\noindent\textbf{Future programme} \\
\noindent
It is proposed that the cooling-demonstration experiment will be
carried out in 2018
\cite{MICEmine:Document:159,MICEmine:Document:160,Bogomilov:2017vwz}.
In order to study the manner in which the properties of beam-energy
absorber materials and lattice parameters determine the
ionization-cooling effect, the collaboration is carrying out
``Step~IV''. 
In Step~IV, a single absorber is sandwiched between two spectrometer
modules allowing control of optical functions at the absorber and
precision measurement of muon parameters upstream and downstream of
the absorber.
The goals of the Step~IV programme are to 
\cite{MICEmine:Document:159}:
\begin{itemize}
  \item Measure the material properties of liquid hydrogen and
    lithium hydride that determine the ionization-cooling performance;
    and
  \item Observe the reduction of normalised transverse emittance.
\end{itemize}
The Step~IV programme will be executed in 2016/17 after which it is
proposed that the experiment be reconfigured for the demonstration of
ionization cooling.
The goals of the cooling-demonstration experiment are to
\cite{MICEmine:Document:159,Bogomilov:2017vwz}:
\begin{itemize}
  \item Observe of transverse-emittance reduction with
    re-acceleration; and
  \item Observe of transverse-emittance reduction and the evolution of
    longitudinal emittance and canonical angular momentum.
\end{itemize}

\subsubsection{RaDIATE}
\label{SubSubSect:SptPrg:AccDev:RaDIATE}

\noindent\textbf{Scientific goals} \\
\noindent
As proton accelerator particle sources become increasingly powerful,
there is a pressing need to understand and to predict the radiation
response of structural window, target and related component materials.
The RaDIATE Collaboration (Radiation Damage In Accelerator Target
Environments) draws on existing expertise in related fields in fission
and fusion materials research to formulate and implement a research
program that will apply the unique combination of facilities and
expertise at participating institutions to a broad range of high-power
accelerator projects of interest to the collaboration (in general,
these projects include neutrino and muon sources neutron spallation
sources, rare isotope ion-beam sources and collimation for
high-intensity accelerator facilities).  
The broad aims of the RaDIATE collaboration are to:
\begin{itemize}
  \item Generate new and useful materials data for application within
    the accelerator and fission/fusion materials communities;
  \item Recruit and develop new scientific and engineering experts who
    can cross the boundaries between these communities; and 
  \item Initiate and coordinate a continuing synergy between research
    in these currently disparate communities, benefiting both
    proton-accelerator applications in science and industry and
    nuclear fission- and fusion-energy technologies.
\end{itemize}
The ultimate ambition is to be able not only to predict operating
lifetimes for as many materials of interest as possible in terms of
integrated proton fluence for the high-energy proton-accelerator
parameter space (e.g. temperature, dose rate, duty factor, dynamic
stress) but also enable the development of radiation-damage and
thermal-shock tolerant materials.
Results of these studies will enable the robust and safe design,
fabrication, and operation of high power/intensity accelerator-target
facilities such as the Long Baseline Neutrino Facility, the neutrino
factory and the muon collider. \\ 

\noindent\textbf{Institutes 11; collaborators 32} \\
\noindent
FNAL, 
Science and Technology Facilities Council,
GSI, Darmstadt,
Oxford University, 
Brookhaven National Laboratory, 
Pacific Northwest National Laboratory, 
Oak Ridge National Laboratory, 
Michigan State University, 
European Spallation Source, 
Los Alamos National Laboratory, 
Argonne National Laboratory,
Centro de Investigaciones Energ\'eticas, Medioambientales y
Tecnol\'ogicas (Center of Energy, Environmental and Technological
Research) \\
In addition, an MOU revision is being prepared to add CERN and J-PARC.\\

\noindent\textbf{Current and future programme} \\
\noindent
RaDIATE related activities are focused on the determination of
critical properties of irradiated materials for application in the
accelerator-target environment, determining optimal
irradiation-environment parameters (e.g. irradiation temperature) for
maximum target-component lifetime, developing application-specific
material-testing methods and verifying models in prototypic loading
environments and educating and developing new experts in the
accelerator-target community.
Major current and near future activities include:
\begin{itemize}
    \item Post-irradiation examination (PIE) of the recovered NuMI
      beryllium primary beam window at Oxford University (2015-16);
    \item High intensity beam test of beryllium, ``BeGrid'', at CERN's
      HiRadMat facility and subsequent PIE at Oxford University
      (2015-16);
    \item Development of a hot-cell compatible fatigue testing machine
      at Fermilab (2015-16);
    \item PIE of the recovered NuMI target, NT-02, graphite fins at
      Pacific Northwest National Laboratory (PNNL) (2016);
    \item Determination of strength model parameters describing
      high-strain-rate mechanical behaviour of beryllium at Southwest
      Research Institute (SwRI) (2016);
    \item Irradiation and subsequent PIE of multiple materials at the
      Brookhaven Linac Isotope Producer (BLIP) at Brookhaven National
      Laboratory (BNL) (2016--2018). PIE will occur at multiple
      institutions. 
      Materials include beryllium, graphite, SiC-coated graphite,
      silicon, aluminium alloys, titanium alloys, molybdenum alloy
      (TZM) and iridium;
    \item Follow-on high intensity beam experiment(s) at CERN's
      HiRadMat facility on beryllium and other target/window candidate
      materials, including materials irradiated in BLIP irradiation
      (2017--2018); and
    \item Triple-beam ion irradiation and subsequent PIE of beryllium
      at the Michigan Ion Beam Laboratory (MIBL) and University of
      Michigan, PNNL, and Oxford University (2017--2019).
\end{itemize}

It is expected that, starting in 2019 and extending to about 2023,
there will be at least one more high-energy proton irradiation run on
multiple materials and several low-energy ion irradiations with
subsequent PIE from both irradiations.
In addition, it is expected that PIE activities will continue on
irradiated materials recovered from operating beam-lines as they become
available. 
The culmination of all studies is expected in 2024--2025 when
material-irradiation data and validated modelling methods will be
available to inform the design, construction and operation of next
generation accelerator-target facilities. 

\subsubsection{MOMENT}
\label{SubSubSect:SptPrg:AccDev:MOMENT}

MOMENT (MuOn-decay MEdium baseline NeuTrino beam facility) has been
proposed in China to carry out the CP-phase measurement using
neutrinos from muon decays.
MOMENT will use a muon beam in a long straight decay channel without
ionization cooling, thus avoiding the some of the technical
difficulties associated with cooling channels and muon acceleration.
A neutrino beam of average energy in the range 200--300\,MeV is
produced to serve a medium baseline experiment at a baseline of 
$\sim 150$\,km.
The comparatively low neutrino energy has the benefit of reducing the
$\pi^0$ background.
Taking the advantage of a strong CW superconducting-linac-development
program in China, the principal goal of which is to serve an
accelerator-driven nuclear reactors system (ADS), the proton driver
for MOMENT also uses a 15\,MW, 1.5\,GeV CW linac that could be a
parasitic program at an ADS experimental facility or a dedicated linac
with much-reduced redundancy compared to that required by ADS
applications.
The concept for a fluidised tungsten granular target is under
development; the more developed mercury-jet target proposed for the
neutrino factory is the ``back-up option''. 
A Gadolinium-doped water Cherenkov detector with a mass of 500\,kt is
being considered.
Different combinations of neutrino sources, including decay-at-rest
neutrinos and pion-decay neutrinos, are also being considered.

A study group with $\sim 40$ members, which is led by IHEP and
includes other Chinese institutions and non-Chinese institutions, is
working on the pre-conceptual design.  

\subsection{Development of software tools}
\label{SubSect:SWnC}

\noindent
{\bf Cross-sections:} \\
In recent years, a large number of experimental results on
neutrino-nucleus cross-sections have become available, resulting in a
steadily intensifying effort to develop and improve Monte Carlo
neutrino event generators such as  Genie\cite{genie}, NEUT\cite{neut},
and NuWro\cite{nuwro}.
Likewise, considerable theoretical activity has been devoted towards a
better understanding of the underlying nuclear- and particle-physics
effects, with a variety of theoretical methods and approaches
(e.g. (super-)scaling, Greens Function Monte Carlo, transport models,
random phase approximation, lattice QCD, effective field theory, etc.)
now being employed with the aim of better explaining and modelling the
measurements.
   
Given the multitude and complexity of the approaches, which often have
disparate ranges of applicability based on what type of neutrino
interaction, kinematic range, and features (e.g. exclusive hadronic
final states versus inclusive lepton kinematics) are targeted, a
persistent challenge has been constructing a consistent overall model
that can capture relevant theoretical developments. 
This has impeded the incorporation of new theoretical developments
into the computational tools commonly used in Monte Carlo event
generators.
    
This issue is exacerbated  by the lack of a unified framework for
formally communicating and implementing these theoretical or
experimental advances, thus requiring  an ad hoc implementation
of each of them by the Monte Carlo developers.
Progress requires the development of a more robust interface between
nuclear- and neutrino-physics experiments, analogous measurements in
electron scattering and photo-production, particle-/nuclear-theory
models/calculations, and Monte Carlo event generators.
To achieve such a goal, in turn, requires coherent, cooperative
interaction between nuclear theorists, particle theorists and the
relevant experimental communities.
There is a growing, currently ongoing activity (e.g. NuSTEC
\cite{Morfin:nuSTEV:2015}) to establish collaborative efforts and to
identify the most pressing questions.
Nonetheless, there is still the need to develop coherent mid-term and
long-term strategies.
A possible strategy is to start working on a universal Monte Carlo
event generator.
The following issues must be addressed in order to ensure the success
of the program: 
\begin{itemize}
  \item The Monte Carlo must get every-day support to answer
    questions, fix bugs, etc.;
  \item Regular training should be organised for users from
    neutrino oscillation experimental groups;
  \item Detailed documentation must be maintained and made
    available to the community; and
  \item Results of new neutrino cross-section measurements should
    be made available in a useful format and promptly included in
    data-analysis tools and global fits. 
\end{itemize}
In addition to event generators, the community will need to ensure
that other simulation frameworks (e.g. Geant4 \cite{Ago03}) that are
used to predict fluxes from neutrino beams and simulating detector
response are  accurate enough to fulfil the increasingly demanding
needs of the next generation of neutrino experiments. \\

\noindent
{\bf Combining different experimental results and global fits:} \\
The importance of combined fits to the data from the various neutrino
experiments, which allows advantage to be taken of the complementary
information on the parameters that the different experiments provide,
has been widely demonstrated in recent years. 
Going forward, the combination of information contained in different
data sets is bound to become more important and more challenging,
especially when it comes to looking for CP-invariance violation and
testing the three-flavour paradigm.
In order to facilitate this process, it is important to develop data
formats and data analyses such that communication between the
experimental and theory communities, and among different experimental
efforts, is optimal.
Future releases of data should not be limited to the final results or
to the experimental data themselves, but also include all the
information (such as effective areas or resolution functions) required
to reproduce the fit outside of the collaboration.
The main goal is to allow apples-to-apples comparisons of different
experimental results and the combination of different data sets in
order to test the validity of different theoretical and
phenomenological models, and to measure parameters, including
mass-squared differences, mixing angles, and CP-odd phases.

\vfill

\subsection{Conclusions and recommendations}
\stepcounter{nuPanel-RM-Conc-Sect}

\noindent
\framebox[\textwidth][l]{
  \parbox[c]{0.98\linewidth}{
        \begin{description}
      \stepcounter{nuPanel-RM-Conc-Sect-Conc}
      \item[\arabic{nuPanel-RM-Conc-Sect}.\arabic{nuPanel-RM-Conc-Sect-Conc}:]
        An appropriate programme of hadro-production and neutrino
        cross-section measurement is required to allow the present and
        next generation of long- and short-baseline experiments to
        achieve their full potential. 
    \end{description}

        \begin{description}
      \stepcounter{nuPanel-RM-Conc-Sect-Conc}
      \item[\arabic{nuPanel-RM-Conc-Sect}.\arabic{nuPanel-RM-Conc-Sect-Conc}:]
        An appropriate programme of hadro-production and neutrino
        cross-section measurement is required to allow the present and
        next generation of long- and short-baseline experiments to
        achieve their full potential. 
      \stepcounter{nuPanel-RM-Conc-Sect-Conc}
      \item[\arabic{nuPanel-RM-Conc-Sect}.\arabic{nuPanel-RM-Conc-Sect-Conc}:]
        Measurements of hadro-production cross sections are critical
        to reducing the systematic error budget of future
        accelerator-based neutrino-oscillation measurements.
        At present, the only experiment that is in operation is
        NA61/SHINE, which is scheduled to complete operation in 2018.
        It is timely to consider the future requirements for
        the hadro-production programme.
        \begin{description}
          \stepcounter{nuPanel-RM-Conc-Sect-Rec}
          \item[\color{BlueViolet} Recommendation \arabic{nuPanel-RM-Conc-Sect}.\arabic{nuPanel-RM-Conc-Sect-Rec}:]
            \textbf{\color{BlueViolet}
              ICFA should encourage careful and timely consideration
              of the requirements for a hadro-production measurement
              programme to follow the present NA61/SHINE including
              possible extensions to the NA61/SHINE programme.
            }
        \end{description}
      \stepcounter{nuPanel-RM-Conc-Sect-Conc}
      \item[\arabic{nuPanel-RM-Conc-Sect}.\arabic{nuPanel-RM-Conc-Sect-Conc}:]
        The present neutrino-nucleus cross-section measurement 
        program is vibrant and appropriate to serve the present 
        generation of experiments over the next five years. 
        The detectors that serve the Short Baseline Neutrino (SBN)
        Program and MINER$\nu$A will support an excellent
        neutrino scattering programme.
      \stepcounter{nuPanel-RM-Conc-Sect-Conc}
      \item[\arabic{nuPanel-RM-Conc-Sect}.\arabic{nuPanel-RM-Conc-Sect-Conc}:]
        In the medium term, the near detectors that are part of the
        DUNE and Hyper-K experiments will take the
        neutrino-interaction programme forward.
        Novel techniques such as the high-resolution tracking detector
        proposed for DUNE and NuPRISM proposed for T2K/Hyper-K have the
        potential to enhance the precision with which muon-neutrino
        interactions are known.
        Robust, timely execution of the near-detector programme at
        DUNE and Hyper-K is essential for the experiments to meet
        their sensitivity goals.
        \begin{description}
          \stepcounter{nuPanel-RM-Conc-Sect-Rec}
          \item[\color{BlueViolet} Recommendation \arabic{nuPanel-RM-Conc-Sect}.\arabic{nuPanel-RM-Conc-Sect-Rec}:]
            \textbf{\color{BlueViolet}
              The near detector programme at DUNE and at Hyper-K
              should be pursued energetically so that
              neutrino-scattering measurements of the requisite
              precision are available to allow the experiments to
              perform to their specified sensitivity.
            }
        \end{description}
    \end{description}

        \begin{description}
      \stepcounter{nuPanel-RM-Conc-Sect-Conc}
      \item[\arabic{nuPanel-RM-Conc-Sect}.\arabic{nuPanel-RM-Conc-Sect-Conc}:]
        Over the coming four to five years, progress in the
        measurement of neutrino-nucleus scattering will be made using,
        for example, the SBN detectors and MINER$\nu$A.
        Substantial theoretical and phenomenological progress in 
        the particle and nuclear physics of neutrino-nucleus
        scattering, beyond the current effort, which is sub-critical,
        will be required over this period.
        \begin{description}
          \stepcounter{nuPanel-RM-Conc-Sect-Rec}
          \item[\color{BlueViolet} Recommendation \arabic{nuPanel-RM-Conc-Sect}.\arabic{nuPanel-RM-Conc-Sect-Rec}:]
            \textbf{\color{BlueViolet}
              ICFA should encourage and promote a vibrant, sustained
              effort towards reliable, precision calculation of
              neutrino-nucleon and neutrino-nucleus scattering. 
              An approach that promotes and exploits the synergy
              between the particle theory, nuclear theory and
              experimental neutrino communities should be adopted.
            }
        \end{description}
    \end{description}

        \begin{description}
      \stepcounter{nuPanel-RM-Conc-Sect-Conc}
      \item[\arabic{nuPanel-RM-Conc-Sect}.\arabic{nuPanel-RM-Conc-Sect-Conc}:]
        The near detectors that form part of the DUNE and Hyper-K
        programmes will take the neutrino-scattering programme forward.
        The detailed specification of the DUNE and Hyper-K near
        detectors will be resolved by the end of the present decade. 
        It will therefore be timely to decide on the long-term
        development of the $\parenbar{\nu}N$-scattering programme
        around $\approx$2020.
        \begin{description}
          \stepcounter{nuPanel-RM-Conc-Sect-Dec}
          \item[\color{RedViolet} Decision point \arabic{nuPanel-RM-Conc-Sect}.\arabic{nuPanel-RM-Conc-Sect-Rec}:]
            \textbf{\color{RedViolet}
              \boldmath{$\approx$}2020: Decide on the future direction
              of the neutrino-nucleus-scattering programme based on
              experimental and theoretical progress and the needs of
              the future neutrino programme.
            }
        \end{description}
    \end{description}

  }
}

\clearpage
\noindent
\framebox[\textwidth][l]{
  \parbox[c]{0.98\linewidth}{
        \begin{description}
      \stepcounter{nuPanel-RM-Conc-Sect-Conc}
      \item[\arabic{nuPanel-RM-Conc-Sect}.\arabic{nuPanel-RM-Conc-Sect-Conc}:]
        The requirements of DUNE and Hyper-K will drive the
        specification of the neutrino-cross-section-measurement
        programme.
        The path forward, beyond the present generation of
        $\parenbar{\nu}N$ scattering experiments, will be determined by
        the degree to which existing techniques (on and off-axis
        near detectors illuminated with pion-decay beams) can deliver
        measurements of the requisite precision.
        \begin{description}
          \stepcounter{nuPanel-RM-Conc-Sect-Rec}
          \item[\color{BlueViolet} Recommendation \arabic{nuPanel-RM-Conc-Sect}.\arabic{nuPanel-RM-Conc-Sect-Rec}:]
            \textbf{\color{BlueViolet}
              The proposed next-generation neutrino-scattering
              experiments, for example nuSTORM, should be evaluated in
              preparation for a decision on the future direction of
              the neutrino-scattering programme to be made
              in \boldmath{$\approx$}2020.
            }
        \end{description}
      \stepcounter{nuPanel-RM-Conc-Sect-Conc}
      \item[\arabic{nuPanel-RM-Conc-Sect}.\arabic{nuPanel-RM-Conc-Sect-Conc}:]
        An exciting programme of detector R\&D is being carried out
        across the world targeted at delivering the technologies
        required by DUNE, Hyper-K and the Short Baseline Neutrino
        Program.
        \begin{description}
          \stepcounter{nuPanel-RM-Conc-Sect-Rec}
          \item[\color{BlueViolet} Recommendation \arabic{nuPanel-RM-Conc-Sect}.\arabic{nuPanel-RM-Conc-Sect-Rec}:]
            \textbf{\color{BlueViolet}
              The present detector R\&D portfolio should be completed.
              Provision should be made for an appropriately resourced
              detector-development programme over the lifetime of the
              next generation of experiments.
            }
        \end{description}
    \end{description}

        \begin{description}
      \stepcounter{nuPanel-RM-Conc-Sect-Conc}
      \item[\arabic{nuPanel-RM-Conc-Sect}.\arabic{nuPanel-RM-Conc-Sect-Conc}:]
        The development of MW-class sources at FNAL and J-PARC are
        critical to the delivery of the experimental programme.
        To go beyond the sensitivity and precision of the next
        generation of accelerator-based experiments is likely to
        require the development of novel accelerator capabilities.
        It is likely that increased international cooperation and
        collaboration will be required to deliver these programmes.
        The MICE experiment and the RaDIATE programme are recognised
        as important contributions to the field, each offering the
        possibility of generating a legacy of enhanced capability.
        \begin{description}
          \stepcounter{nuPanel-RM-Conc-Sect-Rec}
          \item[\color{BlueViolet} Recommendation \arabic{nuPanel-RM-Conc-Sect}.\arabic{nuPanel-RM-Conc-Sect-Rec}:]
            \textbf{\color{BlueViolet}
              Opportunities for international cooperation and/or
              collaboration in the development of MW-class neutrino
              sources should be actively pursued.
            }
          \stepcounter{nuPanel-RM-Conc-Sect-Rec}
          \item[\color{BlueViolet} Recommendation \arabic{nuPanel-RM-Conc-Sect}.\arabic{nuPanel-RM-Conc-Sect-Rec}:]
            \textbf{\color{BlueViolet}
              The MICE experiment should be completed to deliver the
              critical demonstration of ionization cooling.
              ICFA should encourage the timely consideration of the
              accelerator R\&D programme that is required beyond MICE
              to develop the capability to deliver high-brightness
              muon beams.
                    }
        \end{description}
      \stepcounter{nuPanel-RM-Conc-Sect-Conc}
      \item[\arabic{nuPanel-RM-Conc-Sect}.\arabic{nuPanel-RM-Conc-Sect-Conc}:]
         The combination of information in different data sets has
         the potential to lead to insights beyond those which can be
         obtained from any one data set in isolation.  
         The combination of data from a variety of sources is likely
         to be of particular importance in the search for leptonic
	 CP-invariance violation, the search for sterile neutrinos and
         the validation of the three-flavour paradigm. 
         The present trend towards the publication of the information
         required to carry out such ``global'' fits is welcome.
    \end{description}

  }
}

\cleardoublepage
\graphicspath{{05-Reactor/Figures/}}

\section{Measurements using neutrinos from nuclear reactors and
  radioactive sources}
\label{Sect:ReactorSource}

Nuclear reactors play a key role in understanding the properties of
the neutrino. 
They provide an intense source of electron anti-neutrinos with
energies between a few keV and several MeV and therefore allow
precise measurements of oscillation parameters to be made.
The discovery of unknown phenomena through measurements of the
properties of the neutrino may only be possible if the oscillation
parameters are precisely measured. 
In addition, precision measurements will be a sensitive test of the
S$\nu$M framework and will allow searches for physics beyond the
Standard Model to be carried out.
Table \ref{Tab:ReactorSource:ExptList} provides a list of the
experimental collaborations contributing to the reactor-neutrino
oscillation programme.

Precision measurements of the mixing parameters are expected from
reactor experiments that are in operation or that are being proposed.
To date, reactor experiments have determined the values of
$\theta_{13}$ and $\Delta m_{ee}^2$ through the observation of
electron-anti-neutrino disappearance at baselines of 1--2\,km. 
The precision with which the running reactor experiments are
expected to determine parameters of the neutrino-mixing matrix are:
\begin{itemize}
  \item{$\sin^2 2\theta_{13}$:} 
    3$-$4\% at Daya Bay \cite{An:2012eh,An:2012bu},
    10\% at Double Chooz \cite{Abe:2011fz,Abe:2012tg,Abe:2013sxa},
    and 5\% at RENO \cite{Ahn:2010vy,Ahn:2012nd};
    and 
  \item{$\Delta m_{ee}^2$:} 
    $\sim 7 \times 10^{-5}$ eV$^2$ at Daya Bay; and
    $\sim 1 \times 10^{-4}$ eV$^2$ at RENO.
\end{itemize}
It is anticipated that these measurements will be made by $\sim$2017.

For the solar $\Delta m_{21}^2$, the current uncertainties are
determined by KamLAND \cite{Abe:2008aa}.
A reactor experiment with a large liquid-scintillator detector and a
medium-baseline around 50\,km, such as JUNO
\cite{An:2015jdp,Djurcic:2015vqa} and RENO-50 \cite{Seo:2015yqp}, can
provide measurements of $\theta_{12}$,  $\Delta m_{21}^2$ and 
$\Delta m_{ee}^2$ with a precision better than 1\% by $\sim 2025$. 
Combined with results from other experiments for $\theta_{23}$ and
$\theta_{13}$, the unitarity of the neutrino-mixing matrix can be
tested at the  1\% level.
This effort will be valuable to explore physics beyond the Standard
Model.

The large value of $\theta_{13}$ has opened up the possibility of the 
determination of the ordering of the neutrino masses.
With an energy resolution of $\sim 3$\% at 1\,MeV the reactor
experiments JUNO and RENO-50 will be able to determine the mass
hierarchy using the sub-dominant oscillation pattern.
Civil work for the construction of JUNO began in 2015 and JUNO expects
to start data-taking in 2020.
RENO-50 has obtained R\&D funding and will seek to secure construction
funding with the aim of taking data in 2020.
The neutrino-mass hierarchy is expected to be determined with
3$\sigma$--4$\sigma$ significance by JUNO after 6 years of data taking and by
RENO-50 after 10 years of data taking.

While most neutrino-oscillation data fits the three-flavour-mixing
hypothesis, some experiments have reported intriguing
exceptions, including the anomalous disappearance of $\bar{\nu}_e$
and $\nu_e$ produced in reactors and by radioactive sources.
Although these hints currently have only modest statistical
significance, if confirmed, they would be evidence for particles and
interactions beyond the Standard Model.

Various projects will perform searches for sterile neutrinos using
reactor neutrinos over short baselines of between 5\,m and 20\,m.
Compact anti-neutrino sources are available at research reactors while
powerful anti-neutrino fluxes are provided at commercial reactors. 
Challenging background mitigation is necessary because of the
shallow depth at which the experiments operate. 
Two identical detectors at different baselines are desirable to look
for spectral distortions and to be insensitive to uncertainties in the
spectrum of neutrinos produced in the reactor.
The short-baseline reactor experiments that are under consideration or
are being prepared are: DANSS \cite{Danilov:2014vra,Danilov:2013caa}; 
Neutrino-4 \cite{Serebrov:2016wzv}; Nucifer \cite{Boireau:2015dda}; 
NuLat \cite{Lane:2015alq}; Poseidon \cite{Derbin:2012kf}; 
PROSPECT \cite{Ashenfelter:2015uxt}; 
SoLid \cite{SoLID:2013fta,Chen:2014psa}; and 
Stereo \cite{Haser:2016xlb}.
Sterile-neutrino mixing will be explored to a sensitivity of 
$\sin^2 2\theta_{14} \sim 0.01$ and the reactor anti-neutrino anomaly
will be proven or rejected with a significance of 5$\sigma$ before
2020 by multiple experiments.

It is also possible to carry out a sterile-neutrino search using 
neutrinos or anti-neutrinos produced by strong (0.1--10\,MCi)
nuclear-decay sources.
Neutrino radioactive-source experiments can be sensitive to 
$\Delta m^2$ around 1 eV$^2$ due to the low energies (1$-$10\,MeV) of
the neutrinos that are produced. 
Such experiments could observe sterile-neutrino oscillations at
baselines of between 1\,m and 10\,m, i.e., in a single detector or in
several closely separated detectors.
Possible sources include $^{51}$Cr, $^{37}$Ar, $^{144}$Ce, $^{144}$Pr,
$^{37}$Sr and  $^{8}$Li. 
The advantages of using radioactive sources are that the spectra and
flux are precisely known and the neutrino-nucleus scattering cross
sections in the MeV region are also known precisely. 
SOX \cite{Gaffiot:2015fva}, the source experiment at Borexino, plans
to take data in 2016 with a $^{144}$Ce source followed by a $^{51}$Cr
source; in both cases the source will be placed beneath the detector.
The SOX sensitivity to $\sin^2 2\theta_{14}$ is 0.03--0.04 at 95\%
confidence level with two to three years of data taking. 
The BEST \cite{Gavrin:2015aca,Barinov:2016znv} experiment, which it is
proposed will start data taking in the Baksan Underground Laboratory
in 2017--2018, consists of a $^{51}$Cr source at the centre of a
sphere of liquid metal gallium that is surrounded by a second,
cylindrical, volume also of gallium. 
By determining the rate of $^{71}$Ge production in neutrino reactions
in the inner and outer vessels, the BEST sensitivity to the
disappearance of electron neutrinos is at the level of a few percent
with around one year of data taking.
\begin{table}
  \caption{
    Experiments contributing to, or in preparation as part of, the
    reactor-based neutrino-oscillation programme.
  }
  \label{Tab:ReactorSource:ExptList}
  \begin{center}
    \includegraphics[width=0.75\textwidth]{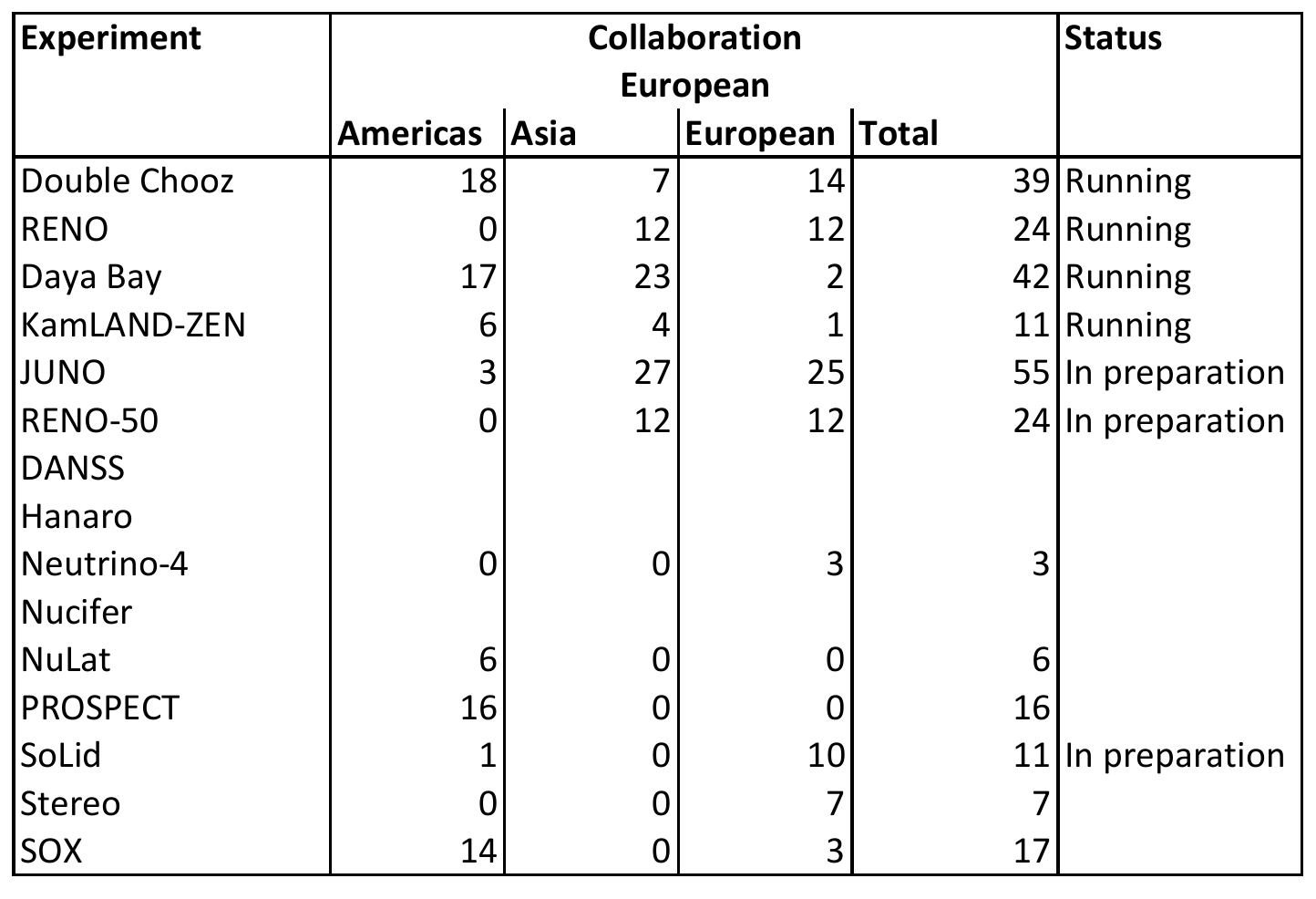}
  \end{center}
\end{table}

\cleardoublepage
\graphicspath{{06-Non-terrestrial/Figures/}}

\section{Non-terrestrial source}
\label{Sect:NonTerre}

In this section the expected oscillation physics results that can be
obtained from present and future experiments studying atmospheric
and solar neutrinos are reviewed.

There are three main categories of detector that are used to study of
atmospheric neutrinos:
\begin{itemize}
  \item Water Cherenkov detectors such as Super-Kamiokande;
  \item Large magnetised iron calorimeter detectors such as INO; and
  \item Large Cherenkov detectors that exploit deep ice or deep sea
  water as the radiating medium.
\end{itemize}
Water Cherenkov detectors continue to be successful in the study of
atmospheric neutrinos over a wide energy range; from a few hundred MeV
to a few TeV.
The sensitivity of this class of detector results from the
photo-detector density with which the sensitive volume is viewed.
The deep ice, or deep sea, detectors have a neutrino-energy threshold
of a few tens of GeV, which is higher than conventional
atmospheric-neutrino detectors, due to the lower density of light
sensors.
However, the physics goals of both categories of detector are to
improve the precision on oscillation parameters and to perform
mass-hierarchy studies.

Atmospheric neutrinos remain an important probe of neutrino
oscillations and provide a sensitive technique for the determination 
of the neutrino-mass hierarchy. 
A difference of up to 20\% in the oscillation probability between the
normal and inverted hierarchy is expected for specific energies and
zenith angles (baselines).
The Earth's density transitions can cause an additional enhancement of
the oscillation signal.
Table \ref{Tab:NonTerre:Source:ExptList} provides a list of the
experimental collaborations contributing to the detection of neutrinos
from non-terrestrial sources.
\begin{table}[b]
  \caption{
    Experiments contributing to, or in preparation as part of, the
    detection of neutrinos from non-terrestrial sources.
  }
  \label{Tab:NonTerre:Source:ExptList}
  \begin{center}
    \includegraphics[width=0.75\textwidth]{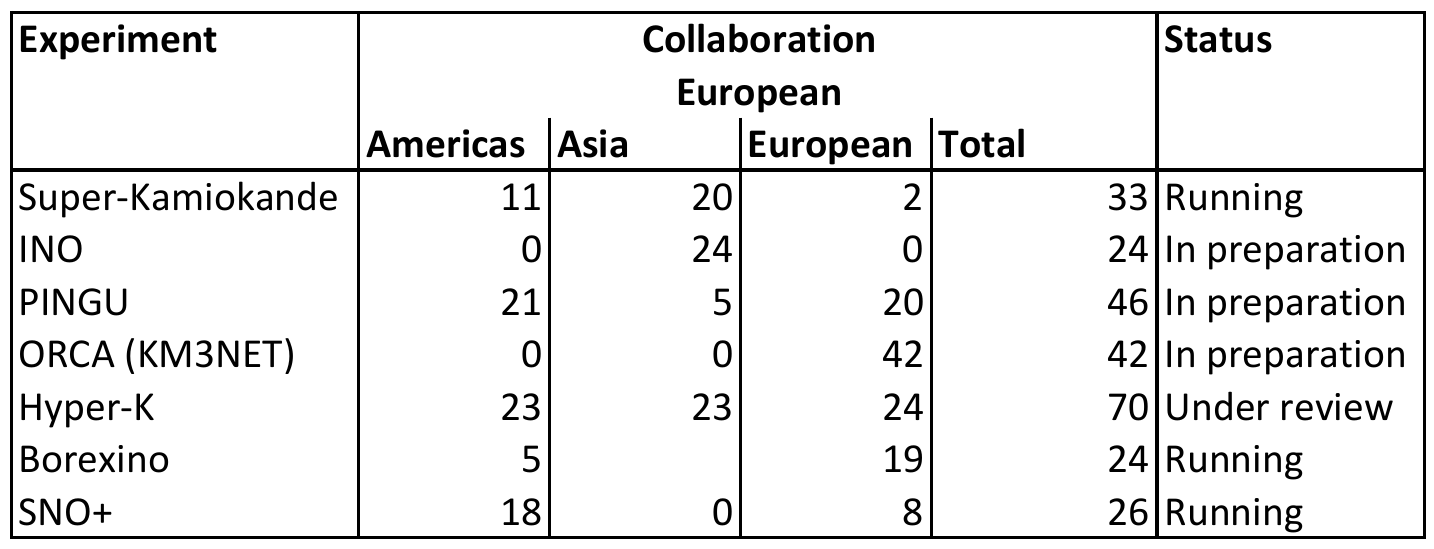}
  \end{center}
\end{table}

\subsection{Atmospheric-neutrino experiments}
\label{SubSect:nuFT:AtmExp}

\subsubsection{Super-Kamiokande}

Super-Kamiokande is a water Cherenkov detector with a fiducial mass of
22\,kt.
It is instrumented with $\sim 12000$ 50-cm PMTs.
Super-Kamiokande has been taking data since 1996 and, in 1998, was the
first experiment to demonstrate the existence of oscillations in the
atmospheric sector \cite{Fukuda:1998mi}.
Since then the Super-Kamiokande collaboration has provided
ground-breaking results on oscillation parameters in the three-flavour
framework from atmospheric-neutrino studies especially on 
$\Delta m^2_{32}$ and $\sin^2\theta_{23}$.
\\

\noindent\textbf{Physics goals} \\
\noindent
It is planned that the Super-Kamiokande experiment (see, for example, 
\cite{Richard:2015aua}) will run for a further ten years.
This will allow the sensitivity to the mass hierarchy to reach
0.7--1.5$\sigma$ to 1--2$\sigma$ depending on the octant.
The sensitivity to the octant should be slightly more than $2\sigma$.

\subsubsection{India Neutrino Observatory Iron Calorimeter (INO-ICAL)}

The main detector to be built at the India Neutrino Observatory (INO)
is the Magnetized Iron Calorimeter (ICAL) composed of 3 modular
17\,kt modules of resistive plate chambers and iron plates for a
total of 52\,kt detector 
\cite{Athar:2006yb,Devi:2014yaa,Indumathi:2015hfa}. 
The detector offers an excellent $\nu / \bar{\nu}$ separation thanks
to the magnetic field of about 1.5\,T that allows the measurement of
the charge and momentum of the muons produced in neutrino
interactions.

\begin{figure}
  \begin{center}                           
    \includegraphics[width=0.75\textwidth]{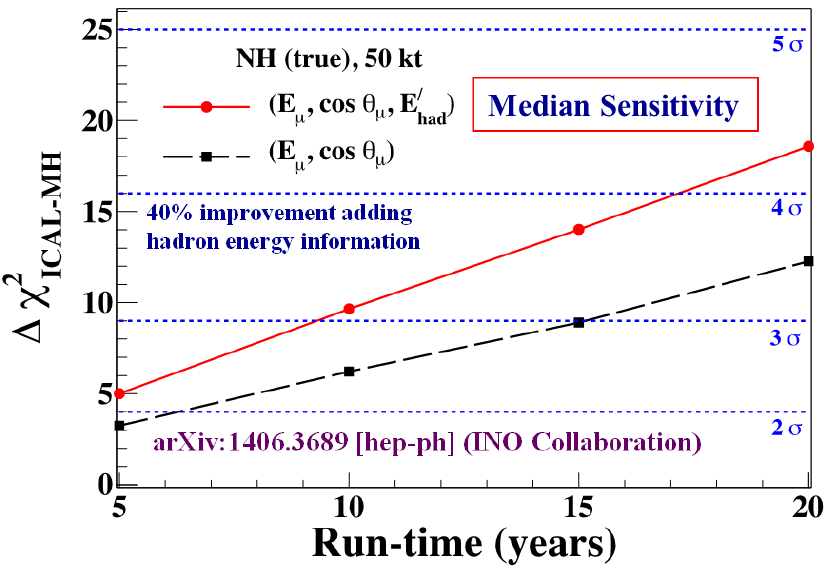}
  \end{center}
  \caption{
    Sensitivity of the ICAL detector to the mass hierarchy as a function
    of the running time for both hierarchies.
  }
  \label{Fig:NonTerre:INO-ICAL}
\end{figure}

The latest study concerning the hierarchy sensitivity of the ICAL
detector, including both muon and associated hadron energy, shows that
a $3\sigma$ sensitivity can be achieved in 10 years of running if the
true values of $\sin^2\theta_{23}$ and $\sin^22\theta_{13}$ are 0.5
and 0.1 respectively (see figure \ref{Fig:NonTerre:INO-ICAL}).
Combining the ICAL results with information from T2K and NO$\nu$A can
give rise to an enhanced sensitivity of $3\sigma$ in 6 years.
Improved precision on the mixing parameters $\Delta m^2_{32}$ and
$\sin^2\theta_{23}$ can be obtained after 10 years of running.
The planning foresees the site infrastructure being developed and the
excavation of the tunnel and cavern taking place over the next three
to four years.

An engineering prototype of one module will be constructed over the
next 2.5 years. 
The construction of the ICAL detector should progress at a rate of
one module per year, which brings the full detector into operation in
about 10 years.

\subsubsection{PINGU and ORCA}

\noindent
{\bf Physics goals}

ORCA and PINGU are designed to measure the flavour oscillations of
atmospheric neutrinos.   
Probing oscillation phenomena in a higher energy range, with stronger
matter effects and different systematic uncertainties provides
complementarity to long-baseline neutrino-beam experiments.
The two detectors are similar in their projected sensitivities, but
differ in that water has a longer scattering length while ice has a
longer attenuation length. 
Both detectors would end up with O(100k) neutrino events per year at
the trigger level. 

\begin{figure}
  \begin{center}
    \includegraphics[width=0.80\textwidth]{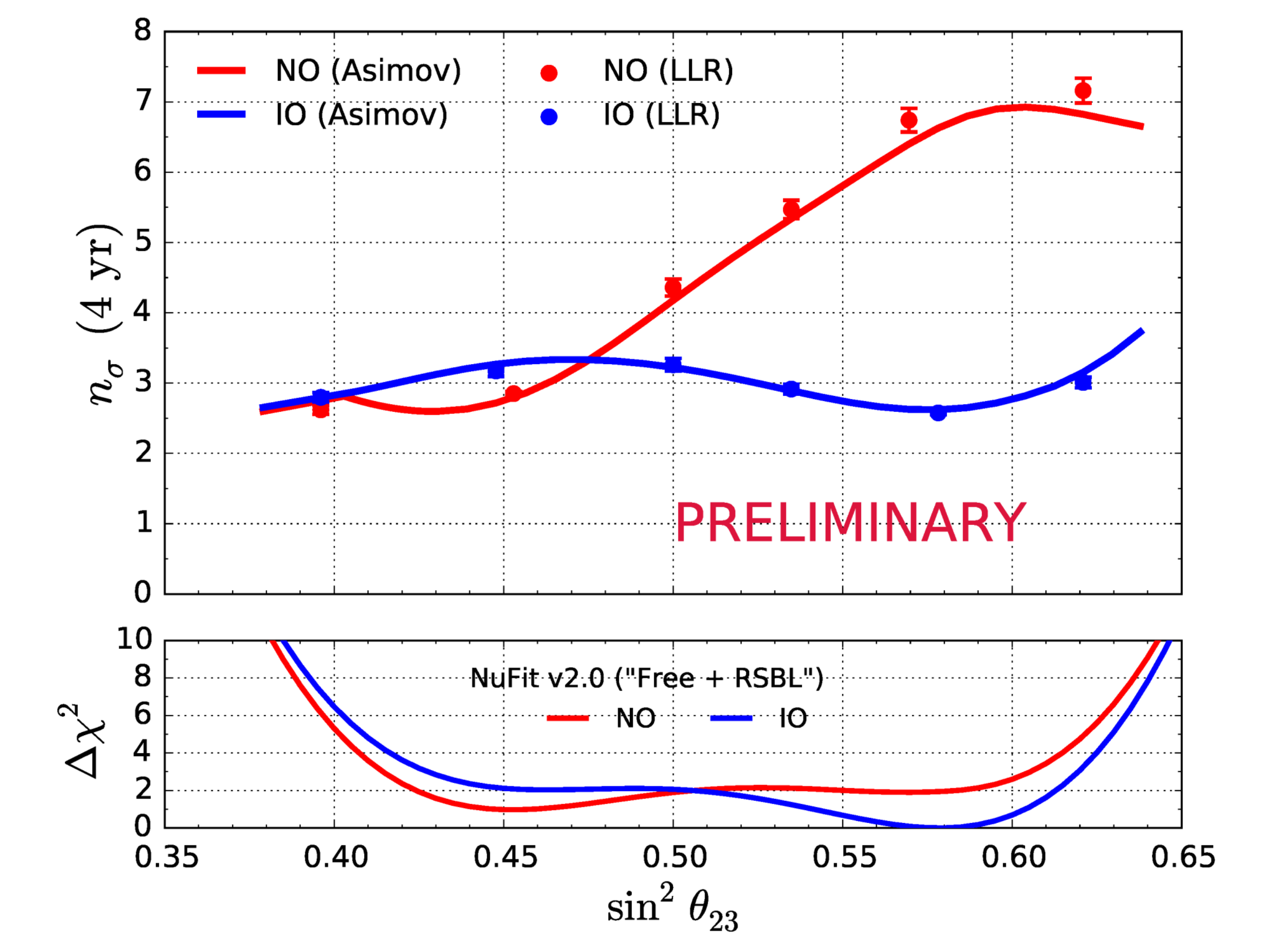}
  \end{center}
  \caption{  
    Expected sensitivity to the mass hierarchy calculated for four
    years of PINGU running time, as a function of the true value of
    $\sin^2 \theta_{23}$ \cite{TheIceCube-Gen2:2016cap}.
  }  
  \label{NMH-PINGU}
\end{figure}

\begin{figure}
  \begin{center}
    \includegraphics[width=0.80\textwidth]{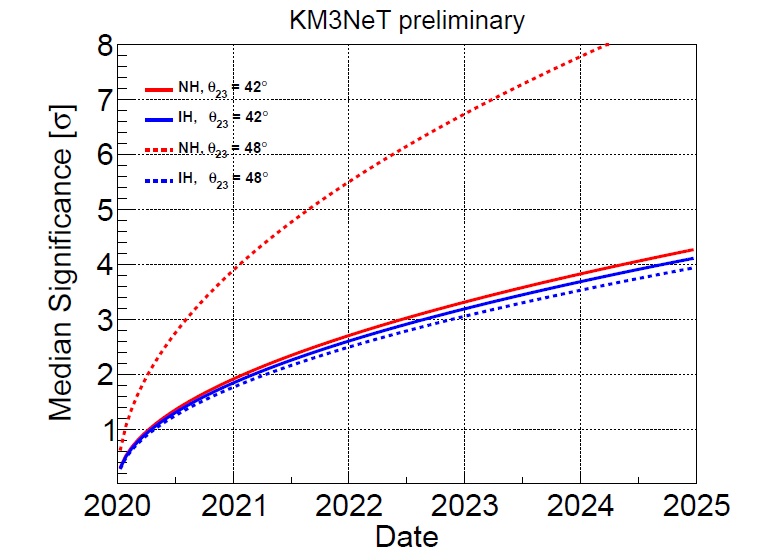}
  \end{center}
  \caption{  
    Expected sensitivity of the ORCA detector to the mass hierarchy as
    a function of running time \cite{Adrian-Martinez:2016fdl}.
  }  
  \label{NMH-ORCA}
\end{figure}

PINGU and ORCA can determine the mass hierarchy with a sensitivity of
$3\sigma$ after 3-4 years (see figures \ref{NMH-PINGU}
and \ref{NMH-ORCA}). 
Also they can measure the atmospheric mixing parameters $\Delta
m^2_{\rm atm}$ and $\theta_{23}$ with a precision comparable to the
NO$\nu$A and T2K experiments using both the muon-neutrino
disappearance and tau neutrino appearance channels.
This will provide a measurement of the tau neutrino appearance rate
with better than $10\%$ precision, a crucial ingredient for tests of
the unitarity of the neutrino-mixing matrix.
Both experiments can probe the octant of the mixing angle
$\theta_{23}$ via matter-resonance effects on neutrinos and
antineutrinos crossing the core and mantle of the Earth, which are
largely independent of $\delta_{CP}$.
Finally, their observations of neutrino oscillations over a wide range
of baselines and energies will provide broad sensitivity to new
physics such as non-standard neutrino interactions (NSI) and sterile neutrinos.

\paragraph*{ORCA: design and next steps}

ORCA (Oscillation Research with Cosmics in the Abyss 
\cite{Adrian-Martinez:2016fdl}) is a deep-sea neutrino detector under
construction in the Mediterranean Sea using the technology developed
for the KM3NeT project.
The geometrical configuration considered is the deployment of 115
vertical ``strings'' with 18 optical modules (OMs) per string at the
KM3NeT-French site in Mediterranean. 
Each OM contains 31 small photomultipliers. 
The corresponding instrumented volume amounts to about 6\,Mt with 2070
optical modules.

Two steps have been identified for the construction of the detector.
The first step corresponds to the deployment of 6--7 strings in ORCA
configuration.
It is funded as Phase\,1 of the project and should demonstrate that
this detection method works in the GeV range.
Following the completion of Phase\,1, the project should go on to
pursue Phase\,2 with 115 strings at the French KM3NeT site that would
be deployed by 2020.

\paragraph*{PINGU: design and next steps}

PINGU (Precision IceCube Next Generation Upgrade 
\cite{TheIceCube-Gen2:2016cap}) is a proposed extension to the IceCube
Observatory that will lower the energy threshold by increasing the
photo-detector density for a portion of the fiducial volume. 
The detector geometry proposed in \cite{TheIceCube-Gen2:2016cap}
comprises 26 new strings with a total of nearly 5000 optical modules
(OMs), deployed in the DeepCore region of the IceCube array and
covering a volume of about 6\,Mt. 
Using innovative OM designs might substantially reduce the number of
OMs.

Some PINGU physics goals (e.g. tau-appearance physics) could be
targeted with 6-8 densely instrumented PINGU-like ``in-fill'' strings
for DeepCore. 
These strings are planned to be deployed in the framework of a staged
approach to IceCube-Gen2, a 5--10\,km$^3$ detector.

\subsubsection{Hyper-K}

\begin{figure}
  \begin{center}                           
    \includegraphics[width=0.45\textwidth]{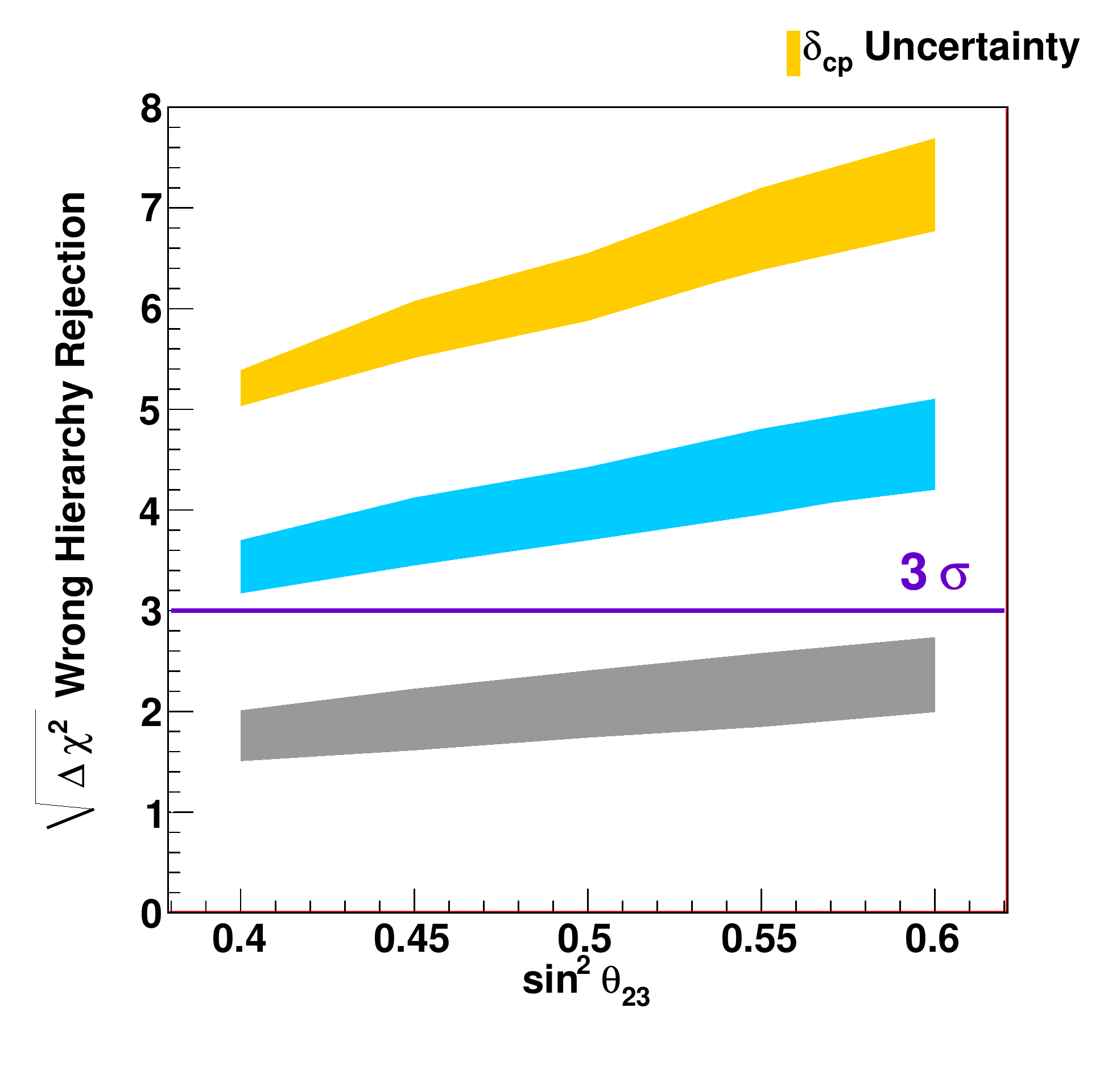}
    \quad \quad \quad \quad 
    \includegraphics[width=0.45\textwidth]{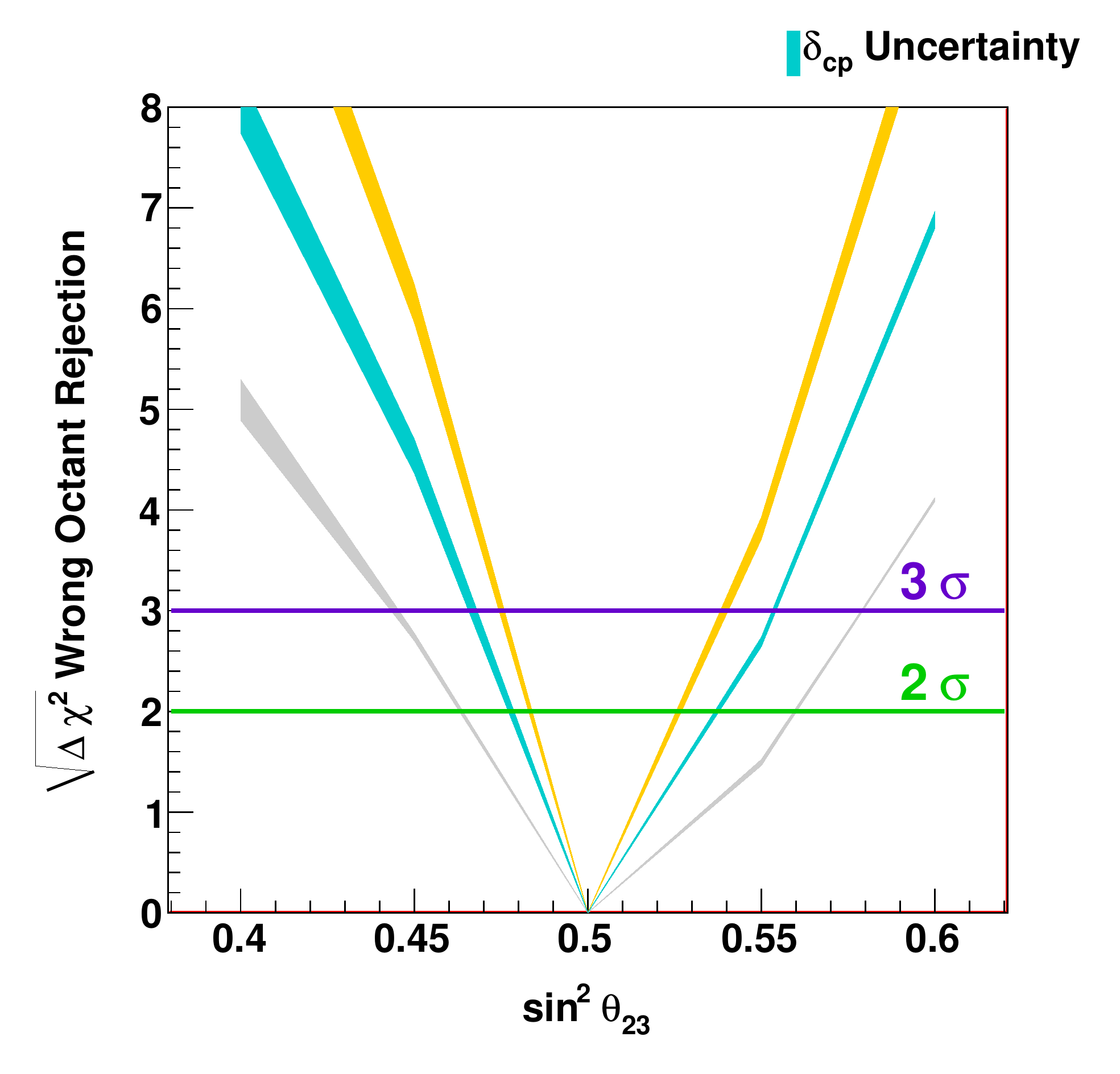}

  \end{center}
  \caption{
    Sensitivity to the mass hierarchy (left) and 
    to $\theta_{23}$ octant (right) 
    from combined analysis of atmospheric neutrino and accelerator 
    data at Hyper-K as a function of true $\sin^2\theta_{23}$ value 
    for three exposures: 1 year (grey), 5 years (blue), 10 years (orange).
    The band width represents the uncertainty of true $\delta_{CP}$ value.
  }
  \label{Fig:NonTerre:HK}
\end{figure}

\noindent\textbf{Physics goals} \\
\noindent
Hyper-K is a large underground water Cherenkov detector with a total
mass of 540\,kt~\cite{Hyper-Kamiokande:2016dsw}. 
It is  comprised of two, 60\,m high, cylindrical tanks which together
have a fiducial volume corresponding to 374\,kt.
The detector is instrumented with approximately 80,000 new 50-cm PMTs
giving a 40\% photo cathode coverage.
The new PMTs have roughly double the photon yield of the PMTs used in
Super-Kamiokande.
After 10 years of data taking, using the atmospheric-neutrino data set
only, the sensitivity to the mass hierarchy will reach $3\sigma$. 
The octant can be determined with a significance greater than
$3\sigma$ for $|\theta_{23} - 45^\circ| > 4^\circ$.
Hyper-K will have improved the sensitivity to mass hierarchy and
$\theta_{23}$ octant by using the J-PARC-accelerator-beam data sets,
which provides a clean measurement of the atmospheric mixing
parameters and therefore provides a precise prediction of the expected
amount of $\nu_\mu \rightarrow \nu_e$ appearance in the resonance
region of upward-going atmospheric neutrinos.
The sensitivity of Hyper-K from the combined analysis is shown in
figure~\ref{Fig:NonTerre:HK}. 
The neutrino mass hierarchy will be determined with $> 3\sigma$
significance after a few years by combination of atmospheric neutrinos
and beam data.
In the combined analysis the $\theta_{23}$ octant can be resolved 
when $|\theta_{23} - 45^\circ|$ is only $2.5^\circ$ in ten years.
 \\

\noindent\textbf{Institutes} \\
The institutes comprising the Hyper-K collaboration were listed in
section \ref{Sect:AccBasedOsc:Next:Hyper-K}. \\

\noindent\textbf{Next steps} \\
The planning is to start excavation in 2018 to start operating the
detector after 2025.

\subsubsection{Baikal experiment}

It has been proposed to set up a large, cubic-km detector in the
Baikal Lake to study atmospheric and astrophysical neutrino fluxes
and, in particular, to map the high-energy neutrino sky in the
Southern Hemisphere, including the region of the galactic
centre \cite{Bednyakov:Ed:2014}. 
The detector will use the water in Lake Baikal, instrumented at depth
with optical sensors that detect the Cherenkov radiation from secondary
particles produced in the interactions of high-energy neutrinos inside
or near the instrumented volume. 
The detector is to be completed by 2020 \cite{Bednyakov:Ed:2014}.

\subsection{Solar-neutrino experiments}

Important questions concerning the nature of solar neutrinos will be
addressed in the coming years by existing and upcoming projects.
The principal topics that will be investigated : low-energy- and
pep-neutrino measurements and studies of the metallicity model using
CNO-neutrino measurements.

\subsubsection{Super-Kamiokande}
The first evidence of solar-neutrino oscillations was obtained in 2001
by comparing Super-Kamiokande
measurements~\cite{Fukuda:2001nj,Fukuda:2001nk} to measurements made by
SNO~\cite{Ahmad:2001an}. 
Solar-neutrino studies have provided precision measurements of 
$\Delta m^2_{21}$ and $\sin^2\theta_{12}$ that are complementary to
measurements of $\Delta m^2_{21}$ and $\sin^2\theta_{12}$ using
reactor anti-neutrinos.
The day/night flux asymmetry due to the effect on the oscillation
probability of the matter that makes up the Earth was observed in
2014~\cite{Renshaw:2013dzu}.  \\

\noindent\textbf{Physics goals} \\
\noindent
The Super-K collaboration seeks to measure the low-energy solar neutrinos to
study region where the solar-matter effect turns on.
This study will provide further direct evidence of solar neutrino
oscillations and allow searches for non-standard physics, such as
flavour-changing neutral current interactions \cite{Friedland:2004pp},
the existence of sterile neutrinos~\cite{deHolanda:2010am} 
and mass varying neutrinos~\cite{GonzalezGarcia:2007ib}, to be made.

\subsubsection{Borexino}

\noindent\textbf{Physics goals} \\
\noindent
Data for a possible measurement of pp neutrinos are being analysed. 
Recent improvements in scintillator re-purification can reduce
backgrounds to lower levels for a possible measurement of CNO
neutrinos. 
The CNO flux is a direct probe of the heavy element composition in the
core of the Sun. \\

\noindent\textbf{Plans for 2015--2018:}\\ 
\noindent
The plans for the Borexino programme over the period 2015--2018 are:
\begin{itemize}
  \item Improved precision of present results of solar pp, pep, $^7$Be
    and $^8$B neutrinos, as well as geo- and reactor-anti-neutrinos; 
  \item CNO neutrinos: improve limit or try to quote a rate (``solar
    metallicity puzzle'');
  \item Sterile neutrino search with $^{144}$Ce anti-neutrino and
    $^{51}$Cr neutrino sources (SOX); and
  \item Supernova neutrinos and anti-neutrinos: BX member of SNEWS,
    keep high duty cycle of 95\% with a low energy threshold.
\end{itemize}

\subsubsection{SNO+}

\noindent\textbf{Physics goals} \\
\noindent
SNO+ is a large liquid scintillator-based experiment located 2\,km
underground at SNOLAB, Sudbury, Canada. 
It reuses the Sudbury Neutrino Observatory detector that consists of a
12\,m diameter acrylic vessel and which will be filled with 
$\sim 780$\,t of ultra-pure liquid scintillator.  
The primary goal of the SNO+ collaboration is to search for the
neutrinoless double-beta decay of $^{130}$Te. 
However, it will have the potential to explore other physics including
low energy pep and CNO solar neutrinos.

With scintillator purity at the Borexino level, the experiment has
sensitivity  to low-energy, CNO, pep and $^8$B neutrinos with unloaded
scintillator.  
If the scintillator purity is improved (by one order magnitude), it
may also be possible to detect pp neutrinos.
$^8$B neutrinos with energy above the $^{130}$Te end-point can be
measured in the Te-loaded-scintillator phase. \\

\noindent\textbf{Programme:} \\
\noindent
The next steps in the SNO+ programme are':
\begin{itemize}
  \item 2015-2016: water commissioning phase;
  \item 2016: scintillator phase; and
  \item 2017: Te loading phase.
\end{itemize}

\cleardoublepage
\section{Non-oscillation programme}
\label{Sect:NonOsc}

To date, all the effects of non-zero neutrino mass have been observed
in neutrino-oscillation experiments. 
Neutrino masses can also manifest themselves in several other physics
observables.
Here we briefly review the three most promising classes of observable
in terms of the sensitivity to physics beyond the Standard Model and
the expected improvements in the near and medium term.

The direct observation of kinematic effects arising from non-zero
neutrino mass is exceptionally challenging because neutrino mass is
tiny compared to the energy scales that pertain in the relevant
particle- and nuclear-physics experiments.
Precise measurements of the shape of the $\beta$-decay spectrum close
to its end point is the most sensitive direct probe of neutrino mass.
End-point measurements are sensitive to the effective
electron-neutrino mass-squared, $m_{\beta}^2=\sum_i|U_{ei}|^2m_i^2$,
where $U_{e i}$ are elements of the electron-row of the neutrino
mixing matrix, $i=1,2,\ldots$ summed over the neutrino mass
eigenvalues $m_i$.
Current measurements of the $\beta$-spectrum of tritium decay
constrain $m_{\beta}^2<4.0$\,eV$^2$ at the 95\% confidence level. 
The Katrin experiment \cite{Osipowicz:2001sq,Mertens:2015ila} aims at
being sensitive to $m_{\beta}^2>0.04$\,eV$^2$ (90\% confidence level
upper limit), hence two orders of magnitude more sensitive to
$m_{\beta}^2$ than the current bound, after three years of
data-taking.
Katrin, which has recently started to take data, is expected to report
results by the end of the decade. 
In order to improve significantly on these bounds using tritium,
qualitatively better experiments are required. 
Project~8 \cite{Monreal:2009za,Oblath:2015pxa}, which just passed its
proof-of-concept milestone, and is currently in the design phase,
appears to offer a path towards $m_{\beta}^2<0.01$\,eV$^2$.
The time-scale for Project 8 is currently uncertain. 
Isotopes other than lithium are being considered, including $^{187}$Re
or $^{163}$Ho, embedded in micro-calorimeters.
These experiments ultimately aim at achieving sensitivities similar to
those of Katrin but they still need to overcome several challenges. 
Competitive results are not expected in the foreseeable future.  

Precision measurements of $\beta$-decay translate into very robust
bounds on neutrino properties. 
While they are also sensitive to certain types of new phenomena
(e.g. new neutrino mass-eigenstates), the relation between the current
bound on $m_{\beta}$ and the known neutrino masses and mixing angles
is well-defined. 
The current neutrino-oscillation data, for example, guarantee that
$m_{\beta}^2>8\times 10^{-5}$\,eV$^2$. 
If the neutrino mass-hierarchy were known to be inverted,
$m_{\beta}^2>2.5\times 10^{-3}$\,eV$^2$ would be guaranteed. 
On the other hand, if Katrin were to observe a signal by the end of
the decade, we would be able to conclude that the three known neutrino
masses were almost exactly degenerate in magnitude, 
$m_1 \simeq m_2 \simeq m_3$.

If the neutrinos are Majorana fermions, lepton-number is not an
exactly conserved quantum number. 
On the flip-side, the observation of lepton-number violation would
reveal that neutrinos are Majorana fermions. 
Neutrino oscillation experiments are, for all practical purposes,
unable to see lepton-number-violating effects. 
The most sensitive probes of lepton-number violation are searches for
neutrinoless double-beta decay. 
If neutrinos are Majorana fermions, neutrino exchange contributes to
the decay width for neutrinoless double-beta decay in a calculable
way.
This contribution is proportional to the effective neutrino mass
$m_{\beta\beta}=|\sum_i U_{ei}^2m_i|$, which also depends on the
Majorana phases (note $U_{e i}^2\neq |U_{e i}|^2$). 
Hence, if active neutrino exchange is the dominant contribution to
neutrinoless double-beta decay, one can translate lower bounds on the
decay lifetime to upper bounds on $m_{\beta\beta}$. 
There are several on-going and planned searches for neutrinoless
double-beta decay, using a variety of isotopes ($^{76}$Ge, $^{130}$Te,
$^{136}$Xe, $^48$Ca, $^82$Se etc.). 
The current generation of experiments aims at being sensitive to
$m_{\beta\beta}\gtrsim 0.1$\,eV by the end of the decade. 
Next-generation experiments, currently under serious consideration and
expected by the middle of the next decade, aim at an
order-of-magnitude-improved sensitivity, $m_{\beta\beta}\gtrsim
0.01$\,eV.

Unlike $\beta$-decay, the connection between neutrinoless
double-beta decay and neutrino properties is more indirect. 
Failure to observe neutrinoless double-beta decay at the level
indicated by $m_{\beta\beta}$ does not necessarily imply that
neutrinos are Dirac fermions, neither is it guaranteed that the
observation of neutrinoless double-beta decay can be directly
translated into a measurement of $m_{\beta\beta}$. 
It is possible, for example, that the lepton-number violating physics
impacts neutrinoless double-beta decay in a way that is not captured
by active neutrino exchange. 
There are also technical computational challenges---the associated
nuclear matrix elements are only known rather poorly---when it comes
to translating the rate for neutrinoless double-beta decay and the
effective neutrino mass. 

Current neutrino oscillation data do not allow one to compute a lower
bound for $m_{\beta\beta}$, even if one assumes the neutrinos are
Majorana fermions.
If the neutrino mass hierarchy were known to be inverted, on the other
hand, $m_{\beta\beta}^2>0.02$\,eV would be guaranteed---and within
reach of next-generation experiments---as long as neutrinos are
Majorana fermions.
More information on $m_{\beta\beta}$ would become available if more
information on the neutrino masses from, e.g., Katrin were
available.

Finally, neutrino masses leave a non-trivial imprint in the large
scale structure of the Universe. 
Neutrinos are one of the relics from the Big Bang and play a role in
the evolution of the Universe.
Hence, cosmic surveys sensitive to the expansion rate of the Universe
at different epochs, or to the formation of structure at different
scales, can be used to measure neutrino properties assuming the
different ingredients that contribute to the thermal history of the
Universe are known and constrained well enough. 
In particular, cosmic surveys are sensitive to the sum of neutrino
masses $\Sigma=\sum_i m_i$. 
Current data translate into $\Sigma<0.26$\,eV at the 95\% confidence
level. 
In the next ten years, improvements to cosmic microwave background
(CMB) lensing data, including polarisation, from CMB observations
(e.g. POLARBEAR \cite{Inoue:2016jbg}, SPTPol \cite{Hanson:2013hsb},
BICEP3 \cite{Ahmed:2014ixy}, etc.), along with next-generation 
large scale structure surveys (DESI
\cite{Aghamousa:2016zmz,Aghamousa:2016sne}, Euclid
\cite{Tereno:2015hja}, LSST \cite{Abate:2012za}) should translate into
sensitivity to $\Sigma\gtrsim 0.015$\,eV.
Given what we currently know about neutrino masses $\Sigma>0.05$\,eV,
next-generation cosmic surveys are ``guaranteed'' to see the effects
of non-zero neutrino masses at, at least, the three-sigma level. 

The relation between cosmic surveys and neutrino properties is quite
indirect. 
If the ingredients that determine the thermal history of the Universe
are different from what we currently anticipate, or if the properties
of these ingredients are not as prescribed by our current
understanding of particle physics, the interpretation of cosmic-survey
data can change significantly. 
Information from all different probes of neutrino masses and neutrino
properties---including oscillations, tritium $\beta$-decay,
neutrinoless double-beta decay---will be required in order to
interpret unambiguously future data and extract the most information
concerning neutrino physics and, potentially, a wealth of new
phenomena.

\cleardoublepage
\setcounter{nuPanel-RM-Conc-Sect}{0}
\setcounter{nuPanel-RM-Conc-Sect-Conc}{0}
\setcounter{nuPanel-RM-Conc-Sect-Rec}{0}
\setcounter{nuPanel-RM-Conc-Sect-Dec}{0}
\graphicspath{{08-Conc-n-rec/Figures/}}

\section{Conclusions and recommendations}
\label{App:Conc-n-rec}

\begin{figure}[h]
  \begin{center}
    \includegraphics[width=1.00\textwidth]{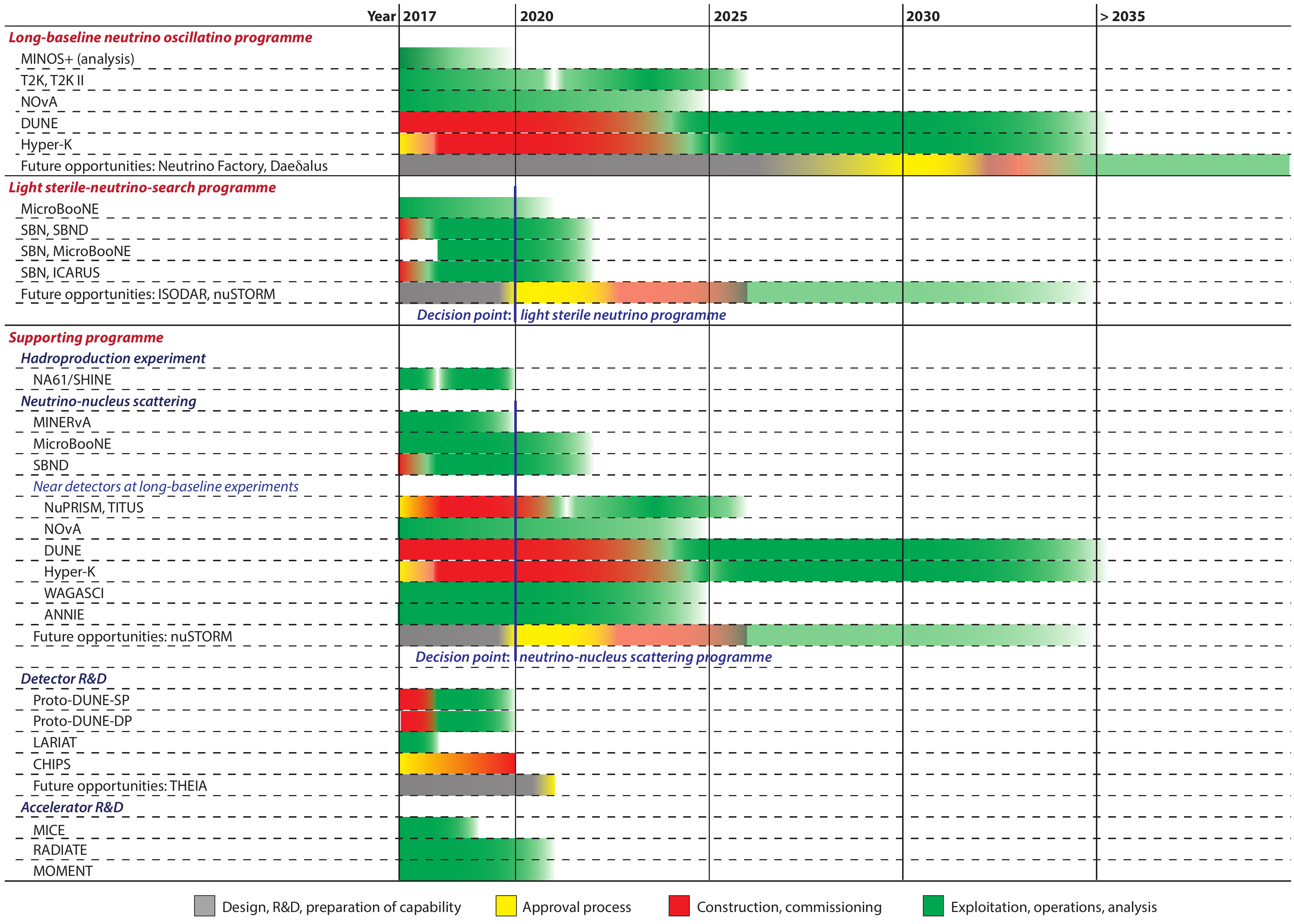}
  \end{center}
  \caption{
    Graphical representation of the roadmap for the accelerator-based
    neutrino program.
    The various experiments and programmes described in sections
    \ref{Sect:AccBasedOsc}, \ref{Sect:Sterile} and \ref{Sect:SBL} of
    this document are listed together with an indication of the
    timeline for the R\&D and approval process, construction and
    commissioning as well as the exploitation, data taking and
    analysis.
    The timeline and dates are approximations, the reader is referred
    to the text for details.
    The transitions between the various stages in the projects are
    indicated.
  }
  \label{Fig:nuPnl-Rm-Grphc}
\end{figure}

Both terrestrial and astrophysical sources are essential to the
elucidation the properties of the neutrino.
In line with its terms of reference~\cite{ICFAnuPanel:ToR:2013}, the
Panel restricted its considerations to the accelerator-based
programme.
A graphical representation of the Panel's roadmap for the
accelerator-based neutrino discovery and measurement programmes
described in sections \ref{Sect:AccBasedOsc} and \ref{Sect:Sterile}
is shown in figure \ref{Fig:nuPnl-Rm-Grphc}.
The supporting programme described in section \ref{Sect:SBL}, which
consists of hadro-production and neutrino-nucleus-scattering
measurements and detector and accelerator R\&D, is also shown.
The decision points identified at $\approx 2020$ for the future
light-sterile-neutrino search and neutrino-nucleus-scattering
programmes are indicated.

The Panel's conclusions, which are restricted by its terms of
reference~\cite{ICFAnuPanel:ToR:2013} to the accelerator-based
programme, are collected below together with the decision points
identified by the Panel and the Panel's recommendations. 
The numbering, ``$n.m$'', is such that $n$ refers to the section of
this document in which the background to the conclusion, decision
point or recommendation may be found. \\

\newpage

\noindent
\textbf{1~~~Introduction}
\vspace{0.5cm}
\stepcounter{nuPanel-RM-Conc-Sect}

\noindent
\framebox[\textwidth][l]{
  \parbox[c]{0.98\linewidth}{
      \parbox[c]{0.98\linewidth}{
    \begin{description}
      \stepcounter{nuPanel-RM-Conc-Sect-Conc}
      \item[\arabic{nuPanel-RM-Conc-Sect}.\arabic{nuPanel-RM-Conc-Sect-Conc}:]
        The neutrino has a tiny mass, much smaller than any other
        fundamental fermion, and its type, or ``flavour'', changes as
        it propagates through space and time.
        These properties imply the existence of new phenomena not
        described by the Standard Model of particle physics and may
        have profound consequences for our understanding of the
        Universe. 
        The tiny neutrino mass seems likely to be related to phenomena
        that occur at very high energy scales, well beyond the reach
        of the present or proposed colliding-beam facilities.
        The study of the neutrino is therefore the study of physics
        beyond the Standard Model and a fundamentally important
        component of the particle-physics programme.
      \stepcounter{nuPanel-RM-Conc-Sect-Conc}
      \item[\arabic{nuPanel-RM-Conc-Sect}.\arabic{nuPanel-RM-Conc-Sect-Conc}:]
        The accelerator-based neutrino programme is global in
        scope, engagement and intellectual contribution.
        Continued and enhanced cooperation in a coherent global
        programme will maximise the impact of each individual
        contribution and of the programme as a whole.
        \begin{description}
          \stepcounter{nuPanel-RM-Conc-Sect-Rec}
          \item[\color{BlueViolet} Recommendation \arabic{nuPanel-RM-Conc-Sect}.\arabic{nuPanel-RM-Conc-Sect-Rec}:]
            \textbf{\color{BlueViolet} The present roadmap document
              should be revised and updated at appropriate intervals.
            }
        \end{description}
    \end{description}
  }

  }
}

\vspace{0.5cm}
\noindent
\textbf{2~~~Accelerator-based long-baseline neutrino-oscillation programme}
\vspace{0.5cm}
\stepcounter{nuPanel-RM-Conc-Sect}

\noindent
\framebox[\textwidth][l]{
  \parbox[c]{0.98\linewidth}{
        \begin{description}
      \stepcounter{nuPanel-RM-Conc-Sect-Conc}
      \item[\arabic{nuPanel-RM-Conc-Sect}.\arabic{nuPanel-RM-Conc-Sect-Conc}:]
        The present accelerator-based long-baseline
        neutrino-oscillation programme is vibrant and has substantial
        discovery potential. 
        \begin{description}
          \stepcounter{nuPanel-RM-Conc-Sect-Rec}
          \item[\color{BlueViolet} Recommendation \arabic{nuPanel-RM-Conc-Sect}.\arabic{nuPanel-RM-Conc-Sect-Rec}:]
            \textbf{\color{BlueViolet} Full exploitation of the present generation of
              experiments should continue thereby maximising their
              discovery potential and the scientific return on
              historical investment.
            }
        \end{description}
    \end{description}

        \begin{description}
      \stepcounter{nuPanel-RM-Conc-Sect-Conc}
      \item[\arabic{nuPanel-RM-Conc-Sect}.\arabic{nuPanel-RM-Conc-Sect-Conc}:]
        The measurements that will be made by the DUNE and the Hyper-K
        collaborations are complementary and the combination of the
        data from the two experiments offers the potential for
        insights beyond those that either experiment can provide on
        its own. 
        \begin{description}
          \stepcounter{nuPanel-RM-Conc-Sect-Rec}
          \item[\color{BlueViolet} Recommendation \arabic{nuPanel-RM-Conc-Sect}.\arabic{nuPanel-RM-Conc-Sect-Rec}:]
            \textbf{\color{BlueViolet} Both the DUNE and the Hyper-K programmes should be
              pursued through the established approval and funding
              processes.
            }
        \end{description}
    \end{description}

  }
}

\noindent
\framebox[\textwidth][l]{
  \parbox[c]{0.98\linewidth}{
        \begin{description}
      \stepcounter{nuPanel-RM-Conc-Sect-Conc}
      \item[\arabic{nuPanel-RM-Conc-Sect}.\arabic{nuPanel-RM-Conc-Sect-Conc}:]
        The sensitivity of long-baseline neutrino-oscillation
        experiments may be enhanced by including events at the second
        oscillation maximum. 
        DUNE may exploit the second oscillation maximum since the LBNF
        broad-band beam will provide an interesting flux at the relevant
        energy.
        The baseline of Hyper-K places the relevant energy
        significantly below the peak of the J-PARC narrow-band 
        beam making it harder for Hyper-K to profit from the second
        oscillation maximum. 
        ESSnuSB has been optimised for the study of the second
        oscillation maximum thereby maximising its sensitivity to
        CP-invariance violation.  
        \begin{description}
          \stepcounter{nuPanel-RM-Conc-Sect-Rec}
          \item[\color{BlueViolet} Recommendation \arabic{nuPanel-RM-Conc-Sect}.\arabic{nuPanel-RM-Conc-Sect-Rec}:]
            \textbf{\color{BlueViolet} The complementarity and
              timeliness of the ESSnuSB proposal should be considered
              in the light of the likely timescales and performance of
              the DUNE programme and, if approved, the Hyper-K programme.
           }
        \end{description}
    \end{description}

  }
}

\vspace{0.5cm}
\noindent
\textbf{3~~~Sterile neutrino searches at accelerators}
\vspace{0.5cm}
\stepcounter{nuPanel-RM-Conc-Sect}

\noindent
\framebox[\textwidth][l]{
  \parbox[c]{0.98\linewidth}{
        \begin{description}
      \stepcounter{nuPanel-RM-Conc-Sect-Conc}
      \item[\arabic{nuPanel-RM-Conc-Sect}.\arabic{nuPanel-RM-Conc-Sect-Conc}:]
        Unambiguous evidence for sterile neutrinos would constitute
        a breakthrough of fundamental significance and would
        revolutionise the field while unambiguous confirmation of the
        short-baseline anomalies would warrant energetic
        investigation.
        \begin{description}
          \stepcounter{nuPanel-RM-Conc-Sect-Rec}
          \item[\color{BlueViolet} Recommendation \arabic{nuPanel-RM-Conc-Sect}.\arabic{nuPanel-RM-Conc-Sect-Rec}:]
            \textbf{\color{BlueViolet}
              The present generation of accelerator-based
              sterile-neutrino-search experiments, including those
              that constitute the SBN Program at FNAL, should be
              exploited so as to maximise their sensitivity.
            }
        \end{description}
      \stepcounter{nuPanel-RM-Conc-Sect-Conc}
      \item[\arabic{nuPanel-RM-Conc-Sect}.\arabic{nuPanel-RM-Conc-Sect-Conc}:]
        Results from the SBN Program and other
        sterile-neutrino-search experiments will be available by
        $\approx$2020.
        It will then be timely to decide on the future direction of
        the accelerator-based sterile-neutrino-search programme.    
        \begin{description}
          \stepcounter{nuPanel-RM-Conc-Sect-Dec}
          \item[\color{RedViolet} Decision point \arabic{nuPanel-RM-Conc-Sect}.\arabic{nuPanel-RM-Conc-Sect-Rec}:]
            \textbf{\color{RedViolet}
              \boldmath{$\approx$}2020: Decide on the future direction
              of the accelerator-based sterile-neutrino-search
              programme.
            }
        \end{description}
    \end{description}

        \begin{description}
      \stepcounter{nuPanel-RM-Conc-Sect-Conc}
      \item[\arabic{nuPanel-RM-Conc-Sect}.\arabic{nuPanel-RM-Conc-Sect-Conc}:]
        Beyond the SBN Program, the way forward will depend on the
        strength of the evidence for sterile neutrinos.
        \begin{description}
          \stepcounter{nuPanel-RM-Conc-Sect-Rec}
          \item[\color{BlueViolet} Recommendation \arabic{nuPanel-RM-Conc-Sect}.\arabic{nuPanel-RM-Conc-Sect-Rec}:]
            \textbf{\color{BlueViolet}
              The sensitivity, cost, schedule and relative strengths
              of the proposed next-generation accelerator-based
              sterile-neutrino-search experiments (IsoDAR, nuSTORM)
              should be evaluated in preparation for a decision to be
              made on the future direction of the
              sterile-neutrino-search programme
              in \boldmath{$\approx$}2020.
              In the mean time, the R\&D programme necessary to
              establish the requisite capabilities should be carried
              out. 
            }
        \end{description}
    \end{description}

  }
}

\vspace{0.5cm}
\noindent
\textbf{4~~~Supporting programme} \\
\vspace{0.5cm}
\stepcounter{nuPanel-RM-Conc-Sect}

\noindent
\framebox[\textwidth][l]{
  \parbox[c]{0.98\linewidth}{
        \begin{description}
      \stepcounter{nuPanel-RM-Conc-Sect-Conc}
      \item[\arabic{nuPanel-RM-Conc-Sect}.\arabic{nuPanel-RM-Conc-Sect-Conc}:]
        An appropriate programme of hadro-production and neutrino
        cross-section measurement is required to allow the present and
        next generation of long- and short-baseline experiments to
        achieve their full potential. 
    \end{description}

        \begin{description}
      \stepcounter{nuPanel-RM-Conc-Sect-Conc}
      \item[\arabic{nuPanel-RM-Conc-Sect}.\arabic{nuPanel-RM-Conc-Sect-Conc}:]
        Over the coming four to five years, progress in the
        measurement of neutrino-nucleus scattering will be made using,
        for example, the SBN detectors and MINER$\nu$A.
        Substantial theoretical and phenomenological progress in 
        the particle and nuclear physics of neutrino-nucleus
        scattering, beyond the current effort, which is sub-critical,
        will be required over this period.
        \begin{description}
          \stepcounter{nuPanel-RM-Conc-Sect-Rec}
          \item[\color{BlueViolet} Recommendation \arabic{nuPanel-RM-Conc-Sect}.\arabic{nuPanel-RM-Conc-Sect-Rec}:]
            \textbf{\color{BlueViolet}
              ICFA should encourage and promote a vibrant, sustained
              effort towards reliable, precision calculation of
              neutrino-nucleon and neutrino-nucleus scattering. 
              An approach that promotes and exploits the synergy
              between the particle theory, nuclear theory and
              experimental neutrino communities should be adopted.
            }
        \end{description}
    \end{description}

        \begin{description}
      \stepcounter{nuPanel-RM-Conc-Sect-Conc}
      \item[\arabic{nuPanel-RM-Conc-Sect}.\arabic{nuPanel-RM-Conc-Sect-Conc}:]
        The near detectors that form part of the DUNE and Hyper-K
        programmes will take the neutrino-scattering programme forward.
        The detailed specification of the DUNE and Hyper-K near
        detectors will be resolved by the end of the present decade. 
        It will therefore be timely to decide on the long-term
        development of the $\parenbar{\nu}N$-scattering programme
        around $\approx$2020.
        \begin{description}
          \stepcounter{nuPanel-RM-Conc-Sect-Dec}
          \item[\color{RedViolet} Decision point \arabic{nuPanel-RM-Conc-Sect}.\arabic{nuPanel-RM-Conc-Sect-Rec}:]
            \textbf{\color{RedViolet}
              \boldmath{$\approx$}2020: Decide on the future direction
              of the neutrino-nucleus-scattering programme based on
              experimental and theoretical progress and the needs of
              the future neutrino programme.
            }
        \end{description}
    \end{description}

  }
}
\clearpage 

\noindent
\framebox[\textwidth][l]{
  \parbox[c]{0.98\linewidth}{

  }
}

\cleardoublepage
\section*{Acknowledgements}

The work presented in this document is the synthesis of many
discussions within the neutrino community, with representatives of
the funding agencies and with laboratory management.
The Panel is grateful for the warm spirit of collaboration that
characterised all of these discussions.
In May 2016, the Panel released a version of the roadmap to promote
discussion of its interim conclusions and to solicit comments on the
report and feedback on the Panel's work.
The Panel thanks all those who have provided comments and feedback all
of which have been used to produce this, the final version of the
roadmap document.
The Panel gratefully acknowledges all the help and support it has
received from the neutrino-physics community and the stakeholders in
the programme during the course of its work.

\cleardoublepage
\bibliographystyle{99-Styles/utphys}
\bibliography{Concatenated-bibliography}

\providecommand{\href}[2]{#2}\begingroup\raggedright\begin{thebibliography}{100}

\bibitem{ICFAnuPanel:ToR:2013}
{The International Committee on Future Accelerators}, ``{ICFA Neutrino Panel:
  terms of reference}.''
  \url{http://www.fnal.gov/directorate/icfa/files/Terms-Of-Reference.pdf},
  2013.

\bibitem{ICFA:nuPanelWWWSite}
{The ICFA Neutrino Panel}, ``{ICFA Neutrino Panel}.''
  \url{http://www.fnal.gov/directorate/icfa/}, 2013.

\bibitem{ICFA:nuPanel:2016:01}
{ICFA Neutrino Panel}, ``{Roadmap for the international, accelerator-based
  neutrino programme; Discussion document}.''
  \url{http://icfa.fnal.gov/wp-content/uploads/2016-05-07-nuPanel-roadmap-Final.pdf},
  2016.

\bibitem{Cao:2014zra}
J.~Cao, A.~de~Gouv\^ea, D.~Duchesneau, R.~Funchal, S.~Geer, {\em et al.},
  ``{Initial report from the ICFA Neutrino Panel},''
\href{http://arxiv.org/abs/1405.7052}{{\ttfamily arXiv:1405.7052
  [physics.acc-ph]}}.

\bibitem{ICFA:WWWSite}
{The International Committee on Future Accelerators}, ``{ICFA WWW site}.''
  \url{http://www.fnal.gov/directorate/icfa/}, 2014.

\bibitem{ICFAnuPanel:Mandate:2013}
{The International Committee on Future Accelerators}, ``{ICFA Neutrino
  Panel}.'' \url{http://www.fnal.gov/directorate/icfa/neutrino_panel.html},
  2013.

\bibitem{DUNE:2015}
{The Deep Underground Neutrino Experiment collaboration}, ``{The Deep
  Underground Neutrino Experiment}.''
  \url{https://web.fnal.gov/collaboration/DUNE/SitePages/Home.aspx}, 2015.

\bibitem{Acciarri:2015uup}
{\bfseries DUNE} Collaboration, R.~Acciarri {\em et al.}, ``{Long-Baseline
  Neutrino Facility (LBNF) and Deep Underground Neutrino Experiment (DUNE)
  Conceptual Design Report Volume 2: The Physics Program for DUNE at LBNF},''
\href{http://arxiv.org/abs/1512.06148}{{\ttfamily arXiv:1512.06148
  [physics.ins-det]}}.

\bibitem{SBN:2014}
{The Short Baseline Neutrino Program}, ``{The Short Baseline Neutrino
  Program}.''
  \url{https://web.fnal.gov/collaboration/sbn/\_layouts/15/start.aspx\#/SitePages/Home.aspx},
  2014.

\bibitem{LBNF:2014}
{The Long Baseline Neutrino Facillity}, ``{The Long Baseline Neutrino
  Facillity}.'' \url{https://web.fnal.gov/project/LBNF/SitePages/Home.aspx},
  2014.

\bibitem{CENF:2015}
{CERN Neutrino Platform}, ``{CERN Neutrino Platform}.''
  \url{http://home.cern/about/experiments/cern-neutrino-platform}, 2015.

\bibitem{Abe:2011ts}
K.~Abe, T.~Abe, H.~Aihara, Y.~Fukuda, Y.~Hayato, {\em et al.}, ``{Letter of
  Intent: The Hyper-Kamiokande Experiment --- Detector Design and Physics
  Potential ---},''
\href{http://arxiv.org/abs/1109.3262}{{\ttfamily arXiv:1109.3262 [hep-ex]}}.

\bibitem{Abe:2015zbg}
{\bfseries Hyper-Kamiokande Proto-} Collaboration, K.~Abe {\em et al.},
  ``{Physics potential of a long-baseline neutrino oscillation experiment using
  a J-PARC neutrino beam and Hyper-Kamiokande},''
  \href{http://dx.doi.org/10.1093/ptep/ptv061}{{\em PTEP} {\bfseries 2015}
  (2015) 053C02},
\href{http://arxiv.org/abs/1502.05199}{{\ttfamily arXiv:1502.05199 [hep-ex]}}.

\bibitem{Hyper-Kamiokande:2016dsw}
{\bfseries Hyper-Kamiokande Proto-} Collaboration, K.~Abe {\em et al.},
  ``{Hyper-Kamiokande Design Report},'' {\em Report No. KEK-PREPRINT-2016-21
  and ICRR-REPORT-701-2016-1} (2016) 1--282.
\url{https://lib-extopc.kek.jp/preprints/PDF/2016/1627/1627021.pdf}.

\bibitem{Cao:2015ita}
J.~Cao {\em et al.}, ``{On the complementarity of Hyper-K and LBNF},''
\href{http://arxiv.org/abs/1501.03918}{{\ttfamily arXiv:1501.03918
  [physics.acc-ph]}}.

\bibitem{T2K:WWW}
{{\bf T2K} Collaboration}, ``{T2K}.'' \url{http://t2k-experiment.org}, 2013.

\bibitem{NOvA:WWW}
{{\bf NOvA} Collaboration}, ``{NOvA Neutrino Experiment}.''
  \url{https://www-nova.fnal.gov}, 2016.

\bibitem{MiCroBooNE:WWW}
{{\bf MicroBooNE} Collaboration}, ``{MicroBooNE}.''
  \url{http://www-microboone.fnal.gov}, 2016.

\bibitem{MINERvA:WWW}
{{\bf MINERvA} Collaboration}, ``{MINERvA: Bringing Neutrinos into Sharp
  Focus}.'' \url{https://minerva.fnal.gov}, 2016.

\bibitem{RECFA:Survey:2010}
{The Restricted European Committee for Future Accelerators}, ``{Survey of
  European Particle Physics 2009}.'' Ecfa/rc/10/388, 2010.

\bibitem{STFC:PPAP:RoadMap:2015}
{STFC Particle Physics Advisory Panel}, ``{Update to the UK Particle Physics
  Roadmap}.'' \url{https://www.stfc.ac.uk/3116.aspx}, 2015.

\bibitem{ECFA:WWW}
{The European Committee for Future Accelerators}, ``{The European Committee for
  Future Accelerators}.'' \url{http://ecfa.web.cern.ch/ecfa/en/Welcome.html},
  2008.

\bibitem{Geesaman:2015fha}
A.~Aprahamian {\em et al.},
``{Reaching for the horizon: The 2015 long range plan for nuclear science},''.

\bibitem{EXO:WWW}
{{\bf EXO} Collaboration}, ``{The EXO Project}.''
  \url{http://www.wipp.energy.gov/science/DBDecay/DBDecayNew2.html}, 2016.

\bibitem{IceCube:WWW}
{{\bf IceCube} Collaboration}, ``{IceCube}.'' \url{https://icecube.wisc.edu},
  2016.

\bibitem{Zuber:2015ita}
{\bfseries HALO} Collaboration, K.~Zuber, ``{HALO, a supernova neutrino
  observatory},''
\href{http://dx.doi.org/10.1016/j.nuclphysbps.2015.06.059}{{\em Nucl. Part.
  Phys. Proc.} {\bfseries 265-266} (2015) 233--235}.

\bibitem{SNOplus:WWW}
{{\bf SNO+} Collaboration}, ``{SNO+}.''
  \url{http://snoplus.phy.queensu.ca/Home.html}, 2016.

\bibitem{MINOS:WWW}
{{\bf MINOS} Collaboration}, ``{The MINOS Experiment and NuMI Beamline}.''
  \url{https://www-numi.fnal.gov}, 2016.

\bibitem{Pontecorvo:1957qd}
B.~Pontecorvo, ``Inverse beta processes and nonconservation of lepton charge,''
{\em Sov. Phys. JETP} {\bfseries 7} (1958) 172--173.

\bibitem{Maki:1962mu}
Z.~Maki, M.~Nakagawa, and S.~Sakata, ``{Remarks on the unified model of
  elementary particles},''
\href{http://dx.doi.org/10.1143/PTP.28.870}{{\em Prog. Theor. Phys.} {\bfseries
  28} (1962) 870--880}.

\bibitem{Agashe:2014kda}
{\bfseries Particle Data Group} Collaboration, K.~Olive {\em et al.}, ``{Review
  of Particle Physics},''
\href{http://dx.doi.org/10.1088/1674-1137/38/9/090001}{{\em Chin.Phys.}
  {\bfseries C38} (2014) 090001}.

\bibitem{Abe:2011ks}
{\bfseries T2K} Collaboration, K.~Abe {\em et al.}, ``{The T2K Experiment},''
  \href{http://dx.doi.org/10.1016/j.nima.2011.06.067}{{\em Nucl.Instrum.Meth.}
  {\bfseries A659} (2011) 106--135},
\href{http://arxiv.org/abs/1106.1238}{{\ttfamily arXiv:1106.1238
  [physics.ins-det]}}.

\bibitem{Abe:2015awa}
{\bfseries T2K} Collaboration, K.~Abe {\em et al.}, ``{Measurements of neutrino
  oscillation in appearance and disappearance channels by the T2K experiment
  with $6.6\cdot 10^{20}$ protons on target},''
  \href{http://dx.doi.org/10.1103/PhysRevD.91.072010}{{\em Phys. Rev.}
  {\bfseries D91} no.~7, (2015) 072010},
\href{http://arxiv.org/abs/1502.01550}{{\ttfamily arXiv:1502.01550 [hep-ex]}}.

\bibitem{T2K2-EOI}
{The T2K collaboration}, ``{Expression of Interest for an Extended Run at T2K
  to $20 \times 10^{21}$\,POT}.'' Private communication, 2016.

\bibitem{Ayres:2004js}
{\bfseries NOvA} Collaboration, D.~Ayres {\em et al.}, ``{NOvA: Proposal to
  build a 30 kiloton off-axis detector to study nu(mu) --> nu(e) oscillations
  in the NuMI beamline},''
\href{http://arxiv.org/abs/hep-ex/0503053}{{\ttfamily arXiv:hep-ex/0503053
  [hep-ex]}}.

\bibitem{Adamson:2016xxw}
{\bfseries NOvA} Collaboration, P.~Adamson {\em et al.}, ``{First measurement
  of muon-neutrino disappearance in NOvA},''
  \href{http://dx.doi.org/http://dx.doi.org/10.1103/PhysRevD.93.051104}{{\em
  Phys. Rev.} {\bfseries D93} no.~5, (2016) 051104},
\href{http://arxiv.org/abs/1601.05037}{{\ttfamily arXiv:1601.05037 [hep-ex]}}.

\bibitem{Adamson:2016tbq}
{\bfseries NOvA} Collaboration, P.~Adamson {\em et al.}, ``{First measurement
  of electron neutrino appearance in a NOvA},''
  \href{http://dx.doi.org/10.1103/PhysRevLett.116.151806}{{\em Phys. Rev.
  Lett.} {\bfseries 116} no.~15, (2016) 151806},
\href{http://arxiv.org/abs/1601.05022}{{\ttfamily arXiv:1601.05022 [hep-ex]}}.

\bibitem{Shanahan:LP15}
{Shanahan, P.}, ``{Long Baseline and Atmospheric Neutrino Experiments}.''
  \url{http://lp2015.ijs.si}, 2015.
\newblock Presented at the XXVII International Symposium on Lepton Photon
  Interactions, Ljubljana, August, 2015.

\bibitem{Abe:2016ero}
{\bfseries Hyper-Kamiokande Proto-} Collaboration, K.~Abe {\em et al.},
  ``{Physics Potentials with the Second Hyper-Kamiokande Detector in Korea},''
\href{http://arxiv.org/abs/1611.06118}{{\ttfamily arXiv:1611.06118 [hep-ex]}}.

\bibitem{Bandyopadhyay:2007kx}
{\bfseries ISS Physics Working Group} Collaboration, A.~Bandyopadhyay {\em et
  al.}, ``{Physics at a future Neutrino Factory and super-beam facility},''
  \href{http://dx.doi.org/10.1088/0034-4885/72/10/106201}{{\em Rept. Prog.
  Phys.} {\bfseries 72} (2009) 106201},
\href{http://arxiv.org/abs/0710.4947}{{\ttfamily arXiv:0710.4947 [hep-ph]}}.

\bibitem{Choubey:2011zzq}
{\bfseries IDS-NF} Collaboration, S.~Choubey {\em et al.}, ``{International
  Design Study for the Neutrino Factory, Interim Design Report},''
\href{http://arxiv.org/abs/1112.2853}{{\ttfamily arXiv:1112.2853 [hep-ex]}}.

\bibitem{Delahaye:2013jla}
J.-P. Delahaye, C.~Ankenbrandt, A.~Bogacz, S.~Brice, A.~Bross, {\em et al.},
  ``{Enabling Intensity and Energy Frontier Science with a Muon Accelerator
  Facility in the U.S.: A White Paper Submitted to the 2013 U.S. Community
  Summer Study of the Division of Particles and Fields of the American Physical
  Society},''
\href{http://arxiv.org/abs/1308.0494}{{\ttfamily arXiv:1308.0494
  [physics.acc-ph]}}.

\bibitem{Bogomilov:2014koa}
M.~Bogomilov, R.~Matev, R.~Tsenov, M.~Dracos, M.~Bonesini, {\em et al.},
  ``{Neutrino factory},''
\href{http://dx.doi.org/10.1103/PhysRevSTAB.17.121002}{{\em Phys.Rev.ST
  Accel.Beams} {\bfseries 17} no.~12, (2014) 121002}.

\bibitem{:1900cvd}
J.~C. Gallardo {\em et al.}, ``{Muon Collider: Feasibility Study},'' 1996.
\newblock \url{http://www.cap.bnl.gov/mumu/pubs/snowmass96/part6.pdf}. Prepared
  for 1996 DPF / DPB Summer Study on New Directions for High Energy Physics
  (Snowmass 96), Snowmass, Colorado, 25 Jun - 12 Jul 1996.

\bibitem{Finley:2000cn}
D.~Finley and N.~Holtkamp, ``A feasibility study of a neutrino source based on
  a muon storage ring,''
{\em Nucl. Instrum. Meth.} {\bfseries A472} (2000) 388--394.

\bibitem{Kuno:2001tb}
Y.~Kuno {\em et al.}, ``A feasibility study of a neutrino factory in japan.''
  {\url{HTTP://WWW-PRISM.KEK.JP/NUFACTJ/}}, May, 2001.

\bibitem{Ozaki:2001}
S.~Ozaki, R.~B. Palmer, M.~S. Zisman, and J.~C. Gallardo, ``{Feasibility
  Study-II of a Muon-Based Neutrino Source},'' Tech. Rep. BNL-52623, Brookhaven
  National Laboratory, Upton, NY, 2001.

\bibitem{Apollonio:2008aa}
{\bfseries ISS Accelerator Working Group} Collaboration, M.~Apollonio {\em et
  al.}, ``{Accelerator design concept for future neutrino facilities},''
  \href{http://dx.doi.org/10.1088/1748-0221/4/07/P07001}{{\em JINST} {\bfseries
  4} (2009) P07001},
\href{http://arxiv.org/abs/0802.4023}{{\ttfamily arXiv:0802.4023
  [physics.acc-ph]}}.

\bibitem{Delahaye:2015yxa}
J.-P. Delahaye {\em et al.}, ``{A Staged Muon Accelerator Facility For Neutrino
  and Collider Physics},'' in {\em {Proceedings, 5th International Particle
  Accelerator Conference (IPAC 2014)}}, p.~WEZA02.
\newblock 2014.
\newblock \href{http://arxiv.org/abs/1502.01647}{{\ttfamily arXiv:1502.01647
  [physics.acc-ph]}}.
\newblock
  \url{{http://inspirehep.net/record/1343158/files/arXiv:1502.01647.pdf}}.

\bibitem{Efthymiopoulos:2008zz}
{\bfseries MERIT} Collaboration, I.~Efthymiopoulos,
  \href{http://dx.doi.org/10.1109/NSSMIC.2008.4775051}{``{MERIT - The high
  intensity liquid mercury target experiment at the CERN PS},''} in {\em
  {Proceedings, 2008 IEEE Nuclear Science Symposium, Medical Imaging Conference
  and 16th International Workshop on Room-Temperature Semiconductor X-Ray and
  Gamma-Ray Detectors (NSS/MIC 2008 / RTSD 2008)}}, pp.~3302--3305.
\newblock 2008.

\bibitem{Machida:2012zz}
S.~Machida, R.~Barlow, J.~Berg, N.~Bliss, R.~Buckley, {\em et al.},
  ``{Acceleration in the linear non-scaling fixed-field alternating-gradient
  accelerator EMMA},''
\href{http://dx.doi.org/10.1038/nphys2179}{{\em Nature Phys.} {\bfseries 8}
  (2012) 243--247}.

\bibitem{Owen:2012zzb}
{\bfseries EMMA} Collaboration, H.~Owen, ``{The EMMA non-scaling FFAG project:
  Implications for intensity frontier accelerators},''
\href{http://dx.doi.org/10.1063/1.3700645}{{\em AIP Conf. Proc.} {\bfseries
  1441} (2012) 655--659}.

\bibitem{MICE-WWW}
{{\bf MICE} Collaboration}, ``{International Muon Ionization Cooling
  Experiment}.'' \url{http://mice.iit.edu}.

\bibitem{Calabretta:2011nr}
L.~Calabretta {\em et al.}, ``{Preliminary Design Study of High-Power H2+
  Cyclotrons for the DAE$\delta$ALUS Experiment},''
\href{http://arxiv.org/abs/1107.0652}{{\ttfamily arXiv:1107.0652
  [physics.acc-ph]}}.

\bibitem{Alonso:2010fy}
{\bfseries DAE$\delta$ALUS} Collaboration, J.~Alonso {\em et al.}, ``{A Study
  of Detector Configurations for the DUSEL CP Violation Searches Combining LBNE
  and DAE$\delta$ALUS},''
\href{http://arxiv.org/abs/1008.4967}{{\ttfamily arXiv:1008.4967 [hep-ex]}}.

\bibitem{Alonso:2010fs}
J.~Alonso, F.~Avignone, W.~Barletta, R.~Barlow, H.~Baumgartner, {\em et al.},
  ``{Expression of Interest for a Novel Search for CP Violation in the Neutrino
  Sector: DAE$\delta$ALUS},''
\href{http://arxiv.org/abs/1006.0260}{{\ttfamily arXiv:1006.0260
  [physics.ins-det]}}.

\bibitem{Scholberg:2012zz}
{\bfseries {\bf DAE$\delta$ALUS}} Collaboration, K.~Scholberg,
  ``{DAE$\delta$ALUS},''
\href{http://dx.doi.org/10.1063/1.3700577}{{\em AIP Conf.Proc.} {\bfseries
  1441} (2012) 435--437}.

\bibitem{Toups:2013dxa}
{\bfseries DAE$\delta$ALUS} Collaboration, M.~Toups, ``{DAE$\delta$ALUS: A Path
  to $\delta_{CP}$ Using Cyclotron Decay-at-Rest Neutrino Sources},''
\href{http://dx.doi.org/10.1016/j.nuclphysbps.2013.04.072}{{\em Nucl. Phys.
  Proc. Suppl.} {\bfseries 237-238} (2013) 117--120}.

\bibitem{Toups:2015pma}
{\bfseries DAE$\delta$ALUS} Collaboration, M.~Toups, ``{DAE$\delta$ALUS: A
  Phased Neutrino Physics Program Using Cyclotron Decay-at-Rest Neutrino
  Sources},''
\href{http://dx.doi.org/10.1016/j.phpro.2014.12.116}{{\em Phys. Procedia}
  {\bfseries 61} (2015) 518--523}.

\bibitem{Adelmann:2012kq}
A.~Adelmann, J.~Alonso, W.~Barletta, R.~Barlow, L.~Bartoszek, {\em et al.},
  ``{Cost-effective Design Options for IsoDAR},''
\href{http://arxiv.org/abs/1210.4454}{{\ttfamily arXiv:1210.4454
  [physics.acc-ph]}}.

\bibitem{Baussan:2013zcy}
{\bfseries ESSnuSB} Collaboration, E.~Baussan {\em et al.}, ``{A Very Intense
  Neutrino Super Beam Experiment for Leptonic CP Violation Discovery based on
  the European Spallation Source Linac: A Snowmass 2013 White Paper},''
\href{http://arxiv.org/abs/1309.7022}{{\ttfamily arXiv:1309.7022 [hep-ex]}}.

\bibitem{Declais:1994su}
Y.~Declais {\em et al.}, ``{Search for neutrino oscillations at 15-meters,
  40-meters, and 95-meters from a nuclear power reactor at Bugey},''
\href{http://dx.doi.org/10.1016/0550-3213(94)00513-E}{{\em Nucl. Phys.}
  {\bfseries B434} (1995) 503--534}.

\bibitem{Aguilar:2001ty}
{\bfseries LSND} Collaboration, A.~Aguilar {\em et al.}, ``{Evidence for
  neutrino oscillations from the observation of $\bar{\nu}_e$ appearance in a
  $\bar{\nu}_\mu$ beam},'' {\em Phys. Rev. D} {\bfseries 64} (2001) 112007,
\href{http://arxiv.org/abs/0104049}{{\ttfamily arXiv:0104049 [hep-ex]}}.

\bibitem{AguilarArevalo:2008rc}
{\bfseries MiniBooNE} Collaboration, A.~A. Aguilar-Arevalo {\em et al.},
  ``{Unexplained Excess of Electron-Like Events From a 1-GeV Neutrino Beam},''
  \href{http://dx.doi.org/10.1103/PhysRevLett.102.101802}{{\em Phys. Rev.
  Lett.} {\bfseries 102} (2009) 101802},
\href{http://arxiv.org/abs/0812.2243}{{\ttfamily arXiv:0812.2243 [hep-ex]}}.

\bibitem{Abdurashitov:2009tn}
{\bfseries SAGE} Collaboration, J.~N. Abdurashitov {\em et al.}, ``{Measurement
  of the solar neutrino capture rate with gallium metal. III: Results for the
  2002--2007 data-taking period},''
  \href{http://dx.doi.org/10.1103/PhysRevC.80.015807}{{\em Phys. Rev.}
  {\bfseries C80} (2009) 015807},
\href{http://arxiv.org/abs/0901.2200}{{\ttfamily arXiv:0901.2200 [nucl-ex]}}.

\bibitem{Kaether:2010ag}
F.~Kaether, W.~Hampel, G.~Heusser, J.~Kiko, and T.~Kirsten, ``{Reanalysis of
  the GALLEX solar neutrino flux and source experiments},''
  \href{http://dx.doi.org/10.1016/j.physletb.2010.01.030}{{\em Phys. Lett.}
  {\bfseries B685} (2010) 47--54},
\href{http://arxiv.org/abs/1001.2731}{{\ttfamily arXiv:1001.2731 [hep-ex]}}.

\bibitem{Mention:2011rk}
G.~Mention, M.~Fechner, T.~Lasserre, T.~A. Mueller, D.~Lhuillier, M.~Cribier,
  and A.~Letourneau, ``{The Reactor Antineutrino Anomaly},''
  \href{http://dx.doi.org/10.1103/PhysRevD.83.073006}{{\em Phys. Rev.}
  {\bfseries D83} (2011) 073006},
\href{http://arxiv.org/abs/1101.2755}{{\ttfamily arXiv:1101.2755 [hep-ex]}}.

\bibitem{Aguilar-Arevalo:2013pmq}
{\bfseries MiniBooNE} Collaboration, A.~A. Aguilar-Arevalo {\em et al.},
  ``{Improved Search for $\bar \nu_\mu \rightarrow \bar \nu_e$ Oscillations in
  the MiniBooNE Experiment},''
  \href{http://dx.doi.org/10.1103/PhysRevLett.110.161801}{{\em Phys. Rev.
  Lett.} {\bfseries 110} (2013) 161801},
\href{http://arxiv.org/abs/1207.4809}{{\ttfamily arXiv:1207.4809 [hep-ex]}}.

\bibitem{Tzanankos:2011zz}
{\bfseries MINOS+} Collaboration, G.~Tzanankos {\em et al.}, ``{MINOS+: a
  Proposal to FNAL to run MINOS with the medium energy NuMI beam}.''
  \url{http://inspirehep.net/record/944685}, 2011.

\bibitem{Chen:2007ae}
{\bfseries MicroBooNE} Collaboration, H.~Chen {\em et al.}, ``{Proposal for a
  New Experiment Using the Booster and NuMI Neutrino Beamlines: MicroBooNE}.''
  \url{http://lss.fnal.gov/archive/test-proposal/0000/fermilab-proposal-0974.shtml},
  2007.

\bibitem{Antonello:2015lea}
{\bfseries LAr1-ND, ICARUS-WA104, MicroBooNE} Collaboration, M.~Antonello {\em
  et al.}, ``{A Proposal for a Three Detector Short-Baseline Neutrino
  Oscillation Program in the Fermilab Booster Neutrino Beam},''
\href{http://arxiv.org/abs/1503.01520}{{\ttfamily arXiv:1503.01520
  [physics.ins-det]}}.

\bibitem{Harada:2016rou}
M.~Harada {\em et al.}, ``{Status Report (22th J-PARC PAC): Searching for a
  Sterile Neutrino at J-PARC MLF (E56, JSNS2)},''
\href{http://arxiv.org/abs/1610.08186}{{\ttfamily arXiv:1610.08186
  [physics.ins-det]}}.

\bibitem{Alonso:2012zv}
{\bfseries DAE$\delta$ALUS} Collaboration, J.~R. Alonso, ``{High Current H2+
  Cyclotrons for Neutrino Physics: The IsoDAR and DAE$\delta$ALUS Projects},''
  \href{http://dx.doi.org/10.1063/1.4802375}{{\em AIP Conf. Proc.} {\bfseries
  1525} (2012) 480--486},
\href{http://arxiv.org/abs/1210.3679}{{\ttfamily arXiv:1210.3679
  [physics.acc-ph]}}.

\bibitem{Shaevitz:2015uar}
{\bfseries IsoDAR/DAE$\delta$ALUS} Collaboration, M.~Shaevitz, ``{Searching for
  Sterile Neutrinos and CP Violation: The IsoDAR and DAE$\delta$ALUS
  Experiments},''
{\em PoS} {\bfseries NEUTEL2015} (2015) 034.

\bibitem{HEPAP:P5:2014}
{Particle Physics Projects Prioritization Panel}, ``{Build for Discovery}.''
  \url{http://science.energy.gov/~/media/hep/hepap/pdf/May\%202014/FINAL_P5_Report_053014.pdf},
  2014.

\bibitem{Adey:2013afh}
D.~Adey, S.~Agarwalla, C.~Ankenbrandt, R.~Asfandiyarov, J.~Back, {\em et al.},
  ``{Neutrinos from Stored Muons nuSTORM: Expression of Interest},''
\href{http://arxiv.org/abs/1305.1419}{{\ttfamily arXiv:1305.1419
  [physics.acc-ph]}}.

\bibitem{Adey:1537983}
D.~Adey, C.~Ankenbrandt, S.~Agarwalla, {\em et al.}, ``Neutrinos from stored
  muons (storm): Expression of interest,'' Tech. Rep. CERN-SPSC-2013-015.
  SPSC-EOI-009, CERN, Geneva, Apr, 2013.

\bibitem{Kyberd:2013ii}
P.~Kyberd {\em et al.}, ``Neutrinos from stored muons proposal to the fermilab
  pac,'' May, 2013.
\newblock
  \url{https://indico.fnal.gov/getFile.py/access?resId=0&materialId=6&confId=6847}.

\bibitem{Kyberd:2013ij}
P.~Kyberd {\em et al.}, ``nu{STORM} costing document,'' May, 2013.
\newblock
  \url{https://indico.fnal.gov/getFile.py/access?resId=0&materialId=8&confId=6847}.

\bibitem{Adey:2013pio}
{\bfseries nuSTORM} Collaboration, D.~Adey {\em et al.}, ``{nuSTORM - Neutrinos
  from STORed Muons: Proposal to the Fermilab PAC},''
\href{http://arxiv.org/abs/1308.6822}{{\ttfamily arXiv:1308.6822
  [physics.acc-ph]}}.

\bibitem{Lackowski:2013ria}
T.~Lackowski, S.~Dixon, R.~Jedziniak, M.~Blewitt, and L.~Fink, ``{nuSTORM
  Project Definition Report},''
\href{http://arxiv.org/abs/1309.1389}{{\ttfamily arXiv:1309.1389
  [physics.ins-det]}}.

\bibitem{Adey:2014rfv}
{\bfseries nuSTORM} Collaboration, D.~Adey {\em et al.}, ``{Light sterile
  neutrino sensitivity at the nuSTORM facility},''
  \href{http://dx.doi.org/10.1103/PhysRevD.89.071301}{{\em Phys.Rev.}
  {\bfseries D89} (2014) 071301},
\href{http://arxiv.org/abs/1402.5250}{{\ttfamily arXiv:1402.5250 [hep-ex]}}.

\bibitem{Adey:2015iha}
D.~Adey, R.~Bayes, A.~Bross, and P.~Snopok, ``{nuSTORM and A Path to a Muon
  Collider},''
\href{http://dx.doi.org/10.1146/annurev-nucl-102014-021930}{{\em Ann. Rev.
  Nucl. Part. Sci.} {\bfseries 65} (2015) 145--175}.

\bibitem{EUmuonDiscussion:2015}
S.~Bertolucci and K.~Long, ``{Discussion of the scientific potential of muon
  beams}.'' \url{https://indico.cern.ch/event/450863/}, 2015.

\bibitem{EUmuonOppoDoc:2015}
S.~Bertolucci and K.~Long, ``{Muon-accelerator based facilities: opportunities
  for science}.''
  \url{https://indico.cern.ch/event/450863/attachments/1189718/1755975/2015-12-11-Muon-discussion-summary.pdf},
  2015.

\bibitem{CERN:PBC:WWW:2017}
J.~Jaeckel, M.~Lamont, and C.~Vallee, ``{Physics Beyond Colliders}.''
  \url{http://pbc.web.cern.ch}, 2017.

\bibitem{Antoniou:2006mh}
{\bfseries NA49-future} Collaboration, N.~Antoniou {\em et al.}, ``{Study of
  hadron production in hadron nucleus and nucleus nucleus collisions at the
  CERN SPS}.'' \url{https://cds.cern.ch/record/995681/files/spsc-2006-034.pdf},
  2006.

\bibitem{Abgrall:2011ae}
{\bfseries NA61/SHINE} Collaboration, N.~Abgrall {\em et al.}, ``{Measurements
  of Cross Sections and Charged Pion Spectra in Proton-Carbon Interactions at
  31 GeV/c},'' \href{http://dx.doi.org/10.1103/PhysRevC.84.034604}{{\em Phys.
  Rev.} {\bfseries C84} (2011) 034604},
\href{http://arxiv.org/abs/1102.0983}{{\ttfamily arXiv:1102.0983 [hep-ex]}}.

\bibitem{Abgrall:2011ts}
{\bfseries NA61/SHINE} Collaboration, N.~Abgrall {\em et al.}, ``{Measurement
  of Production Properties of Positively Charged Kaons in Proton-Carbon
  Interactions at 31 GeV/c},''
  \href{http://dx.doi.org/10.1103/PhysRevC.85.035210}{{\em Phys. Rev.}
  {\bfseries C85} (2012) 035210},
\href{http://arxiv.org/abs/1112.0150}{{\ttfamily arXiv:1112.0150 [hep-ex]}}.

\bibitem{Abgrall:2012pp}
{\bfseries NA61/SHINE} Collaboration, N.~Abgrall {\em et al.}, ``{Pion emission
  from the T2K replica target: method, results and application},''
  \href{http://dx.doi.org/10.1016/j.nima.2012.10.079}{{\em Nucl. Instrum.
  Meth.} {\bfseries A701} (2013) 99--114},
\href{http://arxiv.org/abs/1207.2114}{{\ttfamily arXiv:1207.2114 [hep-ex]}}.

\bibitem{Abgrall:2013wda}
{\bfseries NA61/SHINE} Collaboration, N.~Abgrall {\em et al.}, ``{Measurements
  of production properties of $K_S^0$ mesons and $\Lambda$ hyperons in
  proton-carbon interactions at 31 GeV/ c},''
  \href{http://dx.doi.org/10.1103/PhysRevC.89.025205}{{\em Phys. Rev.}
  {\bfseries C89} no.~2, (2014) 025205},
\href{http://arxiv.org/abs/1309.1997}{{\ttfamily arXiv:1309.1997
  [physics.acc-ph]}}.

\bibitem{Abgrall:2013qoa}
{\bfseries NA61/SHINE} Collaboration, N.~Abgrall {\em et al.}, ``{Measurement
  of negatively charged pion spectra in inelastic p+p interactions at $p_{lab}$
  = 20, 31, 40, 80 and 158 GeV/c},''
  \href{http://dx.doi.org/10.1140/epjc/s10052-014-2794-6}{{\em Eur. Phys. J.}
  {\bfseries C74} no.~3, (2014) 2794},
\href{http://arxiv.org/abs/1310.2417}{{\ttfamily arXiv:1310.2417 [hep-ex]}}.

\bibitem{Abgrall:2014xwa}
{\bfseries NA61} Collaboration, N.~Abgrall {\em et al.}, ``{NA61/SHINE facility
  at the CERN SPS: beams and detector system},''
  \href{http://dx.doi.org/10.1088/1748-0221/9/06/P06005}{{\em JINST} {\bfseries
  9} (2014) P06005},
\href{http://arxiv.org/abs/1401.4699}{{\ttfamily arXiv:1401.4699
  [physics.ins-det]}}.

\bibitem{Aduszkiewicz:2015jna}
{\bfseries NA61/SHINE} Collaboration, A.~Aduszkiewicz {\em et al.},
  ``{Multiplicity and transverse momentum fluctuations in inelastic
  proton-proton interactions at the CERN Super Proton Synchrotron},''
\href{http://arxiv.org/abs/1510.00163}{{\ttfamily arXiv:1510.00163 [hep-ex]}}.

\bibitem{Abgrall:2015hmv}
{\bfseries NA61/SHINE} Collaboration, N.~Abgrall {\em et al.}, ``{Measurements
  of $\pi^\pm$, $K^\pm$, $K^0_S$, $\Lambda$ and proton production in
  proton-carbon interactions at 31 GeV/$c$ with the NA61/SHINE spectrometer at
  the CERN SPS},''
  \href{http://dx.doi.org/http://dx.doi.org/10.1140/epjc/s10052-016-3898-y}{{\em
  Eur. Phys. J.} {\bfseries C76} no.~2, (2016) 84},
\href{http://arxiv.org/abs/1510.02703}{{\ttfamily arXiv:1510.02703 [hep-ex]}}.

\bibitem{Morfin:nuSTEV:2015}
J.~Morfin, ``{NuSTEC (Neutrino Scattering Theory Experiment Collaboration)
  Goals and Strategy }.''
  \url{https://indico.fnal.gov/getFile.py/access?contribId=275\&sessionId=15\&resId=0\&materialId=slides\&confId=8903},
  2015.

\bibitem{Akimov:2015nza}
{\bfseries COHERENT} Collaboration, D.~Akimov {\em et al.}, ``{The COHERENT
  Experiment at the Spallation Neutron Source},''
\href{http://arxiv.org/abs/1509.08702}{{\ttfamily arXiv:1509.08702
  [physics.ins-det]}}.

\bibitem{Bednyakov:Ed:2014}
{V.A.~Bednyakov, D.~Naumov, editors}, ``{The White Book; JINR Neutrino
  Program}.'' \url{http://dlnp.jinr.ru/en/neutrino-research}, 2014.

\bibitem{Bhadra:2014oma}
{\bfseries nuPRISM} Collaboration, S.~Bhadra {\em et al.}, ``{Letter of Intent
  to Construct a nuPRISM Detector in the J-PARC Neutrino Beamline},''
\href{http://arxiv.org/abs/1412.3086}{{\ttfamily arXiv:1412.3086
  [physics.ins-det]}}.

\bibitem{Berns:2013usa}
{\bfseries CAPTAIN} Collaboration, H.~Berns {\em et al.}, ``{The CAPTAIN
  Detector and Physics Program},'' in {\em {Community Summer Study 2013:
  Snowmass on the Mississippi (CSS2013) Minneapolis, MN, USA, July 29-August 6,
  2013}}.
\newblock 2013.
\newblock \href{http://arxiv.org/abs/1309.1740}{{\ttfamily arXiv:1309.1740
  [physics.ins-det]}}.
\newblock \url{http://inspirehep.net/record/1253116/files/arXiv:1309.1740.pdf}.

\bibitem{KutterPawloski:ProtoDUNE:2015}
T.~Kutter and G.~Pawloski, ``{ProtoDUNE}.''
  \url{https://indico.fnal.gov/categoryDisplay.py?categId=454}, 2015.

\bibitem{Nessi:CENF:2015}
M.~Nessi, ``{The CERN Neutrino Platform}.''
  \url{https://indico.cern.ch/event/453274/contribution/14/attachments/1190653/1728543/15-11-20-Nu-Platform-ECFA.pdf},
  2015.

\bibitem{WA105:WWW:2014}
{{\bf WA105} Collaboration}, ``{LBNO-DEMO - Long Baseline Neutrino Observatory
  Demonstration experiment (WA105)}.'' \url{http://wa105.web.cern.ch/wa105/},
  2014.

\bibitem{Cantini:2015naq}
{\bfseries WA105, LAGUNA-LBNO} Collaboration, C.~Cantini {\em et al.},
  ``{Recent R\&D results on LAr LEM TPC and plans for LBNO demonstrators},''
\href{http://dx.doi.org/10.1088/1742-6596/650/1/012011}{{\em J. Phys. Conf.
  Ser.} {\bfseries 650} no.~1, (2015) 012011}.

\bibitem{Adamson:2013xka}
P.~Adamson, J.~Coelho, G.~Davies, J.~Evans, P.~Guzowski, {\em et al.},
  ``{CHerenkov detectors In mine PitS (CHIPS) Letter of Intent to FNAL},''
\href{http://arxiv.org/abs/1307.5918}{{\ttfamily arXiv:1307.5918
  [physics.ins-det]}}.

\bibitem{Vallee:2016xde}
C.~Vallee, ``{Pacific Neutrinos: Towards a High Precision Measurement of CP
  Violation ?},'' 2016.
\newblock
  \url{http://inspirehep.net/record/1494807/files/arXiv:1610.08655.pdf}.

\bibitem{Brunner:2013lua}
J.~Brunner, ``{Counting Electrons to Probe the Neutrino Mass Hierarchy},''
\href{http://arxiv.org/abs/1304.6230}{{\ttfamily arXiv:1304.6230 [hep-ex]}}.

\bibitem{Adrian-Martinez:2016fdl}
{\bfseries KM3Net} Collaboration, S.~Adrian-Martinez {\em et al.}, ``{Letter of
  intent for KM3NeT 2.0},''
  \href{http://dx.doi.org/10.1088/0954-3899/43/8/084001}{{\em J. Phys.}
  {\bfseries G43} no.~8, (2016) 084001},
\href{http://arxiv.org/abs/1601.07459}{{\ttfamily arXiv:1601.07459
  [astro-ph.IM]}}.

\bibitem{Gann:2015fba}
{\bfseries THEIA Interest Group} Collaboration, G.~D. Orebi~Gann, ``{Physics
  Potential of an Advanced Scintillation Detector: Introducing THEIA},''
\href{http://arxiv.org/abs/1504.08284}{{\ttfamily arXiv:1504.08284
  [physics.ins-det]}}.

\bibitem{WsNuPrgInJp:Svoboda:2015}
R.~Svoboda, ``{Water-base liquid scintillator}.''
  \url{http://www-conf.kek.jp/past/ws_nu_prog_in_jp/program/20150804_JPARC_Svoboda.pdf},
  2015.

\bibitem{MICEproposal:2003}
{{\bf MICE} Collaboration}, ``{MICE: An International Muon Ionization Cooling
  Experiment}.'' \url{
  http://mice.iit.edu/micenotes/public/pdf/MICE0021/MICE0021.pdf }, 2003.
\newblock MICE Note 21.

\bibitem{MICE:Note:53:2003}
J.~Wood, ``Scientific approval of mice.''
  \url{http://mice.iit.edu/micenotes/public/pdf/MICE053/MICE0053.pdf}, 2003.
\newblock MICE Note 53.

\bibitem{MICEmine:Document:159}
{MICE Executive Board}, ``Mice report to the mice project board.''
  \url{http://micewww.pp.rl.ac.uk/documents/159}, 2015.

\bibitem{MICEmine:Document:160}
{MICE International Project Office}, ``Resource loaded schedule, costs and
  risks for the completion of the mice project.''
  \url{http://micewww.pp.rl.ac.uk/documents/160}, 2015.

\bibitem{Bogomilov:2017vwz}
{\bfseries MICE} Collaboration, M.~Bogomilov {\em et al.}, ``{Design and
  expected performance of the MICE demonstration of ionization cooling},''
\href{http://arxiv.org/abs/1701.06403}{{\ttfamily arXiv:1701.06403
  [physics.acc-ph]}}.

\bibitem{genie}
C.~Andreopoulos {\em et al.}, ``{The GENIE Neutrino Monte Carlo Generator},''
  \href{http://dx.doi.org/10.1016/j.nima.2009.12.009}{{\em Nucl. Instrum.
  Meth.} {\bfseries A614} (2010) 87--104},
\href{http://arxiv.org/abs/0905.2517}{{\ttfamily arXiv:0905.2517 [hep-ph]}}.

\bibitem{neut}
Y.~Hayato, ``{A neutrino interaction simulation program library NEUT},''
{\em Acta Phys. Polon.} {\bfseries B40} (2009) 2477--2489.

\bibitem{nuwro}
T.~Golan, C.~Juszczak, and J.~T. Sobczyk, ``{Final State Interactions Effects
  in Neutrino-Nucleus Interactions},''
  \href{http://dx.doi.org/10.1103/PhysRevC.86.015505}{{\em Phys. Rev.}
  {\bfseries C86} (2012) 015505},
\href{http://arxiv.org/abs/1202.4197}{{\ttfamily arXiv:1202.4197 [nucl-th]}}.

\bibitem{Ago03}
{\bfseries GEANT4} Collaboration, S.~Agostinelli {\em et al.}, ``Geant4: A
  simulation toolkit,''
{\em Nuclear Instruments and Methods in Physics Research A} {\bfseries 506}
  (2003) 250--303.

\bibitem{An:2012eh}
{\bfseries Daya Bay} Collaboration, F.~An {\em et al.}, ``{Observation of
  electron-antineutrino disappearance at Daya Bay},''
  \href{http://dx.doi.org/10.1103/PhysRevLett.108.171803}{{\em Phys.Rev.Lett.}
  {\bfseries 108} (2012) 171803},
\href{http://arxiv.org/abs/1203.1669}{{\ttfamily arXiv:1203.1669 [hep-ex]}}.

\bibitem{An:2012bu}
{\bfseries Daya Bay} Collaboration, F.~An {\em et al.}, ``{Improved Measurement
  of Electron Antineutrino Disappearance at Daya Bay},''
  \href{http://dx.doi.org/10.1088/1674-1137/37/1/011001}{{\em Chin. Phys.}
  {\bfseries C37} (2013) 011001},
\href{http://arxiv.org/abs/1210.6327}{{\ttfamily arXiv:1210.6327 [hep-ex]}}.

\bibitem{Abe:2011fz}
{\bfseries Double Chooz} Collaboration, Y.~Abe {\em et al.}, ``{Indication for
  the reactor anti-neutrino disappearance in the Double Chooz experiment},''
  \href{http://dx.doi.org/10.1103/PhysRevLett.108.131801}{{\em Phys.Rev.Lett.}
  {\bfseries 108} (2012) 131801},
\href{http://arxiv.org/abs/1112.6353}{{\ttfamily arXiv:1112.6353 [hep-ex]}}.

\bibitem{Abe:2012tg}
{\bfseries Double Chooz} Collaboration, Y.~Abe {\em et al.}, ``{Reactor
  electron antineutrino disappearance in the Double Chooz experiment},''
  \href{http://dx.doi.org/10.1103/PhysRevD.86.052008}{{\em Phys.Rev.}
  {\bfseries D86} (2012) 052008},
\href{http://arxiv.org/abs/1207.6632}{{\ttfamily arXiv:1207.6632 [hep-ex]}}.

\bibitem{Abe:2013sxa}
{\bfseries Double Chooz} Collaboration, Y.~Abe {\em et al.}, ``{First
  Measurement of $\theta_{13}$ from Delayed Neutron Capture on Hydrogen in the
  Double Chooz Experiment},''
  \href{http://dx.doi.org/10.1016/j.physletb.2013.04.050}{{\em Phys.Lett.}
  {\bfseries B723} (2013) 66--70},
\href{http://arxiv.org/abs/1301.2948}{{\ttfamily arXiv:1301.2948 [hep-ex]}}.

\bibitem{Ahn:2010vy}
{\bfseries RENO} Collaboration, J.~Ahn {\em et al.}, ``{RENO: An Experiment for
  Neutrino Oscillation Parameter $\theta_{13}$ Using Reactor Neutrinos at
  Yonggwang},''
\href{http://arxiv.org/abs/1003.1391}{{\ttfamily arXiv:1003.1391 [hep-ex]}}.

\bibitem{Ahn:2012nd}
{\bfseries RENO} Collaboration, J.~Ahn {\em et al.}, ``{Observation of Reactor
  Electron Antineutrino Disappearance in the RENO Experiment},''
  \href{http://dx.doi.org/10.1103/PhysRevLett.108.191802}{{\em Phys.Rev.Lett.}
  {\bfseries 108} (2012) 191802},
\href{http://arxiv.org/abs/1204.0626}{{\ttfamily arXiv:1204.0626 [hep-ex]}}.

\bibitem{Abe:2008aa}
{\bfseries KamLAND} Collaboration, S.~Abe {\em et al.}, ``{Precision
  Measurement of Neutrino Oscillation Parameters with KamLAND},''
  \href{http://dx.doi.org/10.1103/PhysRevLett.100.221803}{{\em Phys.Rev.Lett.}
  {\bfseries 100} (2008) 221803},
\href{http://arxiv.org/abs/0801.4589}{{\ttfamily arXiv:0801.4589 [hep-ex]}}.

\bibitem{An:2015jdp}
{\bfseries JUNO} Collaboration, F.~An {\em et al.}, ``{Neutrino Physics with
  JUNO},'' \href{http://dx.doi.org/10.1088/0954-3899/43/3/030401}{{\em J.
  Phys.} {\bfseries G43} (2016) 030401},
\href{http://arxiv.org/abs/1507.05613}{{\ttfamily arXiv:1507.05613
  [physics.ins-det]}}.

\bibitem{Djurcic:2015vqa}
{\bfseries JUNO} Collaboration, Z.~Djurcic {\em et al.}, ``{JUNO Conceptual
  Design Report},''
\href{http://arxiv.org/abs/1508.07166}{{\ttfamily arXiv:1508.07166
  [physics.ins-det]}}.

\bibitem{Seo:2015yqp}
H.~Seo, ``{Status of RENO-50},''
{\em PoS} {\bfseries NEUTEL2015} (2015) 083.

\bibitem{Danilov:2014vra}
{\bfseries DANSS} Collaboration, M.~Danilov, ``{Sensitivity of DANSS detector
  to short range neutrino oscillations},''
\href{http://arxiv.org/abs/1412.0817}{{\ttfamily arXiv:1412.0817
  [physics.ins-det]}}.

\bibitem{Danilov:2013caa}
{\bfseries DANSS} Collaboration, M.~Danilov, ``{Sensitivity of the DANSS
  detector to short range neutrino oscillations},''
  \href{http://dx.doi.org/10.1016/j.nuclphysbps.2015.09.165}{{\em PoS}
  {\bfseries EPS-HEP2013} (2013) 493},
  \href{http://arxiv.org/abs/1311.2777}{{\ttfamily arXiv:1311.2777
  [physics.ins-det]}}.
[Nucl. Part. Phys. Proc.273-275,1055(2016)].

\bibitem{Serebrov:2016wzv}
A.~P. Serebrov {\em et al.}, ``{Neutrino-4 experiment on search for sterile
  neutrino with multi-section model of detector},''
\href{http://arxiv.org/abs/1605.05909}{{\ttfamily arXiv:1605.05909
  [physics.ins-det]}}.

\bibitem{Boireau:2015dda}
{\bfseries NUCIFER} Collaboration, G.~Boireau {\em et al.}, ``{Online
  Monitoring of the Osiris Reactor with the Nucifer Neutrino Detector},''
  \href{http://dx.doi.org/10.1103/PhysRevD.93.112006}{{\em Phys. Rev.}
  {\bfseries D93} no.~11, (2016) 112006},
\href{http://arxiv.org/abs/1509.05610}{{\ttfamily arXiv:1509.05610
  [physics.ins-det]}}.

\bibitem{Lane:2015alq}
{\bfseries NuLat} Collaboration, C.~Lane {\em et al.}, ``{A new type of
  Neutrino Detector for Sterile Neutrino Search at Nuclear Reactors and Nuclear
  Nonproliferation Applications},''
\href{http://arxiv.org/abs/1501.06935}{{\ttfamily arXiv:1501.06935
  [physics.ins-det]}}.

\bibitem{Derbin:2012kf}
A.~V. Derbin, A.~S. Kayunov, and V.~N. Muratova, ``{Search for Neutrino
  Oscillations at a Research Reactor},''
\href{http://arxiv.org/abs/1204.2449}{{\ttfamily arXiv:1204.2449 [hep-ph]}}.

\bibitem{Ashenfelter:2015uxt}
{\bfseries PROSPECT} Collaboration, J.~Ashenfelter {\em et al.}, ``{The
  PROSPECT Physics Program},''
  \href{http://dx.doi.org/10.1088/0954-3899/43/11/113001}{{\em J. Phys.}
  {\bfseries G43} no.~11, (2016) 113001},
\href{http://arxiv.org/abs/1512.02202}{{\ttfamily arXiv:1512.02202
  [physics.ins-det]}}.

\bibitem{SoLID:2013fta}
{\bfseries SoLID} Collaboration, S.~Collaboration,
``{SoLID Preliminary Conceptual Design Report: Solenoidal Large Intensity
  Device},''.

\bibitem{Chen:2014psa}
{\bfseries SoLID} Collaboration, J.~P. Chen, H.~Gao, T.~K. Hemmick, Z.~E.
  Meziani, and P.~A. Souder, ``{A White Paper on SoLID (Solenoidal Large
  Intensity Device)},''
\href{http://arxiv.org/abs/1409.7741}{{\ttfamily arXiv:1409.7741 [nucl-ex]}}.

\bibitem{Haser:2016xlb}
{\bfseries Stereo} Collaboration, J.~Haser, ``{Search for eV sterile neutrinos
  at a nuclear reactor - the Stereo project},''
\href{http://dx.doi.org/10.1088/1742-6596/718/6/062023}{{\em J. Phys. Conf.
  Ser.} {\bfseries 718} no.~6, (2016) 062023}.

\bibitem{Gaffiot:2015fva}
{\bfseries SOX} Collaboration, J.~Gaffiot, ``{The SOX experiment},''
\href{http://dx.doi.org/10.1016/j.nuclphysbps.2015.06.033}{{\em Nucl. Part.
  Phys. Proc.} {\bfseries 265-266} (2015) 129--131}.

\bibitem{Gavrin:2015aca}
V.~Gavrin {\em et al.}, ``{Current status of new SAGE project with $^{51}$Cr
  neutrino source},''
\href{http://dx.doi.org/10.1134/S1063779615020100}{{\em Phys. Part. Nucl.}
  {\bfseries 46} no.~2, (2015) 131--137}.

\bibitem{Barinov:2016znv}
V.~Barinov, V.~Gavrin, D.~Gorbunov, and T.~Ibragimova, ``{BEST sensitivity to
  O(1) eV sterile neutrino},''
  \href{http://dx.doi.org/10.1103/PhysRevD.93.073002}{{\em Phys. Rev.}
  {\bfseries D93} no.~7, (2016) 073002},
\href{http://arxiv.org/abs/1602.03826}{{\ttfamily arXiv:1602.03826 [hep-ph]}}.

\bibitem{Fukuda:1998mi}
{\bfseries Super-Kamiokande} Collaboration, Y.~Fukuda {\em et al.}, ``{Evidence
  for oscillation of atmospheric neutrinos},''
  \href{http://dx.doi.org/10.1103/PhysRevLett.81.1562}{{\em Phys. Rev. Lett.}
  {\bfseries 81} (1998) 1562--1567},
\href{http://arxiv.org/abs/hep-ex/9807003}{{\ttfamily arXiv:hep-ex/9807003
  [hep-ex]}}.

\bibitem{Richard:2015aua}
{\bfseries Super-Kamiokande} Collaboration, E.~Richard {\em et al.},
  ``{Measurements of the atmospheric neutrino flux by Super-Kamiokande: energy
  spectra, geomagnetic effects, and solar modulation},''
\href{http://arxiv.org/abs/1510.08127}{{\ttfamily arXiv:1510.08127 [hep-ex]}}.

\bibitem{Athar:2006yb}
{\bfseries INO} Collaboration, M.~Athar {\em et al.}, ``{India-based Neutrino
  Observatory: Project Report. Volume I.},''.
\url{http://www.imsc.res.in/~ino/OpenReports/INOReport.pdf}.

\bibitem{Devi:2014yaa}
M.~M. Devi, T.~Thakore, S.~K. Agarwalla, and A.~Dighe, ``{Enhancing sensitivity
  to neutrino parameters at INO combining muon and hadron information},''
  \href{http://dx.doi.org/10.1007/JHEP10(2014)189}{{\em JHEP} {\bfseries 10}
  (2014) 189},
\href{http://arxiv.org/abs/1406.3689}{{\ttfamily arXiv:1406.3689 [hep-ph]}}.

\bibitem{Indumathi:2015hfa}
{\bfseries INO} Collaboration, D.~Indumathi, ``{India-based neutrino
  observatory (INO): Physics reach and status report},''
\href{http://dx.doi.org/10.1063/1.4915571}{{\em AIP Conf. Proc.} {\bfseries
  1666} (2015) 100003}.

\bibitem{TheIceCube-Gen2:2016cap}
{\bfseries IceCube} Collaboration, M.~G. Aartsen {\em et al.}, ``{PINGU: A
  Vision for Neutrino and Particle Physics at the South Pole},''
\href{http://arxiv.org/abs/1607.02671}{{\ttfamily arXiv:1607.02671 [hep-ex]}}.

\bibitem{Fukuda:2001nj}
{\bfseries Super-Kamiokande} Collaboration, S.~Fukuda {\em et al.}, ``{Solar
  B-8 and hep neutrino measurements from 1258 days of Super-Kamiokande data},''
  \href{http://dx.doi.org/10.1103/PhysRevLett.86.5651}{{\em Phys. Rev. Lett.}
  {\bfseries 86} (2001) 5651--5655},
\href{http://arxiv.org/abs/hep-ex/0103032}{{\ttfamily arXiv:hep-ex/0103032
  [hep-ex]}}.

\bibitem{Fukuda:2001nk}
{\bfseries Super-Kamiokande} Collaboration, S.~Fukuda {\em et al.},
  ``{Constraints on neutrino oscillations using 1258 days of Super-Kamiokande
  solar neutrino data},''
  \href{http://dx.doi.org/10.1103/PhysRevLett.86.5656}{{\em Phys. Rev. Lett.}
  {\bfseries 86} (2001) 5656--5660},
\href{http://arxiv.org/abs/hep-ex/0103033}{{\ttfamily arXiv:hep-ex/0103033
  [hep-ex]}}.

\bibitem{Ahmad:2001an}
{\bfseries SNO} Collaboration, Q.~R. Ahmad {\em et al.}, ``Measurement of the
  charged current interactions produced by b-8 solar neutrinos at the sudbury
  neutrino observatory,''
  \href{http://dx.doi.org/10.1103/PhysRevLett.87.071301}{{\em Phys. Rev. Lett.}
  {\bfseries 87} (2001) 071301},
\href{http://arxiv.org/abs/nucl-ex/0106015}{{\ttfamily arXiv:nucl-ex/0106015}}.

\bibitem{Renshaw:2013dzu}
{\bfseries Super-Kamiokande} Collaboration, A.~Renshaw {\em et al.}, ``{First
  Indication of Terrestrial Matter Effects on Solar Neutrino Oscillation},''
  \href{http://dx.doi.org/10.1103/PhysRevLett.112.091805}{{\em Phys. Rev.
  Lett.} {\bfseries 112} no.~9, (2014) 091805},
\href{http://arxiv.org/abs/1312.5176}{{\ttfamily arXiv:1312.5176 [hep-ex]}}.

\bibitem{Friedland:2004pp}
A.~Friedland, C.~Lunardini, and C.~Pena-Garay, ``Solar neutrinos as probes of
  neutrino - matter interactions,'' {\em Phys. Lett.} {\bfseries B594} (2004)
  347,
\href{http://arxiv.org/abs/hep-ph/0402266}{{\ttfamily hep-ph/0402266}}.

\bibitem{deHolanda:2010am}
P.~C. de~Holanda and A.~{\relax Yu}. Smirnov, ``{Solar neutrino spectrum,
  sterile neutrinos and additional radiation in the Universe},''
  \href{http://dx.doi.org/10.1103/PhysRevD.83.113011}{{\em Phys. Rev.}
  {\bfseries D83} (2011) 113011},
\href{http://arxiv.org/abs/1012.5627}{{\ttfamily arXiv:1012.5627 [hep-ph]}}.

\bibitem{GonzalezGarcia:2007ib}
M.~C. Gonzalez-Garcia and M.~Maltoni, ``{Phenomenology with Massive
  Neutrinos},'' \href{http://dx.doi.org/10.1016/j.physrep.2007.12.004}{{\em
  Phys. Rept.} {\bfseries 460} (2008) 1--129},
\href{http://arxiv.org/abs/0704.1800}{{\ttfamily arXiv:0704.1800 [hep-ph]}}.

\bibitem{Osipowicz:2001sq}
{\bfseries KATRIN} Collaboration, A.~Osipowicz {\em et al.}, ``{KATRIN: A next
  generation tritium beta decay experiment with sub-eV sensitivity for the
  electron neutrino mass},''
  \href{http://arxiv.org/abs/hep-ex/0109033}{{\ttfamily arXiv:hep-ex/0109033}}.
arXiv:hep-ex/0109033.

\bibitem{Mertens:2015ila}
{\bfseries KATRIN} Collaboration, S.~Mertens, ``{Status of the KATRIN
  Experiment and Prospects to Search for keV-mass Sterile Neutrinos in Tritium
  $\beta$-decay},''
\href{http://dx.doi.org/10.1016/j.phpro.2014.12.043}{{\em Phys. Procedia}
  {\bfseries 61} (2015) 267--273}.

\bibitem{Monreal:2009za}
B.~Monreal and J.~A. Formaggio, ``{Relativistic Cyclotron Radiation Detection
  of Tritium Decay Electrons as a New Technique for Measuring the Neutrino
  Mass},'' \href{http://dx.doi.org/10.1103/PhysRevD.80.051301}{{\em Phys. Rev.}
  {\bfseries D80} (2009) 051301},
\href{http://arxiv.org/abs/0904.2860}{{\ttfamily arXiv:0904.2860 [nucl-ex]}}.

\bibitem{Oblath:2015pxa}
{\bfseries Project 8} Collaboration, N.~S. Oblath, ``{Project 8: First Results
  \& More},''
{\em PoS} {\bfseries NEUTEL2015} (2015) 046.

\bibitem{Inoue:2016jbg}
{\bfseries POLARBEAR} Collaboration, Y.~Inoue {\em et al.}, ``{POLARBEAR-2: an
  instrument for CMB polarization measurements},''
  \href{http://dx.doi.org/10.1117/12.2231961}{{\em Proc. SPIE Int. Soc. Opt.
  Eng.} {\bfseries 9914} (2016) 99141I},
\href{http://arxiv.org/abs/1608.03025}{{\ttfamily arXiv:1608.03025
  [astro-ph.IM]}}.

\bibitem{Hanson:2013hsb}
{\bfseries SPTpol} Collaboration, D.~Hanson {\em et al.}, ``{Detection of
  B-mode Polarization in the Cosmic Microwave Background with Data from the
  South Pole Telescope},''
  \href{http://dx.doi.org/10.1103/PhysRevLett.111.141301}{{\em Phys. Rev.
  Lett.} {\bfseries 111} no.~14, (2013) 141301},
\href{http://arxiv.org/abs/1307.5830}{{\ttfamily arXiv:1307.5830
  [astro-ph.CO]}}.

\bibitem{Ahmed:2014ixy}
{\bfseries BICEP3} Collaboration, Z.~Ahmed {\em et al.}, ``{BICEP3: a 95GHz
  refracting telescope for degree-scale CMB polarization},''
  \href{http://dx.doi.org/10.1117/12.2057224}{{\em Proc. SPIE Int. Soc. Opt.
  Eng.} {\bfseries 9153} (2014) 91531N},
\href{http://arxiv.org/abs/1407.5928}{{\ttfamily arXiv:1407.5928
  [astro-ph.IM]}}.

\bibitem{Aghamousa:2016zmz}
{\bfseries DESI} Collaboration, A.~Aghamousa {\em et al.}, ``{The DESI
  Experiment Part I: Science,Targeting, and Survey Design},''
\href{http://arxiv.org/abs/1611.00036}{{\ttfamily arXiv:1611.00036
  [astro-ph.IM]}}.

\bibitem{Aghamousa:2016sne}
{\bfseries DESI} Collaboration, A.~Aghamousa {\em et al.}, ``{The DESI
  Experiment Part II: Instrument Design},''
\href{http://arxiv.org/abs/1611.00037}{{\ttfamily arXiv:1611.00037
  [astro-ph.IM]}}.

\bibitem{Tereno:2015hja}
{\bfseries Euclid} Collaboration, I.~Tereno {\em et al.}, ``{Euclid Space
  Mission: building the sky survey},''
  \href{http://dx.doi.org/10.1017/S174392131401093X}{{\em IAU Symp.} {\bfseries
  306} (2015) 379--381},
\href{http://arxiv.org/abs/1502.00903}{{\ttfamily arXiv:1502.00903
  [astro-ph.IM]}}.

\bibitem{Abate:2012za}
{\bfseries LSST Dark Energy Science} Collaboration, A.~Abate {\em et al.},
  ``{Large Synoptic Survey Telescope: Dark Energy Science Collaboration},''
\href{http://arxiv.org/abs/1211.0310}{{\ttfamily arXiv:1211.0310
  [astro-ph.CO]}}.

\end{thebibliography}\endgroup

\cleardoublepage
\appendix
\setcounter{nuPanel-RM-Conc-Sect}{0}

\cleardoublepage
\section{The ICFA Neutrino Panel}
\label{App:ICFA-nu-Panel}

ICFA established the Neutrino Panel with the mandate
\cite{ICFAnuPanel:Mandate:2013}: 
\begin{quote}
  {\it To promote international cooperation in the development of the
    accelerator-based neutrino-oscillation program and to promote
    international collaboration in the development a neutrino factory
    as a future intense source of neutrinos for particle physics
    experiments.
  }
\end{quote}
The membership of the Panel agreed by ICFA at its meeting in February
2013 is shown in table \ref{Tab:PanelMembers}.
The terms of reference for the panel \cite{ICFAnuPanel:ToR:2013} may
be found on the Panel's WWW site \cite{ICFA:nuPanelWWWSite}.
\begin{table}[h]
  \caption{Membership of the ICFA Neutrino Panel.}
  \label{Tab:PanelMembers}
  \begin{center}
    \begin{tabular}{|l|l|}
      \hline
      {\bf Name}      & {\bf Institution}                      \\
      \hline
      J. Cao          & IHEP/Beijing                           \\
      A. de Gouv\^ea  & Northwestern University                \\
      D. Duchesneau   & CNRS/IN2P3                             \\
      S. Geer         & Fermi National Laboratory              \\
      R. Gomes        & Federal University of Goias            \\
      S.B. Kim        & Seoul National University              \\
      T. Kobayashi    & KEK                                    \\
      K. Long (chair) & Imperial College London and STFC       \\
      M. Maltoni      & Universidad Automata Madrid            \\
      M. Mezzetto     & University of Padova                   \\
      N. Mondal       & Tata Institute for Fundamental Resarch \\
      M. Shiozawa     & Tokyo University                       \\
      J. Sobczyk      & Wroclaw University                     \\
      H. A. Tanaka    & University of Toronto, IPP, TRIUMF     \\
      M. Wascko       & Imperial College London                \\
      G. Zeller       & Fermi National Accelerator Laboratory  \\
      \hline
    \end{tabular}
  \end{center}
\end{table}

\end{document}